\documentclass[11pt,a4paper]{article}

\usepackage{amsmath,latexsym,amssymb}
\usepackage{amsthm,etoolbox}
\usepackage{breqn}
\usepackage{makecell}
\usepackage{color}
\usepackage{bm,fancybox,multirow,hhline,indentfirst,bmpsize}
\usepackage{enumitem,comment}
\usepackage[hang,small,bf]{caption}
\usepackage[subrefformat=parens]{subcaption}
\usepackage{ascmac,atbegshi}
\usepackage{natbib,verbatim,here}
\usepackage{booktabs,url}
\usepackage[colorlinks=true,allcolors=blue]{hyperref}
\newcommand{\doi}[1]{\href{https://doi.org/#1}{doi:#1}}

\makeatletter
\def\section{\@startsection {section}{1}{\z@}{-3.5ex plus -1ex minus -.2ex}{2.3 ex plus .2ex}{\large\bf}}
\def\subsection{\@startsection {subsection}{1}{\z@}{-3.5ex plus -1ex minus -.2ex}{2.3 ex plus .2ex}{\large\it}}
\@ifundefined{bibcommenthead}{}{}
\makeatother

\theoremstyle{definition}
\newtheorem{lemma}{Lemma}
\newtheorem{theorem}{Theorem}
\theoremstyle{definition}
\newcommand{\argmin}{\mathop{\rm argmin}}

\selectfont

\allowdisplaybreaks

\def\T{{\rm T}}
\def\E{{\rm E}}

\def\N{{\rm N}}
\def\O{{\rm O}}
\def\o{{\rm o}}
\def\d{{\rm d}}

\def\sgn{{\rm sgn}}
\def\tr{{\rm tr}}
\def\oP{{\rm o}_{\rm P}}
\def\OP{{\rm O}_{\rm P}}

\title{\large\bf Developing an information criterion for spatial data analysis through Bayesian generalized fused lasso}
\author{\normalsize Yuko Kakikawa\\\small Statistical Science Program
Graduate Institute for Advanced Studies, SOKENDAI\bigskip\\\normalsize Yoshiyuki Ninomiya\\\small Department of Fundamental Statistical Mathematics, The Institute of Statistical Mathematics\\\small Statistical Science Program
Graduate Institute for Advanced Studies, SOKENDAI}
\date{}

\begin{document}

\maketitle

\begin{abstract}
In the field of spatial data analysis, spatially varying coefficients (SVC) models, which allow regression coefficients to vary by region and flexibly capture spatial heterogeneity, have continued to be developed in various directions.
Moreover, the Bayesian generalized fused lasso is often used as a method that efficiently provides estimation under the natural assumption that regression coefficients of adjacent regions tend to take the same value.
In most Bayesian methods, the selection of prior distribution is an essential issue, and in the setting of SVC model with the Bayesian generalized fused lasso, determining the complexity of the class of prior distributions is also a challenging aspect, further amplifying the difficulty of the problem.
For example, the widely applicable information criterion (WAIC), which has become standard in Bayesian model selection, does not target determining the complexity.
Therefore, in this study, we adapted a criterion called the prior intensified information criterion (PIIC) to this setting.
Specifically, under an asymptotic setting that retains the influence of the prior distribution, that is, under an asymptotic setting that deliberately does not provide selection consistency, we derived the asymptotic properties of our generalized fused lasso estimator. 
Then, based on these properties, we constructed an information criterion as an asymptotically bias-corrected estimator of predictive risk.
In numerical experiments, we confirmed that PIIC outperforms WAIC in the sense of reducing the predictive risk, and in a real data analysis, we observed that the two criteria give rise to substantially different results.

\medskip


\medskip

\end{abstract}

\section{Introduction}\label{sec:Introduction}

With the development of information technology and the spread of geographic information systems (GIS), spatial data are being observed and stored every day, being utilized in many fields, including environmental science, epidemiology, ecology, and economics.
Spatial data contain positional information, and it is natural to impose similar structures on locations that are geographically close; however, making structures overly similar deprives a model of flexibility.
Therefore, in recent years, it has become common to perform modeling that incorporates appropriate spatial heterogeneity.
Within this trend, the first approach to attract attention was geographically weighted regression (GWR) (\citealt{brunsdon1996geographically}), which performs local ordinary least squares regression using weights given by kernel functions.
However, since this approach considers a different local regression model for each observation point, it has problems in inference in cases of, for example, tests and model selection using the whole dataset.
In this context, \cite{gelfand2003spatial} proposed the spatially varying coefficients (SVC) model, which inherits from GWR the assumption that regression coefficients vary smoothly with the region of the data, and further assumes that these coefficients follow a Gaussian process.
In this paper, the term SVC models refers to not only models using Gaussian processes but also probability models in which regression coefficients vary smoothly according to location.
This SVC model has been developed in various directions; for example, \cite{fahrmeir2011bayesian}, \cite{umlauf2018bamlss}, and \cite{kim2021generalized} proposed methods that capture regression coefficients with splines.
Additionally, for more computationally efficiency than under the assumption of Gaussian processes, as well as the ability to deal with spatial discontinuities and boundary effects to some extent, \cite{assunccao2003space} and \cite{reich2008modeling} assumed Markov random fields.

The abovementioned SVC models are appropriate for settings in which regression coefficients vary continuously. In practice, however, data frequently show that coefficients remain essentially constant, with occasional discontinuous changes, for example when heterogeneity arises across administrative districts.
In such cases, it is natural to represent discontinuous structures with neighboring regions grouped together, letting regression coefficients be common within a group, and allowing variation between groups (\citealt{lawson2000cluster}; \citealt{huang2012mixture}; \citealt{sugasawa2021spatially}).
Moreover, as a method that does not increase computational burden so much even as the number of groups increases, methods using the generalized fused lasso regularization method (\citealt{Hoefling2010path}) have been developed (\citealt{zhao2020solution}; \citealt{li2019spatial}; \citealt{zhong2023sparse}).
In the present paper, based on this method, just as \cite{park2008bayesian} regarded the lasso as a Bayesian method with the Laplace distribution as the prior distribution and called it the Bayesian lasso, we regard the generalized fused lasso as the Bayesian generalized fused lasso and consider how to provide an estimated model closer to the true structure through appropriate selection of the prior distribution.

The standard tool for appropriately selecting prior distributions is an information criterion that is suitable for a Bayesian approach.
Here, putting emphasis on providing models with high predictive performance, we focus on the widely applicable information criterion (WAIC; \citealt{watanabe2010asymptotic}), not the Bayes factor.
The breakthrough in this branch of information criteria was the deviance information criterion (DIC; \citealt{spiegelhalter2002bayesian}); however, WAIC more directly evaluates the Kullback-Leibler divergence between the Bayesian predictive distribution and the true distribution. For this reason, it is included in standard Bayesian textbooks such as \cite{gelman2013bayesian}, and although the mathematical technique for its derivation is difficult, it has come to be the natural choice in applied fields.
In the present setting as well, if it is already decided which distribution to use as the prior and the only problem left is what values the hyperparameters should take, then it should be reasonable to use WAIC.
On the other hand, since WAIC does not have a penalty term for the complexity of the prior distribution, using WAIC to decide which class of prior distribution to use might not be appropriate.
The latter is because WAIC will ultimately select the class of prior distributions that has the highest complexity.

When using the generalized fused lasso for estimation of SVC models, if one assumes a common Laplace distribution for the regression coefficients of all variables, the above concern about selecting the class of prior distributions is not so serious.
However, it is natural to suppose that the variation in regression coefficients differs by variable, so assuming a common Laplace distribution might be inappropriate.
But since assuming a different Laplace distribution for each variable might cause overfitting, it is natural to consider making an adaptive determination from the data about how many Laplace distributions should be used in the prior distribution.
Needless to say, the larger the number of Laplace distributions used, the higher the complexity of the prior distribution.
With the aim of also addressing this problem, \cite{ninomiya2021prior} developed the prior intensified information criterion (PIIC).
WAIC is based on asymptotics in which the influence of prior distributions becomes negligible; that is, WAIC was derived for settings where Bayesian estimators are considered but are closely similar to maximum likelihood estimators. However, since the data size is finite in practice, WAIC might not reflect the properties of Bayesian estimators, which are inevitably influenced by the prior distributions.
Resolving this concern is another purpose of PIIC, so its derivation was obtained for when the order of the logarithm of the prior distribution is $\O(n)$. In addition, a penalty term for the complexity of the prior distributions was also derived.
However, \cite{ninomiya2021prior} only treated the Bayesian lasso experimentally, and verification of whether it is really useful in applied analyses has yet to be made.
In the present paper, with spatial data analysis using SVC models in view, we adapt PIIC for the Bayesian generalized fused lasso.

To adapt PIIC for this purpose, it is necessary to derive the asymptotic properties of generalized fused lasso estimators for when the order of the logarithm of the prior distributions is $\O(n)$.
Here, we modify the approach of \cite{viallon2016robustness}, which derived the selection consistency of generalized fused lasso estimators for when regularization terms are of order $\o(\sqrt{n})$, to fit the $\O(n)$ case.
As a result, although selection consistency does not hold, we can demonstrate fast convergence to zero for the estimators of the regression coefficients in the non-active set and the asymptotic normality for the estimators of the regression coefficients in the active set.
Similar asymptotic properties for lasso estimators were obtained in \cite{ninomiya2016aic}. 
These properties were used for the derivation of AIC for lasso, and later, in \cite{ninomiya2021prior}, for the derivation of PIIC.
The approach in that prior study might also be applicable to generalized fused lasso estimators if one used parameter transformations, which involve overly cumbersome notation, but here we avoid parameter transformations by basing our work on the approach of \cite{viallon2016robustness}.
Thus, in this study, we derived a new version of PIIC without getting involved with AIC.

The remainder of this paper is organized as follows.
First, Section \ref{sec2} introduces SVC models, the generalized fused lasso, and the Bayesian generalized fused lasso, and derives the asymptotic properties of generalized fused lasso estimators deliberately constructed without consistency.
Section \ref{sec3} presents WAIC and then adapts PIIC to SVC models using the Bayesian generalized fused lasso.
Section \ref{sec4} reports numerical experiments, showing that PIIC is superior in the sense of reducing the risk targeted by WAIC and PIIC.
Section \ref{sec5} presents a real data analysis confirming that the two criteria clearly differ, and Section \ref{sec6} summarizes our conclusions.

\section{Model specification and derivation of estimator properties}\label{sec2}

In this section, we consider SVC models, in a broad sense, to be those probability models in which regression coefficients vary smoothly across the regions to which the data are associated with, and present them as representative models used in spatial data analysis.
Then, as the estimation method for SVC models, we discuss the generalized fused lasso regularization method of \cite{Hoefling2010path}. With the construction of a model selection criterion in mind, we derive the asymptotic properties of estimators under a setting different from that of \cite{viallon2016robustness}.
In addition, since the purpose of this paper is the development of Bayesian methods, including the construction of a criterion for them, we also provide a Bayesian interpretation of the generalized fused lasso regularization method.

\subsection{Spatially varying coefficients model}\label{sec:2-1}
For the $i\ (\in{1,\ldots,n})$-th sample, suppose that the response variable $y_i\ (\in\mathbb{R})$, the explanatory variable vector $\tilde{\bm{x}}_i=(\tilde{x}_{i,1},\ldots,\allowbreak\tilde{x}_{i,\tilde{p}})^{\T}\ (\in\mathbb{R}^{\tilde{p}})$, and the indicator variable $\psi_i\ (\in{1,\ldots,M})$ representing which region the sample is associated with are observed, and consider the following SVC model:
\begin{align}
y_{i}=\sum_{m=1}^{M}I(\psi_{i}=m)\tilde{\bm{x}}_{i}^{\T}\bm{\theta}_m+\varepsilon_{i},
\label{eq:svcmodel}
\end{align}
where $\bm{\theta}_m=(\theta_{m,1},\ldots,\theta_{m,\tilde{p}})^{\T}\ (\in\mathbb{R}^{\tilde{p}})$ is the regression coefficient vector for the $m$-th region.
For simplicity, suppose that $\{(\psi_i,\tilde{\bm{x}}_i,\varepsilon_i)\mid i\in\{1,\ldots,n\}\}$ are independent and identically distributed, and that the distribution of $(\psi_i,\tilde{\bm{x}}_i)$ is known.
If the explanatory variables need to be non-random, it suffices to impose conditions on them as in ordinary regression analysis (for example, in the case of the framework for constructing an information criterion for sparse estimation, see \citealt{ninomiya2016aic}). 
Let the error term $\varepsilon_{i}$ follow ${\rm N}(0,\sigma^{2})$ independently of $(\psi_i,\tilde{\bm{x}}_i)$.
The error variance $\sigma^2$ is assumed to be unknown, but since deriving the model selection criterion, which is the main theme of this paper, under the assumption that $\sigma^2$ is unknown causes unnecessary complications, we derive the criterion assuming the error variance is known and then later substitute in an estimator.
Of course, if the variance structure is also made spatio-temporal, the situation becomes different, but we do not deal with that in this paper.
In this setting, letting $\bm{e}_m$ be the $M$-dimensional unit vector whose $m$-th component is $1$, and defining $\bm{x}_i$ as $\bm{e}_{\psi_i}\otimes\tilde{\bm{x}}_i$ and $\bm{\theta}$ as $(\bm{\theta}_1^{\T},\ldots,\bm{\theta}_M^{\T})^{\T}$, we express \eqref{eq:svcmodel} simply as $y_{i}=\bm{x}_{i}^{\T}\bm{\theta}+\varepsilon_{i}$.
Here, $\{\bm{x}_i\mid i\in\{1,\ldots,n\}\}$ are independent and identically distributed.

In this model, we write the joint probability density function of $(y_{i},\bm{x}_i)$, $(2\pi\sigma^2)^{-1/2}\exp\{-(y_i-\bm{x}_i^{\T}\bm{\theta})^2/(2\sigma^2)\}$, as $f(y_{i},\bm{x}_{i}\mid\bm{\theta})$.
In addition, as conditions for constructing the asymptotic theory of sparse estimation for this model, we assume the following.

\begin{itemize}[label=(C1)\ , leftmargin=*]
\item 
The parameter space $\Theta$ of $\bm{\theta}$ is compact, and $\bm{x}_i$ has moments up to the fourth order.
In particular, $\bm{J}\equiv\E[\bm{x}_{i}\bm{x}_{i}^{\T}]$ is a positive definite matrix.
\end{itemize}

\subsection{Generalized fused lasso}\label{sec:2-2}

In the SVC model, regression coefficients corresponding to neighboring regions are set to similar values.
The generalized fused lasso (\citealt{Hoefling2010path}) is a method that efficiently performs estimation in which neighboring regression coefficients tend to have the same value.
Here, when region $m^{\dagger}$ and $m^{\ddagger}$ are adjacent, we designate $(m^{\dagger},m^{\ddagger})$ as an edge, and denote the set of all edges by $\mathcal{E}$.
When defining the graph $(\mathcal{V},\mathcal{E})$ with vertex set $\mathcal{V}=\{1,\ldots,M\}$, the generalized fused lasso is expressed as the following optimization problem:
\begin{align}
& \hat{\bm{\theta}} = (\hat{\theta}_{1,1},\ldots,\hat{\theta}_{1,\tilde{p}},\hat{\theta}_{2,1},\ldots,\hat{\theta}_{2,\tilde{p}},\ldots,\hat{\theta}_{M,1},\ldots,\hat{\theta}_{M,\tilde{p}})
\notag \\
& \equiv \argmin_{\bm{\theta}\in\Theta}\Bigg\{-\sum_{i=1}^n \log f(y_i,\bm{x}_i\mid\bm{\theta})+n \sum_{j=1}^{\tilde{p}}\lambda_{1,j}\sum_{m\in\mathcal{V}}|\theta_{m,j}|+n \sum_{j=1}^{\tilde{p}}\lambda_{2,j}\sum_{(m^{\dagger},m^{\ddagger})\in\mathcal{E}}|\theta_{m^{\dagger},j}-\theta_{m^{\ddagger},j}|\Bigg\},
\label{eq:generalized fused lasso_theta}
\end{align}
where $\lambda_{1,j}\ (j\in\{1,\ldots,\tilde{p}\})$ and $\lambda_{2,j}\ (j\in\{1,\ldots,\tilde{p}\})$ are regularization parameters.
This optimization problem includes a penalty term $\lambda_{2,j}|\theta_{m^{\dagger},j}-\theta_{m^{\ddagger},j}|$ such that the difference between the regression coefficients $\theta_{m^{\dagger},j}$ and $\theta_{m^{\ddagger},j}$ corresponding to neighboring regions $(m^{\dagger},m^{\ddagger})\in\mathcal{E}$ is likely to be estimated as $0$.
It also includes a penalty term $\lambda_{1,j}|\theta_{m,j}|$ such that all regression coefficients $\theta_{m,j}$ are likely to be estimated as $0$, and variable selection is performed simultaneously.

Usually, the regularization term of the generalized fused lasso is given without an $n$. 
However, for example, in constructing an asymptotic theory for providing an information criterion, it is unreasonable to assume an asymptotic setting in which the influence of the regularization term is small so that selection consistency is derived (see \citealt{ninomiya2016aic}). 
Therefore, a regularization term with order $\O(n)$ is considered.
As a preparatory result for the later lemma and theorems, the following are derived from (C1).

\begin{itemize}[label=(R1)\ , leftmargin=*]
\item
For each $\bm{\theta}$, there exists a convex and differentiable function $h(\bm{\theta})$ such that $-\sum_{i=1}^n\log \allowbreak f(y_i,\bm{x}_i\mid\bm{\theta})/n \xrightarrow{\rm p} h(\bm{\theta})$.
\end{itemize}
\begin{itemize}[label=(R2)\ , leftmargin=*]
\item
For $h(\bm{\theta})$ given in (R1),
$\sum_{i=1}^n {\rm E}[-\partial\log f(y_i,\bm{x}_i\mid\bm{\theta})/\partial\bm{\theta}]/n-\partial h(\bm{\theta}) / \partial \bm{\theta} = \bm{0}$.
\end{itemize}
\begin{itemize}[label=(R3)\ , leftmargin=*]
\item
$\sum_{i=1}^n(\partial\log f(y_i,\bm{x}_i\mid\bm{\theta})/\partial\bm{\theta}-\E[\partial\log f(y_i,\bm{x}_i\mid\bm{\theta})/\partial\bm{\theta}]) / \sqrt{n} \xrightarrow{\rm d} \N(\bm{0}, \bm{J})$.
\end{itemize}
The proofs are easy; they can be obtained, for example, by simplifying the proofs in \cite{ninomiya2016aic}.

Next, we investigate the asymptotic properties of the generalized fused lasso estimator $\hat{\bm{\theta}}$.
For simplicity of the following discussion, we set $\lambda_{1,1}=\ldots=\lambda_{1,\tilde{p}}=\lambda_1$ and $\lambda_{2,1}=\ldots=\lambda_{2,\tilde{p}}=\lambda_2$.
To avoid notational complexity, we newly denote $(\theta_{1,1},\ldots,\theta_{1,\tilde{p}},\theta_{2,1},\ldots, \allowbreak \theta_{2,\tilde{p}},\ldots,\theta_{M,1},\allowbreak\ldots,\theta_{M,\tilde{p}})$ as $(\xi_{1},\ldots,\xi_{p})$, and accordingly change the notation of the parameter space from $\Theta$ to $\Xi$.
In addition, when $j^{\dagger}>j^{\ddagger}$ and $\xi_{j^{\dagger}}$ and $\xi_{j^{\ddagger}}$ are coefficient parameters corresponding to the same variable and the corresponding regions are adjacent, that is, when $|\xi_{j^{\dagger}}-\xi_{j^{\ddagger}}|$ appears in the penalty term in \eqref{eq:generalized fused lasso_theta}, the pair of indices $(j^{\dagger},j^{\ddagger})$ is regarded as an edge, and we denote the set of all edges by $E$.
When defining graph $(V,E)$ with vertex set $V=\{1,\ldots,p\}$, \eqref{eq:generalized fused lasso_theta} is rewritten as 
\begin{align}
\hat{\bm{\xi}} = (\hat{\xi}_1,\ldots,\hat{\xi}_p) \equiv \argmin_{\bm{\xi}\in\Xi}\Bigg\{-\sum_{i=1}^n\log f(y_i,\bm{x}_i\mid\bm{\xi})+n\lambda_1\sum_{j\in V}|\xi_{j}|+n\lambda_2\sum_{(j^{\dagger},j^{\ddagger})\in E}|\xi_{j^{\dagger}}-\xi_{j^{\ddagger}}|\Bigg\}.
\label{eq:generalized fused lasso}
\end{align}
Since this estimator depends on $\bm{\lambda}=(\lambda_1,\lambda_2)$, we could write it as $\hat{\bm{\xi}}_{\bm{\lambda}}$, but for simplicity, we will write $\hat{\bm{\xi}}$ until it becomes essential to emphasize the dependence on $\bm{\lambda}$.
Regarding the first-order asymptotics, the following lemma is easily obtained, where $\bm{\xi}^*=(\xi_1^*,\ldots,\xi_p^*)$ is the minimizer of $-\E[\log f(y,\bm{x}\mid\bm{\xi})]+\lambda_1\sum_{j\in V}|\xi_{j}|+n\lambda_2\sum_{(j^{\dagger},j^{\ddagger})\in E}|\xi_{j^{\dagger}}-\xi_{j^{\ddagger}}|$ with respect to $\bm{\xi}$.

\begin{lemma}
\label{lemma1}
Under condition (C1), the generalized fused lasso estimator $\hat{\bm{\xi}}$ converges in probability to $\bm{\xi}^*$.
\end{lemma}

\noindent
For the proof, see Appendix \ref{lemma1proof}.

Next, we consider the second-order asymptotics.
Let $\mathcal{J}^{(1)}=\{j:\xi_{j}^*=0\}$ and $\mathcal{J}^{(2)}=\{j:\xi_j^*\neq 0\}$, and consider the set of connected components in the graph $(V^*,E^*)$ consisting of the vertex set $V^*\equiv\mathcal{J}^{(2)}$ and the edge set $E^*\equiv\{(j^{\dagger},j^{\ddagger})\in E: \xi_{j^{\dagger}}^*=\xi_{j^{\ddagger}}^*\}$.
An isolated point not connected to any edge is regarded as a one-connected component.
Then, from each connected component whose vertex values are nonzero, one representative vertex is arbitrarily chosen, and the set of all representatives is denoted as $\mathcal{J}^{(3)}\ (\subset\mathcal{J}^{(2)})$.
That is, if $j\in\mathcal{J}^{(2)}$, then $\exists j^{\dagger}\in\mathcal{J}^{(3)};\ \xi_j^*=\xi_{j^{\dagger}}^*$, and furthermore, if $j^{\ddagger}\neq j^{\dagger}$, then $\xi_j^*\neq\xi_{j^{\ddagger}}^*$.
For any vector $\bm{c}=(c_j)_{j\in\{1,\ldots,p\}}$, let $\bm{c}^{(\ell)}$ denote the vector $(c_j)_{j\in\mathcal{J}^{(\ell)}}$ ($\ell\in\{1,2,3\}$), and for any matrix $\bm{C}=(C_{j^{\dagger}j^{\ddagger}})_{j^{\dagger}\in\{1,\ldots,p\},j^{\ddagger}\in\{1,\ldots,p\}}$, let $\bm{C}^{(\ell^{\dagger}\ell^{\ddagger})}$ denote the matrix $(C_{j^{\dagger}j^{\ddagger}})_{j^{\dagger}\in\mathcal{J}^{(\ell^{\dagger})},j^{\ddagger}\in\mathcal{J}^{(\ell^{\ddagger})}}$ ($\ell^{\dagger},\ell^{\ddagger}\in\{1,2,3\}$).
In addition, we define
\begin{align}
\bm{A} \equiv (I(\xi_{j^{\dagger}}^*=\xi_{j^{\ddagger}}^*))_{j^{\dagger}\in\{1,\ldots,p\},j^{\ddagger}\in\{1,\ldots,p\}},
\label{defA}
\end{align}
which is a matrix whose components are all either $0$ or $1$, and we will use $\bm{A}^{(32)}$ and $\bm{A}^{(23)}$ later.
Here, in addition to (C1), we assume the following condition.
\begin{itemize}[label=(C2)\ , leftmargin=*]
\item
$\bm{\xi}^*$ exists in the interior of $\Xi$.
In addition, when $\bm{\xi}^*$ gives a sparse solution, that is, when $\xi_j^*$ becomes $0$ or $\xi_{j^{\dagger}}^*-\xi_{j^{\ddagger}}^*$ becomes $0$, then $\bm{\lambda}=(\lambda_1,\lambda_2)$ is defined so that $\bm{\xi}^*$ is not the solution of the minimization problem with the absolute value operator removed, such as $\argmin_{\bm{\xi}\in\Xi}\{h(\bm{\xi})+\lambda_1\sum_{j\in V}(\pm\xi_j)+\lambda_2\sum_{(j^{\dagger},j^{\ddagger})\in E}|\xi_{j^{\dagger}}-\xi_{j^{\ddagger}}|\}$ or $\argmin_{\bm{\xi}\in\Xi}[h(\bm{\xi})+\lambda_1\sum_{j\in V}|\xi_j|+\lambda_2\sum_{(j^{\dagger},j^{\ddagger})\in E}\{\pm(\xi_{j^{\dagger}}-\xi_{j^{\ddagger}})\}]$.
\end{itemize}
The objective function minimized by $\bm{\xi}^*$ is convex with respect to $\bm{\xi}$, and therefore if $\Xi$ is sufficiently large, the first sentence of condition (C2) can be expected to hold.
Furthermore, since the second sentence of condition (C2) is satisfied for almost all $\bm{\lambda}=(\lambda_1,\lambda_2)$, it is only a technical condition, and we do not consider it when actually applying the method to data.
We now are prepared to prove the following asymptotic property.

\begin{theorem}
Under conditions (C1) and (C2), for the generalized fused lasso estimator $\hat{\bm{\xi}}$, it holds that
\begin{align*}
& n(\hat{\bm{\xi}}^{(1)}-\bm{\xi}^{*(1)}) \xrightarrow{\rm p} \bm{0},
\\
& j^{\dagger}\in\mathcal{J}^{(2)} \quad \Rightarrow \quad \exists j^{\ddagger}\in\mathcal{J}^{(3)};\ \sqrt{n}(\hat{\xi}_{j^{\dagger}}-\hat{\xi}_{j^{\ddagger}})\xrightarrow{\rm p}0,
\\
& \sqrt{n}(\hat{\bm{\xi}}^{(3)}-\bm{\xi}^{*(3)}) \xrightarrow{\rm d}\N(\bm{0},(\bm{A}^{(32)}\bm{J}^{(22)}\bm{A}^{(23)})^{-1}).
\end{align*}
\label{th1}
\end{theorem}

\noindent
For the proof, see Appendix \ref{theorem1proof}.

\subsection{Bayesian generalized fused lasso}\label{sec:2-3}

\cite{park2008bayesian} regarded the lasso regularization method as a Bayesian method with the Laplace distribution as the prior distribution, and the generalized fused lasso regularization method can be regarded in the same way.
Specifically, we assume the Laplace distribution with the location parameter $0$ and the scale parameter $1/\lambda_{1,j}$ or $1/\lambda_{2,j}$ for the regression coefficients and the differences between regression coefficients as
\begin{align}
\pi_{\rm gfl}(\bm{\theta};\bm{\lambda}) \propto \exp\Bigg(-n\sum_{j=1}^{\tilde{p}}\lambda_{1,j}\sum_{m\in\mathcal{V}}|\theta_{m,j}|-n\sum_{j=1}^{\tilde{p}}\lambda_{2,j}\sum_{(m^{\dagger},m^{\ddagger})\in\mathcal{E}}|\theta_{m^{\dagger},j}-\theta_{m^{\ddagger},j}|\Bigg).
\label{prior1}
\end{align}
As in Section \ref{sec:2-2}, when we assume the setting $\lambda_{1,1}=\ldots=\lambda_{1,\tilde{p}}=\lambda_1$ and $\lambda_{2,1}=\ldots=\lambda_{2,\tilde{p}}=\lambda_2$, and if we newly express $\bm{\theta}=(\theta_{1,1},\ldots,\theta_{1,\tilde{p}},\theta_{2,1},\ldots, \theta_{2,\tilde{p}},\ldots,\theta_{M,1},\allowbreak\ldots,\theta_{M,\tilde{p}})$ as $\bm{\xi}=(\xi_{1},\ldots,\xi_{p})$, then this prior distribution can be written as
\begin{align}
\pi_{\rm gfl}(\bm{\xi};\bm{\lambda}) \propto \exp\Bigg(-n\lambda_1\sum_{j\in V}|\xi_j|-n\lambda_2\sum_{(j^{\dagger},j^{\ddagger})\in E}|\xi_{j^{\dagger}}-\xi_{j^{\ddagger}}|\Bigg).
\label{prior2}
\end{align}
The maximum a posteriori estimator for the regression coefficients assuming the prior distribution \eqref{prior1} or \eqref{prior2} is equivalent to the generalized fused lasso estimator expressed by \eqref{eq:generalized fused lasso_theta} or \eqref{eq:generalized fused lasso}, respectively.
What were treated as regularization parameters in the generalized fused lasso become hyperparameters of the Laplace distribution in the Bayesian generalized fused lasso.

In the SVC model of this paper, the complexity of the model changes depending on whether the values of $\lambda_{2,j}$ in \eqref{prior1} are varied for each $j$ and how many different values are used.
If we are interested in considering two Bayesian models with extremely different complexities, then the following two models (Models 1 and 2) are suitable, so only those two models are treated in the numerical experiments and real data analysis of this paper for simplicity.
In the first model, Model 1, is as specified above; that is, the prior distribution is as in \eqref{prior1} with settings $\lambda_{1,1}=\cdots=\lambda_{1,\tilde{p}}=\lambda_1$ and $\lambda_{2,1}=\cdots=\lambda_{2,\tilde{p}}=\lambda_2$, which corresponds to the prior distribution \eqref{prior2}.
This setting assumes that regression coefficients for all the explanatory variables follow a common Laplace distribution, and that the differences between adjacent coefficients also do so.
In the second model, Model 2, the prior distribution of \eqref{prior1} is used as is. 
This means that the regression coefficients and the differences between adjacent regression coefficients are assumed to follow different Laplace distributions for all explanatory variables.
Obviously, the latter model has greater complexity.
For data in which how all variables affect the outcome is assumed to have the same spatial structure, Model 1 is suitable, whereas Model 2 is more appropriate for data in which how the effects differ depends on the variable type.
This means that which model is suitable depends on the data. To provide an analytical method that improves performance by changing the model according to the data, the construction of an appropriate model selection method is a key point.

\section{Construction of information criterion}\label{sec3}

AIC (\citealt{Aka73}) is one of the most commonly used information criteria. In the Bayesian approach, information criteria designed to reduce the divergence between the true and estimated structures have also come to be widely used in recent years.
In this section, we first introduce WAIC (\citealt{watanabe2010asymptotic}) as a representative of such information criteria.
We then discuss PIIC (\citealt{ninomiya2021prior}), which attempts to resolve some concerns of WAIC, and adapt it to the SVC model with the Bayesian generalized fused lasso.

\subsection{Widely applicable information criterion}\label{sec:3-1}

Assume a probability density function $f(\cdot,\cdot\mid\bm{\xi})$ and a prior distribution $\pi_n(\cdot;\bm{\lambda})$, and consider independent random vectors satisfying
\begin{align*}
(y_1,\bm{x}_1),\ldots,(y_n,\bm{x}_n),(\tilde{y}_1,\tilde{\bm{x}}_1),\ldots,(\tilde{y}_n,\tilde{\bm{x}}_n)\sim f(\cdot,\cdot\mid\bm{\xi}),\qquad\bm{\xi}\sim\pi_n(\cdot;\bm{\lambda}).
\end{align*}
Here, $(y_{i},\bm{x}_{i})$ is the $i$-th sample that gives data by realization, and $(\tilde{y}_{i},\tilde{\bm{x}}_{i})$ is a copy of $(y_{i},\bm{x}_{i})$ that appears as a notional random vector in deriving the information criterion.
Let $\bm{y}=(y_1,\ldots,y_n)^{\T}$ and $\bm{X}=(\bm{x}_1,\ldots,\bm{x}_n)^{\T}$, and denote by $\E_{\bm{\xi}\mid\bm{y},\bm{X};\bm{\lambda}}[\cdot]$ the expectation based on the posterior distribution $f(\bm{y},\bm{X}\mid\bm{\xi})\pi_n(\bm{\xi};\bm{\lambda})/\int_{\bm{\Xi}} f(\bm{y},\bm{X}\mid\bm{\xi})\pi_n(\bm{\xi};\bm{\lambda})\d\bm{\xi}$ of $\bm{\xi}$.
Then the Bayesian predictive distribution $f(\cdot,\cdot\mid\bm{y},\bm{X};\bm{\lambda})$ can be written as $\E_{\bm{\xi}\mid\bm{y},\bm{X};\bm{\lambda}}[f(\cdot,\cdot\mid \bm{\xi})]$.

As an index for evaluating predictive accuracy, WAIC, which is an information criterion based on the Bayesian predictive distribution, uses the Kullback-Leibler divergence between the true distribution and the Bayesian predictive distribution minus a constant. This quantity is the so-called expected log-loss of the Bayesian predictive distribution, $-\sum_{i=1}^n\E_{\tilde{y}_i,\tilde{\bm{x}}_i}[\log \allowbreak f(\tilde{y}_i,\tilde{\bm{x}}_i\mid\bm{y},\bm{X};\bm{\lambda})]$.
In practice, this evaluation index cannot be explicitly calculated, and therefore $-\sum_{i=1}^n\log f(y_i,\bm{x}_i\mid\bm{y},\bm{X};\bm{\lambda})$ is considered as an initial estimator.
However, this initial estimator uses $(y_i,\bm{x}_i)$ instead of $(\tilde{y}_i,\tilde{\bm{x}}_i)$, and hence it underestimates the target.
Then, by regarding an asymptotic evaluation of the expectation
\begin{align}
& \sum_{i=1}^n\log f(y_i,\bm{x}_i\mid\bm{y},\bm{X};\bm{\lambda}) - \sum_{i=1}^n\log f(y_i,\bm{x}_i\mid\bm{\xi}^*)
\notag \\
& - \sum_{i=1}^n\log f(\tilde{y}_i,\tilde{\bm{x}}_i\mid\bm{y},\bm{X};\bm{\lambda}) +\sum_{i=1}^n\log f(\tilde{y}_i,\tilde{\bm{x}}_i\mid\bm{\xi}^*),
\label{eq:bias0}
\end{align}
as asymptotic bias, an asymptotic bias correction of $-\sum_{i=1}^n\log f(y_i,\bm{x}_i\mid\bm{y},\bm{X};\bm{\lambda})$ can be achieved.
As a result, WAIC is obtained in the form of an initial estimator plus the following simple penalty term:
\begin{align*}
\sum_{i=1}^n\E_{\bm{\xi}\mid\bm{y},\bm{X};\bm{\lambda}}[\{\log f(y_i,\bm{x}_i\mid\bm{\xi})\}^2]
-\sum_{i=1}^n\{\E_{\bm{\xi}\mid\bm{y},\bm{X};\bm{\lambda}}[\log f(y_i,\bm{x}_i\mid\bm{\xi})]\}^2.
\end{align*}
Note that the expectation of WAIC asymptotically coincides with the expected log-loss of the Bayesian predictive distribution.

\subsection{Prior intensified information criterion}\label{sec:3-2}

Despite WAIC being a powerful tool for the selection of Bayesian models, the following two problems should be considered when applying it to the selection of SVC models using Bayesian regularization methods.
First, WAIC adopts bias correction based on the second-order term of the asymptotic expansion without making the prior distribution depend on $n$. Then, there arises the problem that the influence of the prior distribution, which appears as higher-order terms, is not sufficiently reflected.
That is, WAIC is targeted at Bayesian estimation that is closely similar to maximum likelihood estimation, and in particular, it might be poorly compatible with Bayesian estimation that induces sparsity.
This is significant because emphasis of such estimation is placed on sparseness in reality, whereas WAIC is based on asymptotics close to that of maximum likelihood estimation that does not induce sparsity.
The second problem is that WAIC always selects models with prior distributions having more hyperparameters when comparing models, such as those discussed in Section \ref{sec:2-3}, which employ prior distributions from classes with different degrees of freedom.

PIIC was proposed to solve both these problems.
The prior distribution was assumed to depend on $n$, thereby strengthening the influence of the prior distribution on the estimation, in order to solve the first problem.
Described concretely using the $\pi_{\rm gfl}(\bm{\xi};\bm{\lambda})$ term defined in \eqref{prior2}, PIIC assumes $\pi_n(\bm{\xi};\bm{\lambda})\propto\pi_{\rm gfl}(\bm{\xi};\bm{\lambda})$, whereas WAIC assumes $\pi_n(\bm{\xi};\bm{\lambda})\propto\pi_{\rm gfl}(\bm{\xi};\bm{\lambda})^{1/n}$.
In addition, to solve the second problem, a penalty term for hyperparameters was added.
However, since \cite{ninomiya2021prior} did not consider the Bayesian generalized fused lasso, we developed PIIC for the SVC model with the Bayesian generalized fused lasso, as will be presented in this section.

PIIC is the asymptotic bias-corrected statistic of $-\sum_{i=1}^n\log f(y_i,\bm{x}_i\mid\bm{y},\bm{X};\bm{\lambda})$, sharing the same concept as WAIC, and it asymptotically evaluates the expectation of \eqref{eq:bias0}.
Here, we avoid a discussion on moment convergence, and define the expectation of the weak limit of \eqref{eq:bias0} as the asymptotic bias.
With that in mind, we henceforth asymptotically evaluate \eqref{eq:bias0}.
To proceed simultaneously with the expansion for $(y_i,\bm{x}_i)$ and the expansion for $(\tilde{y}_i,\tilde{\bm{x}}_i)$, we adopt the notation $(\breve{y}_i,\breve{\bm{x}}_i)$, which denotes either $(y_i,\bm{x}_i)$ or $(\tilde{y}_i,\tilde{\bm{x}}_i)$.
The predictive distribution $\log f(y_i,\bm{x}_i\mid\bm{y},\bm{X};\bm{\lambda})$ or $\log f(\tilde{y}_i,\tilde{\bm{x}}_i\mid\bm{y},\bm{X};\bm{\lambda})$ in \eqref{eq:bias0} is expressed as
\begin{align*}
\log\int f(\breve{y}_i,\breve{\bm{x}}_i\mid\bm{\xi})f(\bm{y},\bm{X}\mid\bm{\xi})\pi_{\rm gfl}(\bm{\xi};\bm{\lambda}){\rm d}\bm{\xi}-\log\int f(\bm{y},\bm{X}\mid\bm{\xi})\pi_{\rm gfl}(\bm{\xi};\bm{\lambda}){\rm d}\bm{\xi}.
\end{align*}
For these two integrals, by applying higher-order Laplace approximation (\citealt{Tierney01031986}), we can express them as follows using the function $h(y,\bm{x}\mid\bm{\xi})$:
\begin{align}
\log f(\breve{y}_i,\breve{\bm{x}}_i\mid\bm{y},\bm{X};\bm{\lambda}) = \log f(\breve{y}_i,\breve{\bm{x}}_i\mid\hat{\bm{\xi}}) + \frac{1}{n}h(\breve{y}_i,\breve{\bm{x}}_i\mid\hat{\bm{\xi}}) + \oP\Big(\frac{1}{n}\Big) = \log f(\breve{y}_i,\breve{\bm{x}}_i\mid\hat{\bm{\xi}}) + \OP\Big(\frac{1}{n}\Big). 
\label{key1}
\end{align}
As a result, \eqref{eq:bias0} can be rewritten as
\begin{align}
&\sum_{i=1}^n\log f(y_i,\bm{x}_i\mid\hat{\bm{\xi}})+\frac{1}{n}\sum_{i=1}^{n}h(y_i,\bm{x}_i\mid\hat{\bm{\xi}})-\sum_{i=1}^{n}\log f(y_i,\bm{x}_i\mid\bm{\xi}^*)
\notag \\
&-\sum_{i=1}^n\log f(\tilde{y}_i,\tilde{\bm{x}}_i\mid\hat{\bm{\xi}})-\frac{1}{n}\sum_{i=1}^{n}h(\tilde{y}_i,\tilde{\bm{x}}_i\mid\hat{\bm{\xi}})+\sum_{i=1}^{n}\log f(\tilde{y}_i,\tilde{\bm{x}}_i\mid\bm{\xi}^*)
+\oP(1). 
\label{eq:fzz2}
\end{align}
Since the expectation of $h(y_i,\bm{x}_i\mid\hat{\bm{\xi}})$ and the expectation of $h(\tilde{y}_i,\tilde{\bm{x}}_i\mid\hat{\bm{\xi}})$ are asymptotically equal, the sum of the second and fifth terms of \eqref{eq:fzz2} is $\oP(1)$.
In addition, when we perform a Taylor expansion on the sum of the first, third, fourth, and sixth terms, the terms involving $\hat{\bm{\xi}}^{(1)}$ become $\oP(1)$ from the first equation in Theorem \ref{th1}, $n\hat{\bm{\xi}}^{(1)}=\oP(1)$, and therefore the sum can be written as
\begin{align}
\sqrt{n}(\hat{\bm{\xi}}^{(2)}-\bm{\xi}^{*(2)})^{\T}\frac{1}{\sqrt{n}}\sum_{i=1}^n\bigg\{\frac{\partial}{\partial\bm{\xi}^{(2)}}\log f(y_i,\bm{x}_i\mid\bm{\xi}^*)-\frac{\partial}{\partial\bm{\xi}^{(2)}}\log f(\tilde{y}_i,\tilde{\bm{x}}_i\mid\bm{\xi}^*)\bigg\}
+\oP(1). 
\label{eq:taylorlap}
\end{align}
Here, note that the second-order terms of the Taylor expansion cancel each other out, between those from $f(y_i,\bm{x}_i\mid\bm{\xi}^*)$ and those from $f(\tilde{y}_i,\tilde{\bm{x}}_i\mid\bm{\xi}^*)$.
Removing $\oP(1)$, we can rewrite this as
\begin{align*}
& \sum_{j\in\mathcal{J}^{(2)}}\sqrt{n}(\hat{\xi}_{j,\bm{\lambda}}-\xi_{j,\bm{\lambda}}^*)\Bigg[\frac{1}{\sqrt{n}}\sum_{i=1}^n\bigg\{\frac{\partial}{\partial\xi_j}\log f(y_i,\bm{x}_i\mid\bm{\xi}^*)-\frac{\partial}{\partial\xi_j}\log f(\tilde{y}_i,\tilde{\bm{x}}_i\mid\bm{\xi}^*)\bigg\}\Bigg]
\notag \\
& = \sum_{j^{\dagger}\in\mathcal{J}^{(3)}}\sum_{j^{\ddagger}:\xi_{j^{\ddagger}}^*=\xi_{j^{\dagger}}^*}\sqrt{n}(\hat{\xi}_{j^{\ddagger}}-\xi_{j^{\ddagger}}^*)\frac{1}{\sqrt{n}}\sum_{i=1}^n\bigg\{\frac{\partial}{\partial\xi_{j^{\ddagger}}}\log f(y_i,\bm{x}_i\mid\bm{\xi}^*)-\frac{\partial}{\partial\xi_{j^{\ddagger}}}\log f(\tilde{y}_i,\tilde{\bm{x}}_i\mid\bm{\xi}^*)\bigg\}.
\end{align*}
Then, applying the second equation of Theorem \ref{th1}, we ultimately find that \eqref{eq:taylorlap} becomes
\begin{align*}
\sum_{j^{\dagger}\in\mathcal{J}^{(3)}}\sqrt{n}(\hat{\xi}_{j^{\dagger}}-\xi_{j^{\dagger}}^*)\sum_{j^{\ddagger}:\xi_{j^{\ddagger}}^*=\xi_{j^{\dagger}}^*}\frac{1}{\sqrt{n}}\sum_{i=1}^n\bigg\{\frac{\partial}{\partial\xi_{j^{\ddagger}}}\log f(y_i,\bm{x}_i\mid\bm{\xi}^*)-\frac{\partial}{\partial\xi_{j^{\ddagger}}}\log f(\tilde{y}_i,\tilde{\bm{x}}_i\mid\bm{\xi}^*)\bigg\}+\oP(1).
\end{align*}
Let us now consider the weak limit of the above expression.
First, from (R3), there exist $\bm{s}^{(2)}$ and its copy $\tilde{\bm{s}}^{(2)}$, which are $|\mathcal{J}^{(2)}|$-dimensional sub-vectors of a random vector $\bm{s}$ following $\N(\bm{0},\bm{J})$ and themselves follow $\N(\bm{0},\bm{J}^{(22)})$. Then, it holds that
\begin{align*}
\frac{1}{\sqrt{n}}\sum_{i=1}^n\bigg\{\frac{\partial}{\partial\bm{\xi}^{(2)}}\log f(y_i,\bm{x}_i\mid\bm{\xi}^*)-\frac{\partial}{\partial\bm{\xi}^{(2)}}\log f(\tilde{y}_i,\tilde{\bm{x}}_i\mid\bm{\xi}^*)\bigg\} \stackrel{\d}{\to} \bm{s}^{(2)}+\tilde{\bm{s}}^{(2)}.
\end{align*}
Here, we are using the fact that the expectations of the first and second terms inside the sum are equal.
Therefore, using the matrix $\bm{A}$ defined in \eqref{defA}, we obtain
\begin{align*}
\Bigg\{\sum_{j^{\ddagger}:\xi_{j^{\ddagger}}^*=\xi_{j^{\dagger}}^*}\frac{1}{\sqrt{n}}\sum_{i=1}^n\bigg\{\frac{\partial}{\partial\xi_{j^{\ddagger}}}\log f(y_i,\bm{x}_i\mid\bm{\xi}^*)-\frac{\partial}{\partial\xi_{j^{\ddagger}}}\log f(\tilde{y}_i,\tilde{\bm{x}}_i\mid\bm{\xi}^*)\bigg\}\Bigg\}_{j^{\dagger}\in\mathcal{J}^{(3)}} \stackrel{\d}{\to} \bm{A}^{(32)}(\bm{s}^{(2)}+\tilde{\bm{s}}^{(2)}).
\end{align*}
In addition, from \eqref{base} in the Appendix, which is the basis of the third equation in Theorem \ref{th1}, it is known that $\sqrt{n}(\hat{\xi}_{j^{\dagger}}-\xi_{j^{\dagger}}^*)_{j^{\dagger}\in\mathcal{J}^{(3)}}$ converges in distribution to $(\bm{A}^{(32)}\bm{J}^{(22)}\allowbreak\bm{A}^{(23)})^{-1}\bm{A}^{(32)}\bm{s}^{(2)}$.
As can be imagined from these, \eqref{eq:taylorlap} converges in distribution to 
\begin{align}
\bm{s}^{(2)\T}\bm{A}^{(23)}(\bm{A}^{(32)}\bm{J}^{(22)}\bm{A}^{(23)})^{-1}\bm{A}^{(32)}(\bm{s}^{(2)}+\tilde{\bm{s}}^{(2)}).
\label{climit}
\end{align}
The expectation of this is
\begin{align*}
& \E[\tr\{(\bm{A}^{(32)}\bm{J}^{(22)}\bm{A}^{(23)})^{-1}\bm{A}^{(32)}(\bm{s}^{(2)}+\tilde{\bm{s}}^{(2)})\bm{s}^{(2)\T}\bm{A}^{(23)}\}] 
\notag \\
& = \tr\{(\bm{A}^{(32)}\bm{J}^{(22)}\bm{A}^{(23)})^{-1}\bm{A}^{(32)}\E[\bm{s}^{(2)}\bm{s}^{(2)\T}]\bm{A}^{(23)}\} = \tr(\bm{I}_{|\mathcal{J}^{(3)}|}) = |\mathcal{J}^{(3)}|,
\end{align*}
and therefore the following theorem holds.
\begin{theorem}
Under conditions (C1) and (C2), the asymptotic bias of $-\sum_{i=1}^n\log f(y_i,\bm{x}_i\mid\bm{y},\bm{X};\bm{\lambda})$ with respect to $-\sum_{i=1}^n\E_{\tilde{y}_i,\tilde{\bm{x}}_i}[\log \allowbreak f(\tilde{y}_i,\tilde{\bm{x}}_i\mid\bm{y},\bm{X};\bm{\lambda})]$ is $|\mathcal{J}^{(3)}|$.
\label{th2}
\end{theorem}
\noindent
Based on Theorem \ref{th2}, we propose the following information criterion:
\begin{align*}
{\rm PIIC1} = -\sum_{i=1}^n\log f(y_i,\bm{x}_i\mid\bm{y},\bm{X};\bm{\lambda}) + |\hat{\mathcal{J}}^{(3)}|.
\end{align*}
Here, for the graph consisting of the vertex set $\{j\in V:\hat{\xi}_j\neq 0\}$ and the edge set $\{(j^{\dagger},j^{\ddagger})\in E: \hat{\xi}_{j^{\dagger}}=\hat{\xi}_{j^{\ddagger}}\}$, $\hat{\mathcal{J}}^{(3)}$ is the collection obtained by arbitrarily choosing one vertex as a representative from each connected component. It follows trivially that $|\hat{\mathcal{J}}^{(3)}|$ is a consistent estimator of $|\mathcal{J}^{(3)}|$.
For the selection of $\bm{\lambda}$, it suffices to use a $\hat{\bm{\lambda}}$ that minimizes PIIC1.

Unlike WAIC, PIIC1 incorporates the influence of a prior distribution that introduces sparsity to the Bayesian generalized fused lasso estimator.
However, as with WAIC, PIIC1 always selects the model with the larger dimension of $\bm{\lambda}$ when comparing models that use prior distributions with hyperparameters $\bm{\lambda}$ of different dimensions.
Therefore, by making a bias evaluation that also takes into account the fact that $\bm{\lambda}$ is selected from the data and adding a penalty term that affects the selection of the dimension of hyperparameters, it is made possible to perform appropriate model selection even when there are candidate prior distributions with different complexities.
Since the selection of $\bm{\lambda}$ becomes the topic, in the following part of this section, we attach $\bm{\lambda}$ to estimators and their limits, denoting them by $\hat{\bm{\xi}}_{\bm{\lambda}}$ and $\bm{\xi}_{\bm{\lambda}}^*$, respectively.
In addition, as regularity conditions, we add the following, with $\bm{\lambda}^*$ being the minimizer with respect to $\bm{\lambda}$ of $-\E[\log f(y,\bm{x}\mid\bm{\xi}_{\bm{\lambda}}^*)]$.
\begin{itemize}[label=(C3)\ , leftmargin=*]
\item
$\bm{\lambda}^*$ exists in the interior of the compact set $\Lambda$, which is the parameter space of $\bm{\lambda}$, and $\log f(y,\bm{x}\mid\bm{\xi}_{\bm{\lambda}}^*)$ is of class C$^2$ in a neighborhood of $\bm{\lambda}=\bm{\lambda}^*$.
\end{itemize}
In view of the fact that PIIC1 can be written as
\begin{align*}
& -\sum_{i=1}^n\log f(y_i,\bm{x}_i\mid\bm{y},\bm{X};\bm{\lambda})\{1+\oP(1)\}
\notag \\
& = -\sum_{i=1}^n\log f(y_i,\bm{x}_i\mid\hat {\bm{\xi}}_{\bm{\lambda}})\{1+\oP(1)\} = -\sum_{i=1}^n\log f(y_i,\bm{x}_i\mid\bm{\xi}^*_{\bm{\lambda}})\{1+\oP(1)\}
\end{align*}
from \eqref{key1}, letting $\tilde{\bm{\lambda}}$ be the minimizer with respect to $\bm{\lambda}$ of $-\sum_{i=1}^n\log f(y_i,\bm{x}_i\mid\bm{\xi}^*_{\bm{\lambda}})$ and taking $-\sum_{i=1}^n\E_{\tilde{y}_i,\tilde{\bm{x}}_i} \allowbreak [\log f(\tilde{y}_i,\tilde{\bm{x}}_i\mid\bm{y},\bm{X};\tilde{\bm{\lambda}})]$ as the target, we substitute $\tilde{\bm{\lambda}}$ for $\bm{\lambda}$ in the first and third terms of \eqref{eq:bias0} as in 
\begin{align}
& \sum_{i=1}^n\log f(y_i,\bm{x}_i\mid\bm{y},\bm{X};\tilde{\bm{\lambda}}) - \sum_{i=1}^n\log f(y_i,\bm{x}_i\mid\bm{\xi}^*_{\bm{\lambda}^*})
\notag \\
& - \sum_{i=1}^n\log f(\tilde{y}_i,\tilde{\bm{x}}_i\mid\bm{y},\bm{X};\tilde{\bm{\lambda}}) -\sum_{i=1}^n\log f(\tilde{y}_i,\tilde{\bm{x}}_i\mid\bm{\xi}^*_{\bm{\lambda}^*}),
\label{eq:bias0_lambda}
\end{align}
and asymptotically evaluate this to derive the information criterion.
First, since $\tilde{\bm{\lambda}}-\bm{\lambda}^*=\oP(1)$ from the uniform law of large numbers, we perform a Taylor expansion of $\bm{0}=\sum_{i=1}^n\partial\log f(y_i,\bm{x}_i\mid\bm{\xi}_{\tilde{\bm{\lambda}}}^*)/\partial\bm{\lambda}$ around $\tilde{\bm{\lambda}}=\bm{\lambda}^*$.
Then, letting
\begin{align}
\bm{J}_1(\bm{\lambda}) \equiv \E\bigg[-\frac{\partial^2}{\partial\bm{\lambda}\partial\bm{\lambda}^{\T}}\log f(y_i,\bm{x}_i\mid\bm{\xi}_{\bm{\lambda}}^*)\bigg],
\end{align}
and
\begin{align*}
\bm{J}_2(\bm{\lambda}) \equiv \E\bigg[\bigg\{\frac{\partial}{\partial\bm{\lambda}}\log f(y_i,\bm{x}_i\mid\bm{\xi}_{\bm{\lambda}}^*)\bigg\}\bigg\{\frac{\partial}{\partial\bm{\lambda}}\log f(y_i,\bm{x}_i\mid\bm{\xi}_{\bm{\lambda}}^*)\bigg\}^{\T}\bigg],
\end{align*}
we obtain
\begin{align}
\tilde{\bm{\lambda}}-\bm{\lambda}^* = \bm{J}_1(\bm{\lambda}^*)^{-1}\frac{1}{n}\sum_{i=1}^n\frac{\partial}{\partial\bm{\lambda}}\log f(y_i,\bm{x}_i\mid\bm{\xi}_{\bm{\lambda}^*}^*)\{1+\oP(1)\},
\label{eq:xi-xi}
\end{align}
where $\bm{J}_2(\bm{\lambda})$ has been defined for later use.
When \eqref{eq:bias0_lambda} is asymptotically evaluated using \eqref{key1} and \eqref{eq:xi-xi}, what is obtained is an expression in which $\bm{\lambda}^*$ is substituted for $\bm{\lambda}$ in \eqref{eq:bias0}, with the addition of
\begin{align*}
& \Bigg\{\frac{1}{\sqrt{n}}\sum_{i=1}^n\frac{\partial}{\partial\bm{\lambda}^{\T}}\log f(y_i,\bm{x}_i\mid\bm{\xi}_{\bm{\lambda}^*}^*)\Bigg\} \bm{J}_1(\bm{\lambda}^*)^{-1} \Bigg\{\frac{1}{\sqrt{n}}\sum_{i=1}^n\frac{\partial}{\partial\bm{\lambda}}\log f(y_i,\bm{x}_i\mid\bm{\xi}_{\bm{\lambda}^*}^*)\Bigg\} \{1+\oP(1)\}
\notag \\
& - \Bigg\{\frac{1}{\sqrt{n}}\sum_{i=1}^n\frac{\partial}{\partial\bm{\lambda}^{\T}}\log f(\tilde{y}_i,\tilde{\bm{x}}_i\mid\bm{\xi}_{\bm{\lambda}^*}^*)\Bigg\} \bm{J}_1(\bm{\lambda}^*)^{-1} \Bigg\{\frac{1}{\sqrt{n}}\sum_{i=1}^n\frac{\partial}{\partial\bm{\lambda}}\log f(y_i,\bm{x}_i\mid\bm{\xi}_{\bm{\lambda}^*}^*)\Bigg\} \{1+\oP(1)\}.
\end{align*}
Therefore, letting $\bm{s}_1$ and $\bm{s}_2$ be random vectors independently following $\N(\bm{0},\bm{J}_2(\bm{\lambda}^*))$, we find that the weak limit of \eqref{eq:bias0_lambda} is that of \eqref{climit} with the addition of
\begin{align*}
\bm{s}_1^{\T}\bm{J}_1(\bm{\lambda}^*)^{-1}\bm{s}_1 - \bm{s}_2^{\T}\bm{J}_1(\bm{\lambda}^*)^{-1}\bm{s}_1.
\end{align*}
The expectation of the weak limit is
\begin{align*}
|\mathcal{J}^{(3)}| + \tr\{\bm{J}_1(\bm{\lambda}^*)^{-1}\E[\bm{s}_1\bm{s}_1^{\T}]\} + \E[\bm{s}_2^{\T}]\bm{J}_1(\bm{\lambda}^*)^{-1}\E[\bm{s}_1] = |\mathcal{J}^{(3)}| + \tr\{\bm{J}_1(\bm{\lambda}^*)^{-1}\bm{J}_2(\bm{\lambda}^*)\},
\end{align*}
and therefore the following theorem holds.
\begin{theorem}
Under conditions (C1), (C2), and (C3), the asymptotic bias of $-\sum_{i=1}^n\log f(y_i,\bm{x}_i\mid\bm{y},\bm{X},\tilde{\bm{\lambda}})$ with respect to $-\sum_{i=1}^n\E_{\tilde{y}_i,\tilde{\bm{x}}_i}[\log \allowbreak f(\tilde{y}_i,\tilde{\bm{x}}_i\mid\bm{y},\bm{X};\tilde{\bm{\lambda}})]$ is $|\mathcal{J}^{(3)}|+\tr\{\bm{J}_1(\bm{\lambda}^*)^{-1}\bm{J}_2(\bm{\lambda}^*)\}$.
\label{th3}
\end{theorem}
\noindent
In evaluating $\bm{J}_1(\bm{\lambda}^*)$ and $\bm{J}_2(\bm{\lambda}^*)$ of Theorem \ref{th3}, based on \eqref{key1}, we use
\begin{align*}
\hat{\bm{J}}_1(\hat{\bm{\lambda}}) \equiv -\frac{1}{n}\sum_{i=1}^n\frac{\partial^2}{\partial\bm{\lambda}\partial\bm{\lambda}^{\T}}\log f(y_i,\bm{x}_i\mid\bm{y},\bm{X};\hat{\bm{\lambda}})
\end{align*}
and
\begin{align*}
\hat{\bm{J}}_2(\hat{\bm{\lambda}}) \equiv \frac{1}{n}\sum_{i=1}^n\frac{\partial}{\partial\bm{\lambda}}\log f(y_i,\bm{x}_i\mid\bm{y},\bm{X};\hat{\bm{\lambda}})\frac{\partial}{\partial\bm{\lambda}^{\T}}\log f(y_i,\bm{x}_i\mid\bm{y},\bm{X};\hat{\bm{\lambda}}).
\end{align*}
As a result, we propose the following information criterion with an added penalty term for hyperparameters:
\begin{align*}
{\rm PIIC2} \equiv -\sum_{i=1}^n\log f(y_i,\bm{x}_i\mid\bm{y},\bm{X};\hat{\bm{\lambda}}) + |\hat{\mathcal{J}}^{(3)}| + \tr\{\hat{\bm{J}}_1(\hat{\bm{\lambda}})^{-1} \hat{\bm{J}}_2(\hat{\bm{\lambda}})\}.
\end{align*}

\section{Numerical experiments}\label{sec4}

\begin{figure}[t]
\centering
\begin{minipage}[b]{0.24\columnwidth}
    \centering
    \includegraphics[width=0.9\columnwidth]{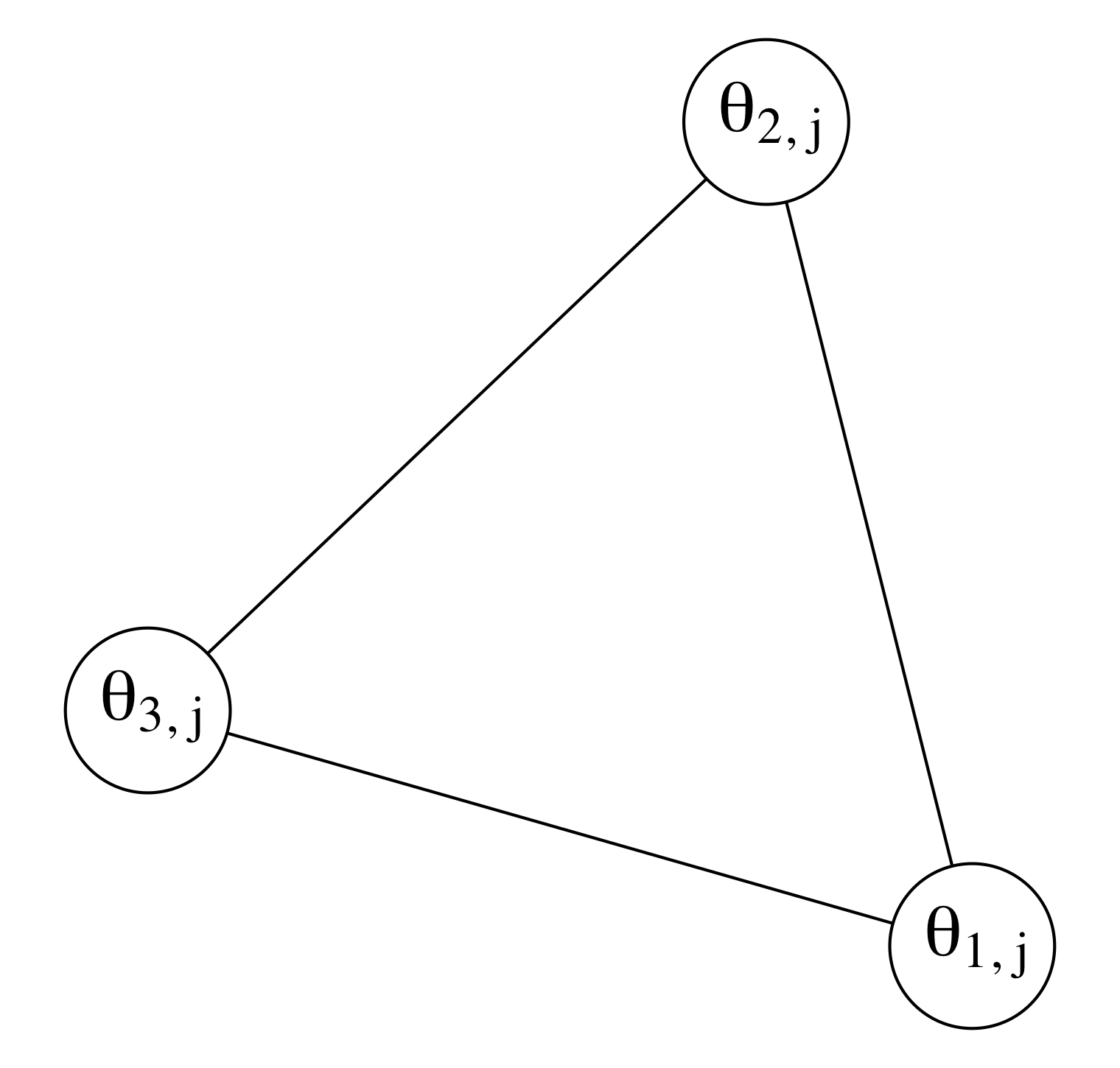}
    \subcaption{Graph 1}
\end{minipage}
\begin{minipage}[b]{0.24\columnwidth}
    \centering
    \includegraphics[width=0.9\columnwidth]{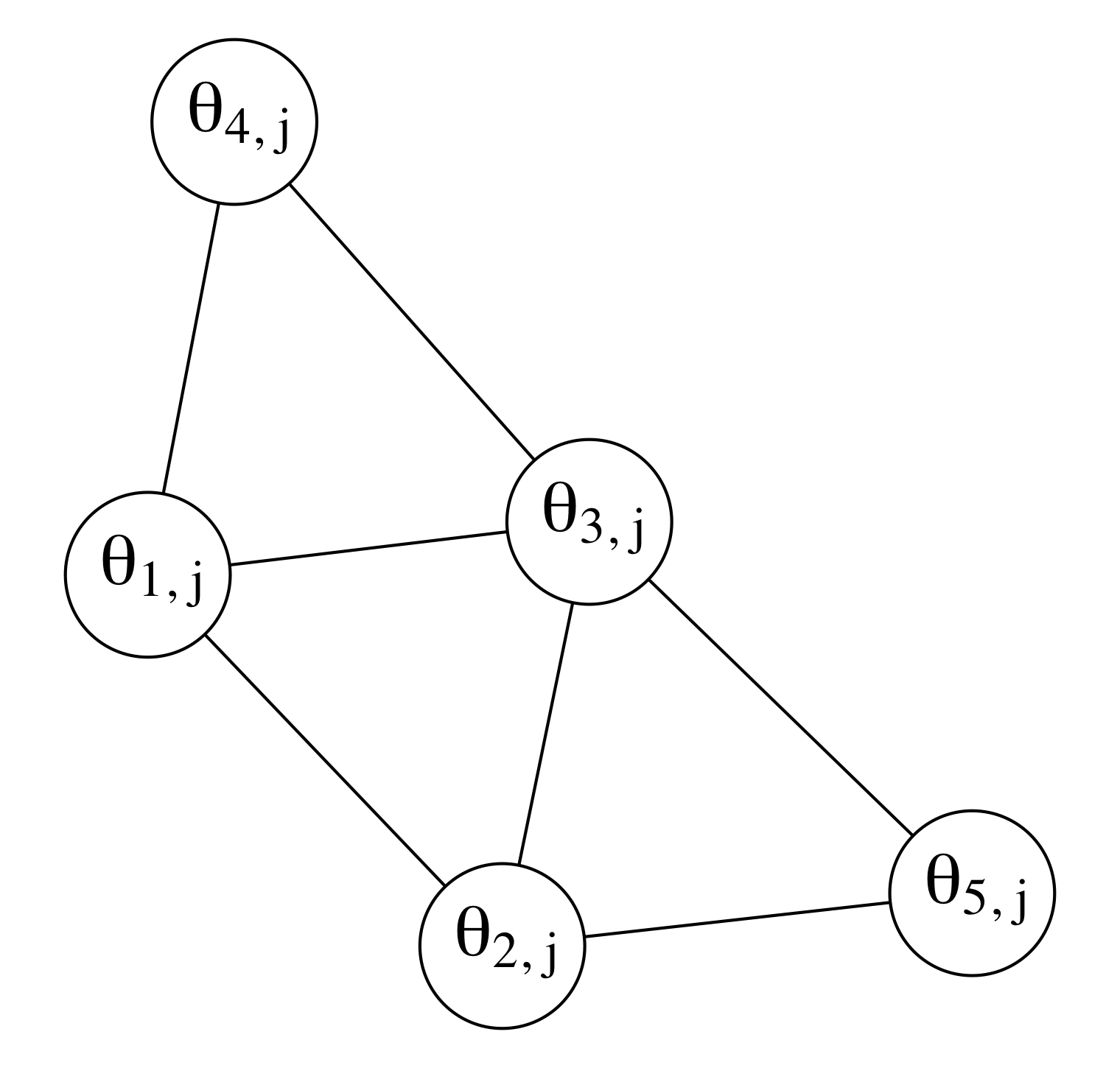}
    \subcaption{Graph 2}
\end{minipage}
\begin{minipage}[b]{0.24\columnwidth}
    \centering
    \includegraphics[width=0.9\columnwidth]{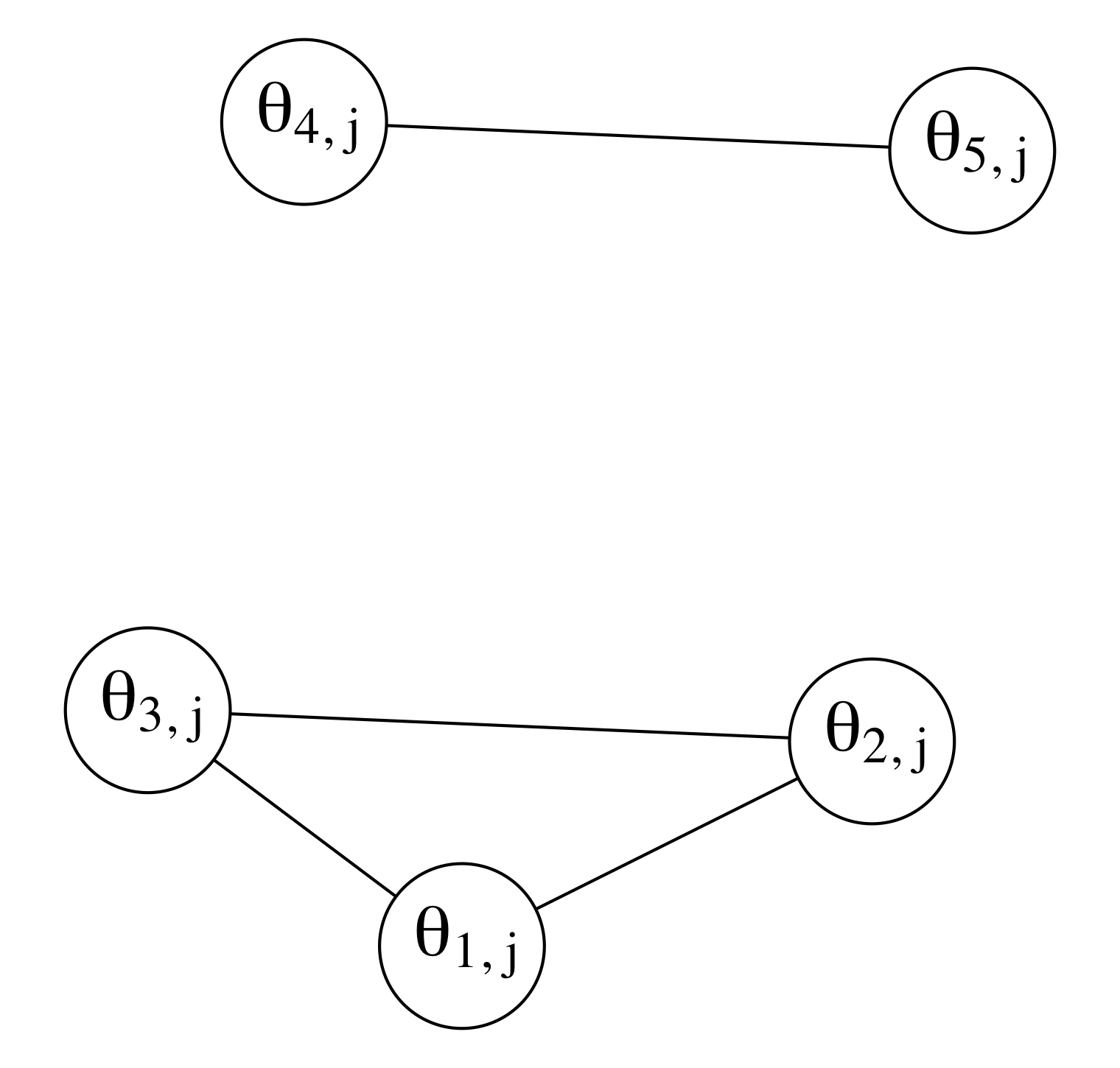}
    \subcaption{Graph 3}
\end{minipage}
\begin{minipage}[b]{0.24\columnwidth}
    \centering
    \includegraphics[width=0.9\columnwidth]{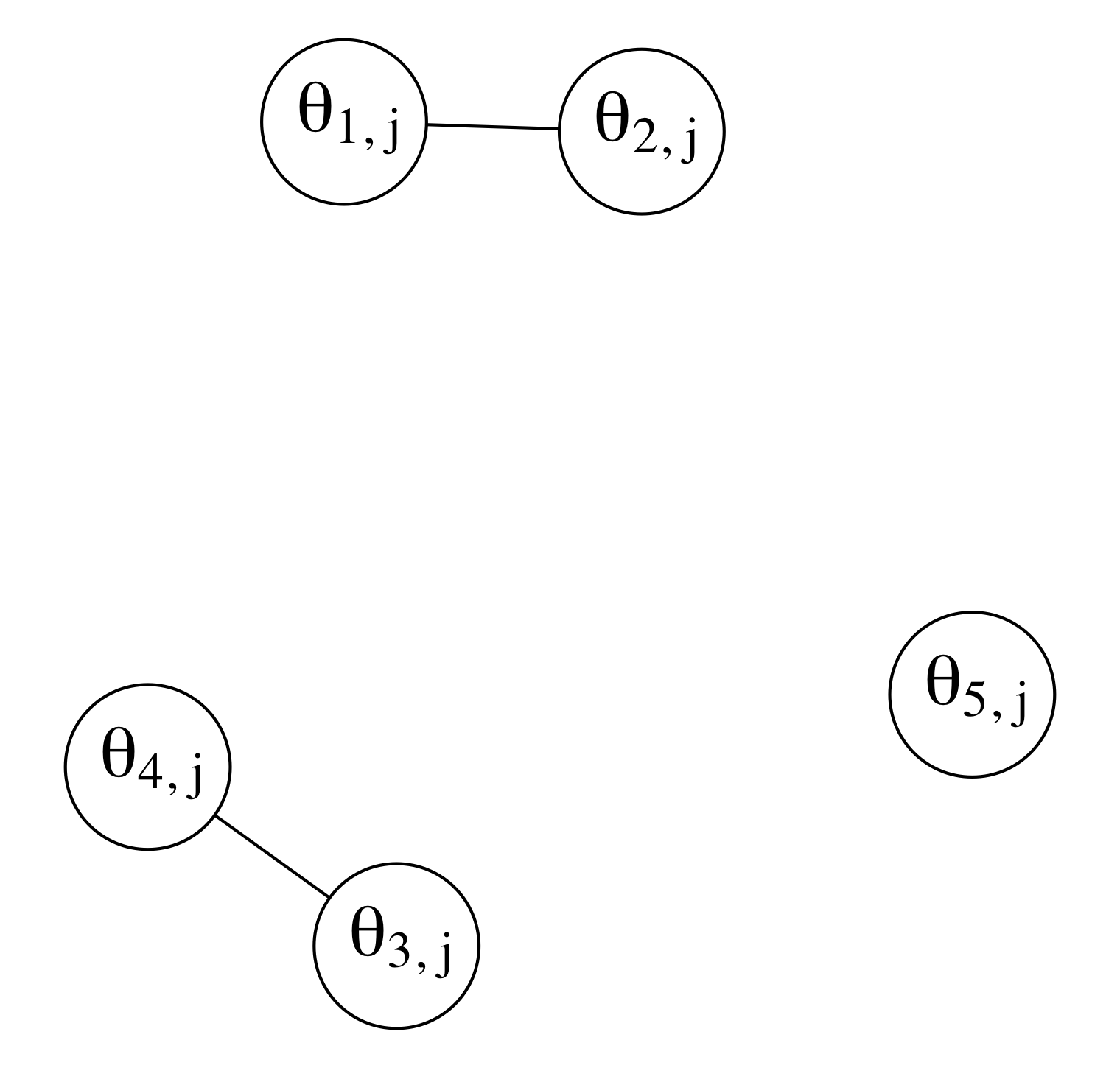}
    \subcaption{Graph 4}
\end{minipage}

\bigskip

\begin{minipage}[b]{0.24\columnwidth}
    \centering
    \includegraphics[width=0.9\columnwidth]{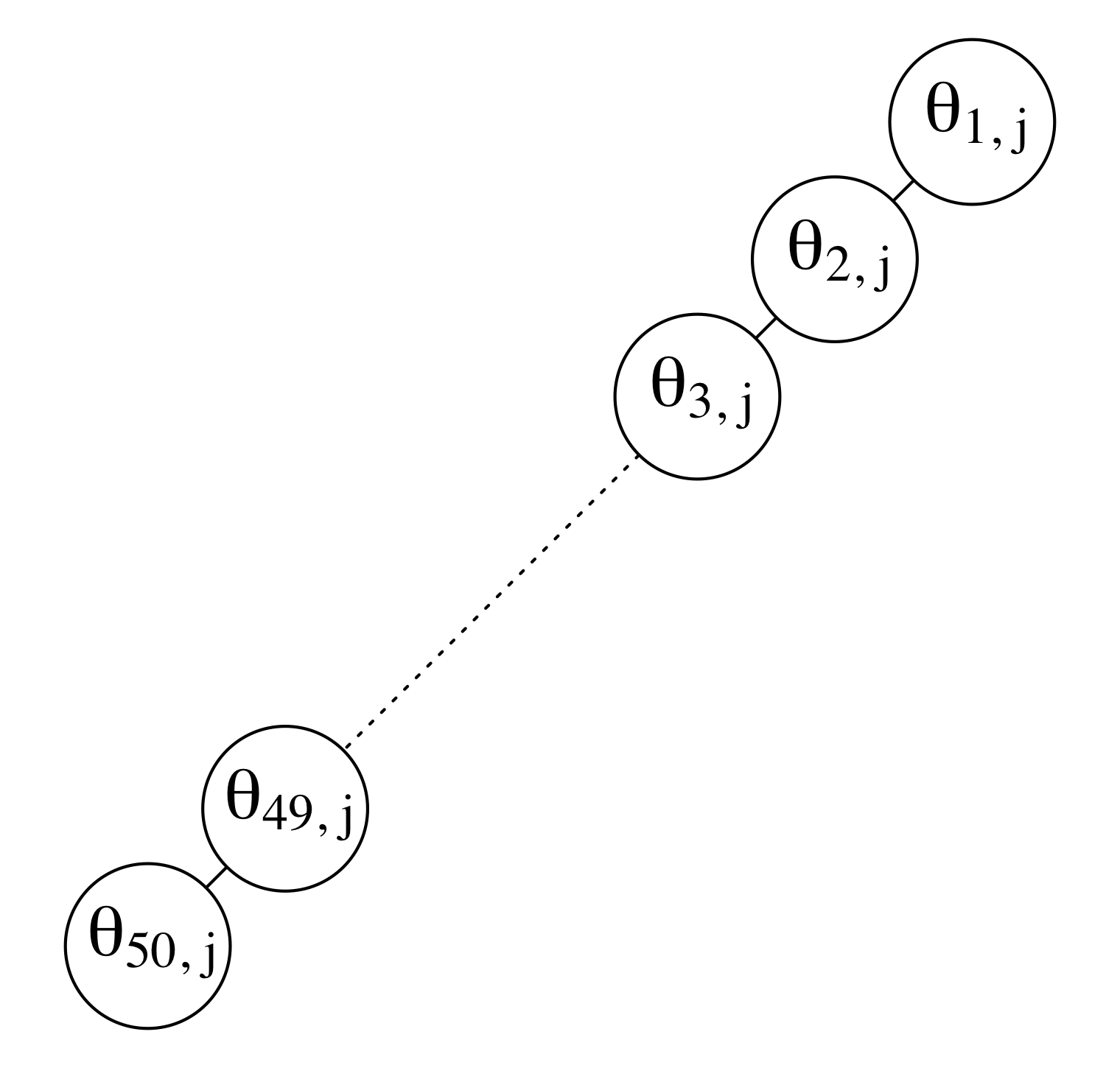}
    \subcaption{Graph 5}
\end{minipage}
\begin{minipage}[b]{0.24\columnwidth}
    \centering
    \includegraphics[width=0.9\columnwidth]{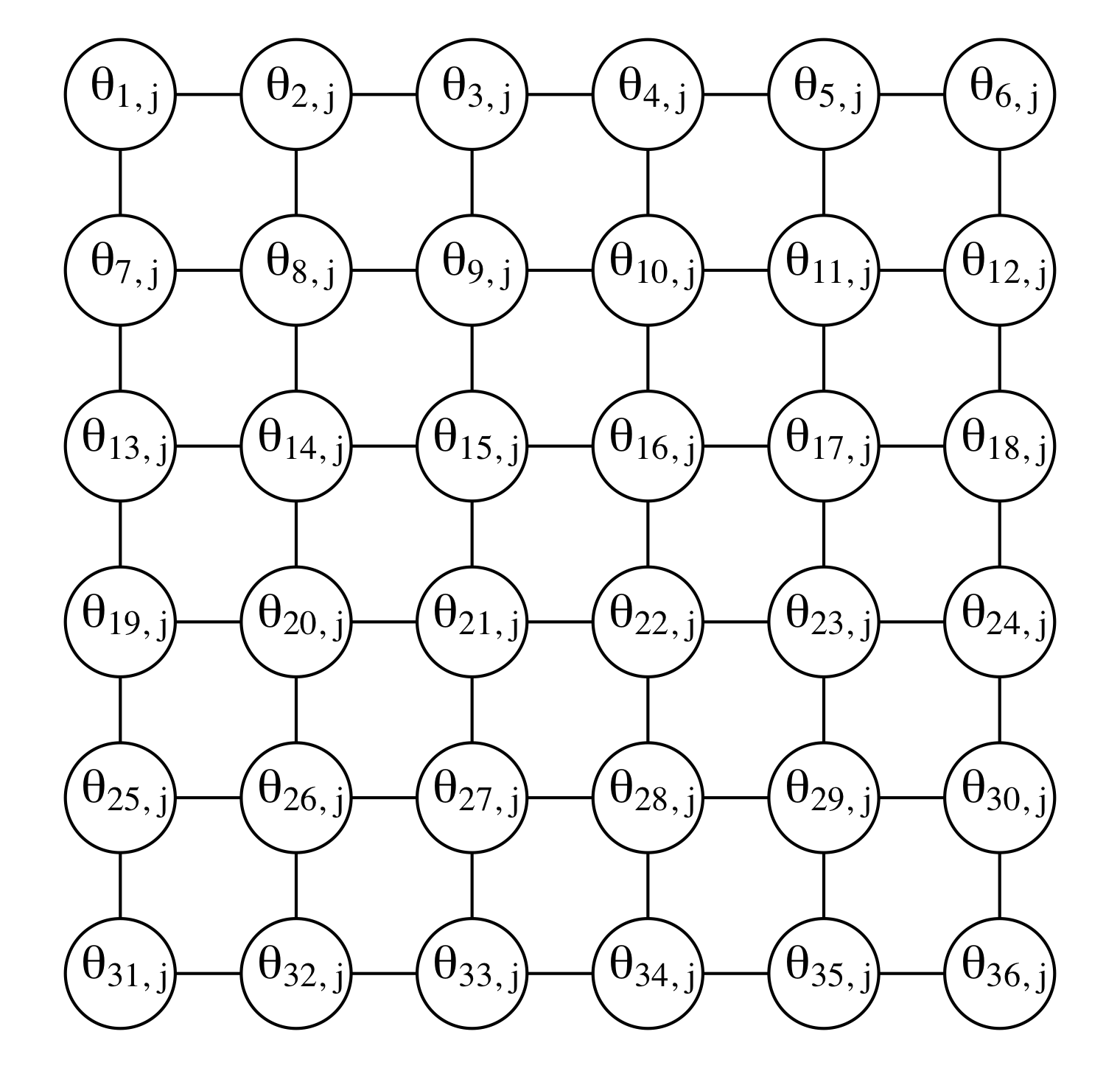}
    \subcaption{Graph 6}
\end{minipage}
\begin{minipage}[b]{0.24\columnwidth}
    \centering
    \includegraphics[width=0.9\columnwidth]{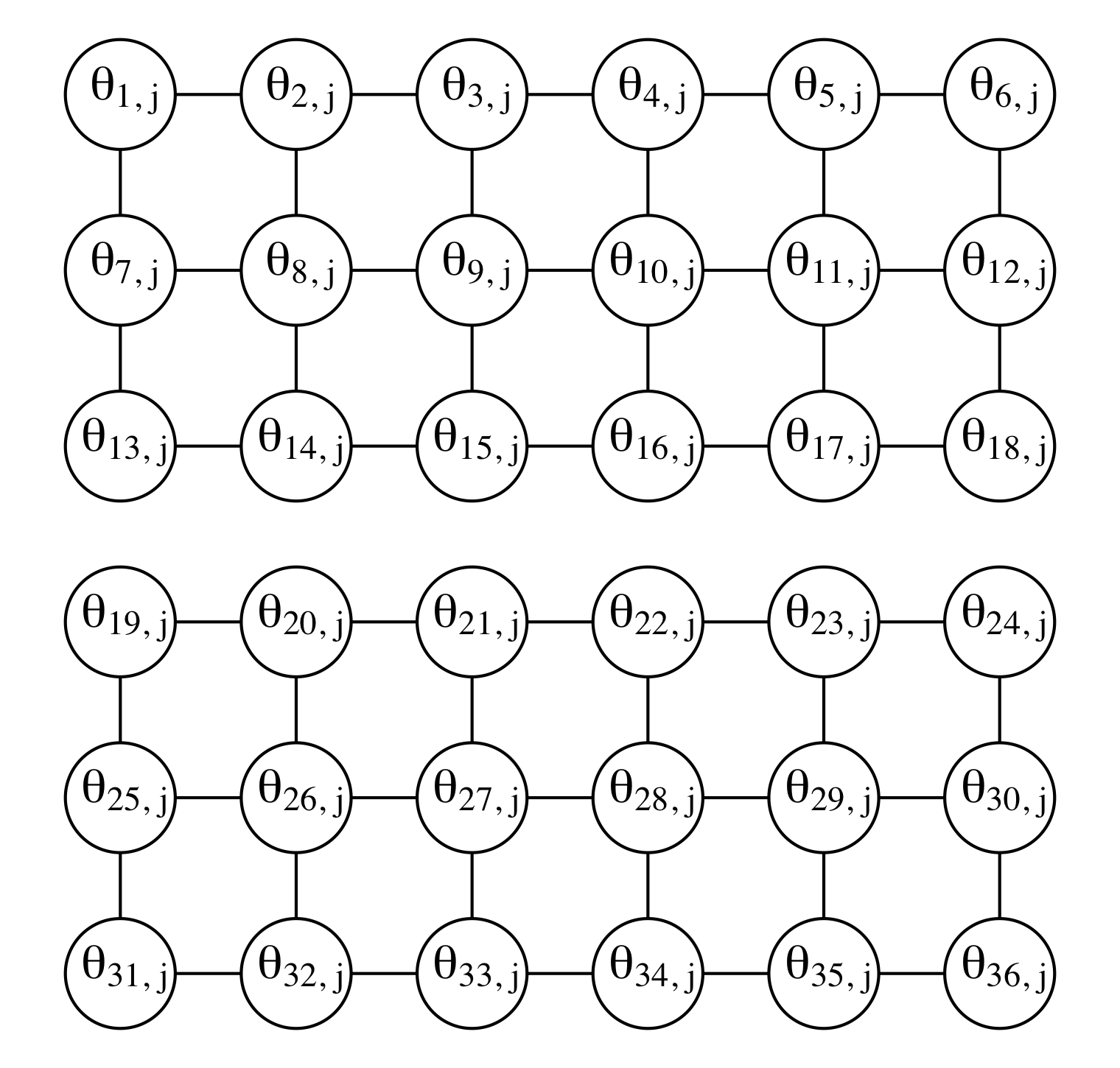}
    \subcaption{Graph 7}
\end{minipage}
\begin{minipage}[b]{0.24\columnwidth}
    \centering
    \includegraphics[width=0.9\columnwidth]{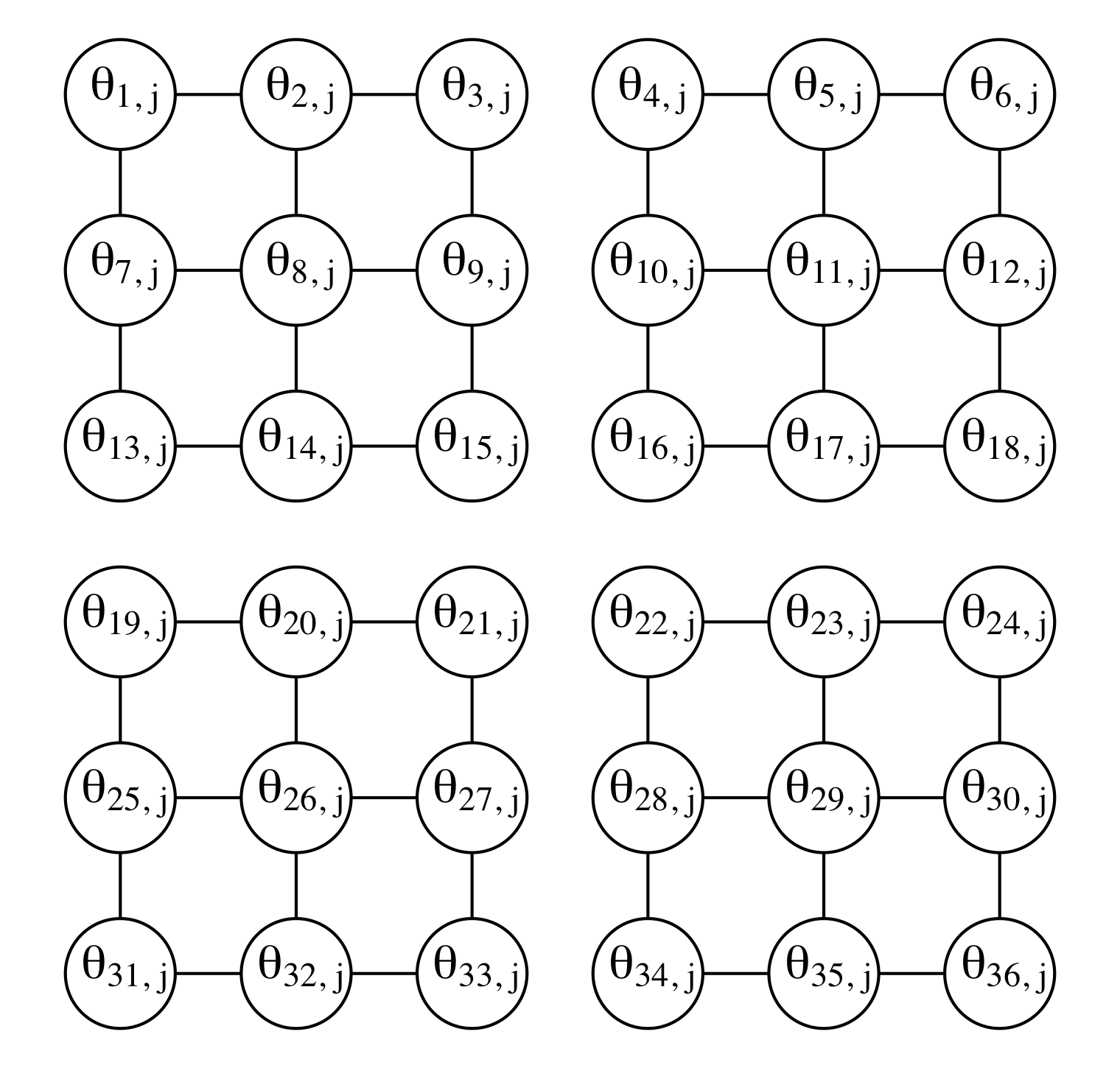}
    \subcaption{Graph 8}
\end{minipage}
\caption{Graphs representing regional adjacency relationships in SVC model in \eqref{eq:svcmodel}.}
\label{fig:graph}
\end{figure}

In order to evaluate how the developed method performs compared with the existing method, we conducted numerical experiments as follows.
In the SVC model \eqref{eq:svcmodel}, we set $\tilde{p}=3$, appropriately specify $\bm{\theta}=(\bm{\theta}_1^{\T},\bm{\theta}_2^{\T},\bm{\theta}_3^{\T})^{\T}$, independently generate $\psi_i\sim{\rm Multi}(1,\bm{p}_{\psi})$, $\tilde{x}_i\sim{\rm N}(0,5)$ and $\varepsilon_i\sim{\rm N}(0,\sigma^2)$, and then construct $y_i$.
Here, $\sigma^2$ is taken to be $1.0$, $1.5$, or $2.0$.
In this numerical experiment, the data are generated without including an intercept term, and the estimation of the intercept is not performed.
Moreover, in order to focus on evaluating the performance of information criteria for the regularization to fuse adjacent regression coefficients, we do not impose the regularization for variable selection that shrinks regression coefficients to zero.
As the graph $(\mathcal{V},\mathcal{E})$ representing the regions and their adjacency in the SVC model, we consider Graphs 1--8 in Figure \ref{fig:graph}, and the experiments using these are referred to as Cases 1--8, respectively.
Two regions connected by an edge are regarded as adjacent.
For each of Cases 1--8, we consider two settings for the sample size and the true values of the parameters, which are denoted by Setting 1 and Setting 2.
Specifically, the detailed setups for Cases 1--4 are summarized in Table \ref{tab:1}, for Cases 5 and 6 in Table \ref{tab:2}, and for Cases 7 and 8 in Table \ref{tab:3}.

\begin{table}[t]
\centering
\caption{Simulation setups for Cases 1--4.}
\label{tab:1}
\scalebox{0.85}{
\begin{tabular}{|c|c|c|c|c|}
\hline
 & \multicolumn{2}{c|}{Case 1} & \multicolumn{2}{c|}{Case 2} \\
\cline{2-5}
 & Setting 1 & Setting 2 & Setting 1 & Setting 2 \\
\hline
$n$ & 20 & 35 & 45 & 50 \\
\hline
$M$ & \multicolumn{2}{c|}{3}&  \multicolumn{2}{c|}{5} \\
\hline
$\bm{p}_{\psi}$ & $(1/3,1/3,1/3)$ & $(1/3,1/6,1/2)$ & $(1/5,\dots,1/5)$ & $(1/10,3/10,2/10,2/10,2/10)$ \\
\hline
$\bm{\theta}_1$ & \multicolumn{2}{c|}{$(1.0,1.0,1.0)^{\T}$} & \multicolumn{2}{c|}{$(1.0,1.0,1.0,1.0,1.0)^{\T}$} \\
\hline
$\bm{\theta}_2$ & \multicolumn{2}{c|}{$(2.0,-2.0,-3.5)^{\T}$} & \multicolumn{2}{c|}{$(2.0,2.0,1.5,2.5,1.5)^{\T}$} \\
\hline
$\bm{\theta}_3$ & \multicolumn{2}{c|}{$(3.0,-3.0,1.5)^{\T}$} & \multicolumn{2}{c|}{$(3.0,-2.5,-3.0,0.5,-0.5)^{\T}$} \\
\hline
\hline
 & \multicolumn{2}{c|}{Case 3} & \multicolumn{2}{c|}{Case 4} \\
\cline{2-5}
 & Setting 1 & Setting 2 & Setting 1 & Setting 2 \\
\hline
$n$ & 45 & 50 & 45 & 50 \\
\hline
$M$ & \multicolumn{2}{c|}{5}& \multicolumn{2}{c|}{5} \\
\hline
$\bm{p}_{\psi}$ & $(1/5,\dots,1/5)$ & $(1/10,3/10,2/10,2/10,2/10)$ & $(1/5,\dots,1/5)$ & $(1/10,2/10,1/10,2/10,4/10)$ \\
\hline
$\bm{\theta}_1$ & \multicolumn{2}{c|}{$(1.0,1.0,1.0,-1.5,-1.5)^{\T}$} & \multicolumn{2}{c|}{$(1.0,1.0,-1.5,-1.5,5.0)^{\T}$} \\
\hline
$\bm{\theta}_2$ & \multicolumn{2}{c|}{$(2.0,-2.0,-3.5,0.5,-0.5)^{\T}$} & \multicolumn{2}{c|}{$(2.0,-2.0,-2.5,-0.5,-3.5)^{\T}$} \\
\hline
$\bm{\theta}_3$ & \multicolumn{2}{c|}{$(3.0,-3.0,1.5,-1,-2.5)^{\T}$} & \multicolumn{2}{c|}{$(3.0,-3.0,-1.0,0.5,-5.0)^{\T}$} \\
\hline
\end{tabular}
}
\end{table}

\begin{table}[t!]
\centering
\caption{Simulation setups for Cases 5 and 6. $\bm{a}_m$ denotes an $m$-dimensional vector with all components $a$.}
\label{tab:2}
\scalebox{0.85}{
\begin{tabular}{|c|>{\centering\arraybackslash}m{7cm}|>{\centering\arraybackslash}m{9.5cm}|}
\hline
 & \multicolumn{2}{c|}{Case 5} \\
\cline{2-3}
 & Setting 1 & Setting 2 \\
\hline
$n$ & \multicolumn{2}{c|}{400}\\
\hline
$M$ & \multicolumn{2}{c|}{50}\\
\hline
$\bm{p}_{\psi}$ & \multicolumn{2}{c|}{$(1/50,\dots,1/50)$}\\
\hline
$\bm{\theta}_1$ & $\bm{1.0}_{50}^{\T}$ &
$\bm{\theta}_1 = A_{\beta_1} z_{\beta_1},\;
A_{\beta_1} =
\mathbf{1}_{50},\;
z_{\beta_1}\sim {\rm N}(0,3)$\\
\hline
$\bm{\theta}_2$ & $(\bm{-2.0}_{10}^{\T}, \bm{-1.5}_{10}^{\T}, \bm{-1.0}_{10}^{\T}, \bm{0.5}_{10}^{\T}, \bm{1.5}_{10}^{\T})^{\T}$ &
$\bm{\theta}_2 = A_{\beta_2} z_{\beta_2},\;
A_{\beta_2} =
\mathbf{1}_{50},\;
z_{\beta_2}\sim {\rm N}(0,3)$\\
\hline
$\bm{\theta}_3$ & 
$(\bm{-3.0}_{5}^{\T}, \bm{-2.5}_{5}^{\T}, \bm{-2.0}_{5}^{\T}, \bm{-1.5}_{5}^{\T}, \bm{-1.0}_{5}^{\T}, \newline
\bm{0.5}_{5}^{\T}, \bm{1.5}_{5}^{\T}, \bm{2.0}_{5}^{\T}, \bm{2.5}_{5}^{\T}, \bm{3.0}_{5}^{\T})^{\T}$ &
$\bm{\theta}_3=z_{\beta_{3}},\;z_{\beta_{3}} \sim {\rm N}(\mathbf{0}, \Sigma_{\beta_3}),\; \Sigma_{\beta_3} =
\underbrace{
\begin{pmatrix}
5 & 1.5 & \cdots & 1.5\\
1.5 & 5 & \cdots & 1.5\\
\vdots & \vdots & \ddots & \vdots\\
1.5 & 1.5 & \cdots & 5
\end{pmatrix}}_{50\times50}$
\\
\hline
\hline
 & \multicolumn{2}{c|}{Case 6} \\
\cline{2-3}
 & Setting 1 & Setting 2 \\
\hline
$n$ & \multicolumn{2}{c|}{400}\\
\hline
$M$ & \multicolumn{2}{c|}{36}\\
\hline
$\bm{p}_{\psi}$ & \multicolumn{2}{c|}{$(1/36,\dots,1/36)$}\\
\hline
$\bm{\theta}_1$ & $\bm{2.0}_{36}^{\T}$ &
$\bm{\theta}_1 = A_{\beta_1} z_{\beta_1},\;
A_{\beta_1} =
\mathbf{1}_{50},\;
z_{\beta_1}\sim {\rm N}(0,3)$\\
\hline
$\bm{\theta}_2$ & 
$(\bm{-2.0}_{2}^{\T}, \bm{-1.5}_{2}^{\T}, \bm{-1.0}_{2}^{\T}, \bm{-2.0}_{2}^{\T}, \newline
\bm{-1.5}_{2}^{\T}, \bm{-1.0}_{2}^{\T}, \bm{-2.0}_{2}^{\T}, \bm{-1.5}_{2}^{\T}, \newline 
\bm{-1.0}_{2}^{\T}, \bm{0.5}_{2}^{\T}, \bm{1.5}_{2}^{\T}, \bm{2.0}_{2}^{\T}, \bm{0.5}_{2}^{\T}, \newline
\bm{1.5}_{2}^{\T}, \bm{2.0}_{2}^{\T}, \bm{0.5}_{2}^{\T}, \bm{1.5}_{2}^{\T}, \bm{2.0}_{2}^{\T})^{\T}$ &
$\bm{\theta}_2 = A_{\beta_2} z_{\beta_2},\;
A_{\beta_2} =
\mathbf{1}_{50},\;
z_{\beta_2}\sim {\rm N}(0,3)$\\
\hline
$\bm{\theta}_3$ & 
$(\bm{-3.0}_{2}^{\T}, \bm{-2.5}_{2}^{\T}, \bm{-2.0}_{2}^{\T}, \bm{-3.0}_{2}^{\T}, \newline
\bm{-2.5}_{2}^{\T}, \bm{-2.0}_{2}^{\T}, \bm{-1.5}_{2}^{\T}, \bm{-1.0}_{2}^{\T}, \newline
\bm{0.5}_{2}^{\T}, \bm{-1.5}_{2}^{\T}, \bm{-1.0}_{2}^{\T}, \bm{0.5}_{2}^{\T}, \bm{1.0}_{2}^{\T}, \newline
\bm{2.0}_{2}^{\T}, \bm{2.5}_{2}^{\T}, \bm{1.0}_{2}^{\T}, \bm{2.0}_{2}^{\T}, \bm{2.5}_{2}^{\T})^{\T}$ &
$\bm{\theta}_3=z_{\beta_{3}},\;z_{\beta_{3}} \sim {\rm N}(\mathbf{0}, \Sigma_{\beta_3}),\; \Sigma_{\beta_3} =
\underbrace{
\begin{pmatrix}
5 & 1.5 & \cdots & 1.5\\
1.5 & 5 & \cdots & 1.5\\
\vdots & \vdots & \ddots & \vdots\\
1.5 & 1.5 & \cdots & 5
\end{pmatrix}}_{36\times36}$
\\
\hline
\end{tabular}
}
\end{table}

\begin{table}[p]
\caption{Simulation setups for Cases 7 and 8. $\bm{a}_m$ denotes an $m$-dimensional vector with all components $a$.}
\label{tab:3}
\centering
\scalebox{0.85}{
\begin{tabular}{|c|>{\centering\arraybackslash}m{6cm}|>{\centering\arraybackslash}m{10.5cm}|}
\hline
 & \multicolumn{2}{c|}{Case 7} \\
\cline{2-3}
 & Setting 1 & Setting 2 \\
\hline
$n$ & \multicolumn{2}{c|}{400}\\
\hline
$M$ & \multicolumn{2}{c|}{36}\\
\hline
$\bm{p}_{\psi}$ & \multicolumn{2}{c|}{$(1/36,\dots,1/36)$}\\
\hline
$\bm{\theta}_1$ & $(\bm{-2.0}_{18}^{\T}, \bm{2.0}_{18}^{\T})^{\T}$ &
$\bm{\theta}_1 = A_{\beta_1} z_{\beta_1},\;
A_{\beta_1} =
\begin{pmatrix}
\mathbf{1}_{18} & \mathbf{0}_{18} \\
\mathbf{0}_{18} & \mathbf{1}_{18}
\end{pmatrix},\;
z_{\beta_1}\sim {\rm N}(0,3I_2)$\\
\hline
$\bm{\theta}_2$ & 
$(\bm{-2.0}_{2}^{\T},\bm{-1.5}_{2}^{\T}, \bm{-1.0}_{2}^{\T},\bm{-2.0}_{2}^{\T},\newline
\bm{-1.5}_{2}^{\T},\bm{-1.0}_{2}^{\T},\bm{-2.0}_{2}^{\T}, \bm{-1.5}_{2}^{\T},\newline
\bm{-1.0}_{2}^{\T},\bm{1.0}_{2}^{\T},\bm{1.5}_{2}^{\T},\bm{2.0}_{2}^{\T},\bm{1.0}_{2}^{\T},\newline
\bm{1.5}_{2}^{\T},\bm{2.0}_{2}^{\T},\bm{1.0}_{2}^{\T}, \bm{1.5}_{2}^{\T},\bm{2.0}_{2}^{\T})^{\T}$ &
$\bm{\theta}_2 = A_{\beta_2} z_{\beta_2},\; 
A_{\beta_2} =
\begin{pmatrix}
\mathbf{1}_{18} & \mathbf{0}_{18} \\
\mathbf{0}_{18} & \mathbf{1}_{18}
\end{pmatrix},\;
z_{\beta_2}\sim {\rm N}(0,3I_2)$ \\
\hline
$\bm{\theta}_3$ &  $(\bm{-3.0}_{3}^{\T},\bm{-2.5}_{3}^{\T},\bm{-3.0}_{3}^{\T},\bm{-2.5}_{3}^{\T},\newline
\bm{-2.0}_{3}^{\T},\bm{-1.5}_{3}^{\T},\bm{0.5}_{3}^{\T},\bm{1.0}_{3}^{\T},\newline
\bm{0.5}_{3}^{\T},\bm{1.0}_{3}^{\T},\bm{1.5}_{3}^{\T}, \bm{2.0}_{3}^{\T})^{\T}$ &
$\bm{\theta}_3 = z_{\beta_3},\; z_{\beta_3}\sim {\rm N}(\bm{0},\Sigma_{\beta_3}),\;
\Sigma_{\beta_3} =
\begin{pmatrix}
\Sigma_1 & 0 \\
0 & \Sigma_2
\end{pmatrix},\newline
\Sigma_1 = 
\underbrace{
\begin{pmatrix}
5 & 0.5 & \cdots & 0.5 \\
0.5 & 5 & \cdots & 0.5 \\
\vdots & \vdots & \ddots & \vdots \\
0.5 & 0.5 & \cdots & 5
\end{pmatrix}}_{18\times18} ,
\Sigma_2 = 
\underbrace{
\begin{pmatrix}
3 & 0.9 & \cdots & 0.9 \\
0.9 & 3 & \cdots & 0.9 \\
\vdots & \vdots & \ddots & \vdots \\
0.9 & 0.9 & \cdots & 3
\end{pmatrix}}_{18\times18}$
\\
\hline
\hline
 & \multicolumn{2}{c|}{Case 8} \\
\cline{2-3}
 & Setting 1 & Setting 2 \\
\hline
$n$ & \multicolumn{2}{c|}{400}\\
\hline
$M$ & \multicolumn{2}{c|}{36} \\
\hline
$\bm{p}_{\psi}$ & \multicolumn{2}{c|}{$(1/36,\dots,1/36)$} \\
\hline
$\bm{\theta}_1$ &  
$(\bm{-2.0}_{3}^{\T}, \bm{-1.0}_{3}^{\T},\bm{-2.0}_{3}^{\T}, \bm{-1.0}_{3}^{\T},\newline
\bm{-2.0}_{3}^{\T},\bm{-1.0}_{3}^{\T},\bm{1.0}_{3}^{\T}, \bm{2.0}_{3}^{\T},\newline
\bm{1.0}_{3}^{\T},\bm{2.0}_{3}^{\T},\bm{1.0}_{3}^{\T},\bm{2.0}_{3}^{\T})^{\T}$ &
$\bm{\theta}_1 = A_{\beta_1} z_{\beta_1},\; A_{\beta_1} =
\begin{pmatrix}
\mathbf{1}_{9} & \mathbf{0}_{9} & \mathbf{0}_{9} & \mathbf{0}_{9} \\
\mathbf{0}_{9} & \mathbf{1}_{9} & \mathbf{0}_{9} & \mathbf{0}_{9} \\
\mathbf{0}_{9} & \mathbf{0}_{9} & \mathbf{1}_{9} & \mathbf{0}_{9} \\
\mathbf{0}_{9} & \mathbf{0}_{9} & \mathbf{0}_{9} & \mathbf{1}_{9} 
\end{pmatrix},\; z_{\beta_1}\sim {\rm N}(0,0.5I_4)$ \\
\hline
$\bm{\theta}_2$ & 
$(
\bm{-2.0}_{2}^{\T}, -1.5, \bm{-1.0}_{2}^{\T}, -0.5, \newline
\bm{-2.0}_{2}^{\T}, -1.5, \bm{-1.0}_{2}^{\T}, -0.5, \newline
\bm{-2.0}_{2}^{\T}, -1.5, \bm{-1.0}_{2}^{\T}, -0.5, \newline
\bm{0.5}_{2}^{\T}, 1.0, \bm{1.5}_{2}^{\T}, 2.0, \bm{0.5}_{2}^{\T}, 1.0, \newline
\bm{1.5}_{2}^{\T}, 2.0, \bm{0.5}_{2}^{\T}, 1.0, \bm{1.5}_{2}^{\T}, 2.0
)^{\T}$ &
$\bm{\theta}_2 = A_{\beta_2} z_{\beta_2},\; A_{\beta_2} =
\begin{pmatrix}
\mathbf{1}_{9} & \mathbf{0}_{9} & \mathbf{0}_{9} & \mathbf{0}_{9} \\
\mathbf{0}_{9} & \mathbf{1}_{9} & \mathbf{0}_{9} & \mathbf{0}_{9} \\
\mathbf{0}_{9} & \mathbf{0}_{9} & \mathbf{1}_{9} & \mathbf{0}_{9} \\
\mathbf{0}_{9} & \mathbf{0}_{9} & \mathbf{0}_{9} & \mathbf{1}_{9} 
\end{pmatrix},\; z_{\beta_2}\sim {\rm N}(0,0.5I_4)$ \\
\hline
$\bm{\theta}_3$ & $(\bm{-3.0}_{3}^{\T},\bm{-1.5}_{3}^{\T},\bm{-2.5}_{3}^{\T},\bm{-1.0}_{3}^{\T},\newline
\bm{-2.0}_{3}^{\T},\bm{-0.5}_{3}^{\T},\bm{0.5}_{3}^{\T},\bm{2.0}_{3}^{\T},\newline
\bm{1.0}_{3}^{\T},\bm{2.5}_{3}^{\T},\bm{1.5}_{3}^{\T},\bm{3.0}_{3}^{\T})^{\T}$ &
$\bm{\theta}_3 = z_{\beta_3},\; \;z_{\beta_3}\sim {\rm N}(\bm{0},\Sigma_{\beta_3}),\newline
\Sigma_{\beta_3} =
\begin{pmatrix}
\Sigma & 0 & 0 & 0\\
0 & \Sigma & 0 & 0\\
0 & 0 & \Sigma & 0\\
0 & 0 & 0 & \Sigma
\end{pmatrix},\;
\Sigma =
\underbrace{\begin{pmatrix}
5 & 0.5 & \cdots & 0.5 \\
0.5 & 5 & \cdots & 0.5 \\
\vdots & \vdots & \ddots & \vdots \\
0.5 & 0.5 & \cdots & 5
\end{pmatrix}}_{9\times9}$
\\
\hline
\end{tabular}
}
\end{table}

For each experimental setting, we generate 100 datasets and compute four information criteria: WAIC (WAIC1) and PIIC1 for Model 1 explained in Section \ref{sec:2-3}, and WAIC (WAIC2) and PIIC2 for the combination of Model 1 and Model 2.
Then, when using WAIC1 and PIIC1, we select the hyperparameter of Model 1 and construct the Bayesian predictive distribution, whereas when using WAIC2 and PIIC2, we select not only the hyperparameters but also between Model 1 and Model 2 and construct the Bayesian predictive distribution, to compare them.
It should be noted, however, that whereas PIIC2 attempts to select the more appropriate of Model 1 and Model 2, WAIC2 eventually always selects Model 2.

Model 1 is the model in which the prior distribution of the differences of adjacent regression coefficients is given by a Laplace distribution with a common hyperparameter for all explanatory variables, so the number of hyperparameters is one in that case.
In contrast, Model 2 is the model in which the prior distribution is given by Laplace distributions with different hyperparameters for each explanatory variable. Specifically, the number of hyperparameters is 3, giving it a larger complexity than Model 1.
As the concrete procedure of hyperparameter selection, in Model 1, we simply generate 20 candidate values and search for the optimal one.
In Model 2, we sequentially focus on one hyperparameter at a time, generating 20 candidate values for it and then searching for the optimal value, and repeat this until a predetermined level of convergence is achieved.

As an evaluation measure for the information criteria, we use as risk the Kullback-Leibler divergence between the predictive distribution obtained and the true distribution, that is, $\E_{\tilde{y}_i,\tilde{\bm{x}}_i}[-\log f(\tilde{y}_i,\tilde{\bm{x}}_i\mid \bm{y},\bm{X};\bm{\lambda})]$ or $\E_{\tilde{y}_i,\tilde{\bm{x}}_i}[-\log f(\tilde{y}_i,\tilde{\bm{x}}_i\mid \bm{y},\bm{X};\hat{\bm{\lambda}})]$.
More precisely, the expectation is evaluated by its empirical version; that is, the expectation operator is replaced with $n^{-1}\sum_{i=1}^n$.
As a reference measure, we also report the rates in which WAIC yields a smaller or larger risk than PIIC.

Table \ref{tab:4} summarizes the results for Cases 1--4.
In Cases 1--4, for almost all experimental settings, the risk given by PIIC1 is smaller than that given by WAIC1.
This suggests that the asymptotic setting on which PIIC1 is based is more appropriate; nevertheless, the difference between them is small.
This agrees with the experimental results of \cite{ninomiya2021prior}, which show that although the superiority of PIIC1 over WAIC1 is clear for non-sparse estimation, PIIC1 has only a slight superiority for sparse estimation.
On the other hand, for all experimental settings, the risk given by PIIC2 is clearly smaller than that given by WAIC2.
The reason for the clear difference is that in Cases 1--4, the number of vertices in the graph is small, and simpler models tend to perform better, but WAIC2 always selects the more complex model.
In fact, when comparing PIIC1 and PIIC2, PIIC2 always yields smaller risk; however, the difference is not large, indicating that little benefit is gained from allowing the selection of the more complex model.

\begin{table}[p]
\begin{center}
\caption{Comparison of information criteria (Cases 1--4). Each value for WAIC1, PIIC1, WAIC2, and PIIC2 is the average of evaluations of Kullback-Leibler divergence between the true and Bayesian predictive distributions. For pairs WAIC1 and PIIC1, and WAIC2 and PIIC2, the smaller of the two values is shown in bold. Each vector for Rate 1 (Rate 2) indicates the rates at which the evaluation for WAIC1 (WAIC2) is smaller than, equal to, and larger than the evaluation for PIIC1 (PIIC2).}
\label{tab:4}
\begin{tabular}{ccccccccc}
\toprule
Setting&$\sigma^2$&Case&WAIC1&PIIC1&Rate1&WAIC2&PIIC2&Rate2\\
\hline
\multirow{12}{*}{1}&\multirow{4}{*}{$1.0$}&1&1.314&$\bm{1.311}$&(3,91,6)&1.515&$\bm{1.293}$&(28,0,72)\\
&&2&0.875&$\bm{0.868}$&(0,94,6)&1.033&$\bm{0.864}$&(16,0,84)\\
&&3&0.992&$\bm{0.991}$&(1,95,4)&1.136&$\bm{0.982}$&(24,2,74)\\
&&4&1.097&$\bm{1.091}$&(0,94,6)&1.228&$\bm{1.088}$&(27,0,73)\\
\cline{2-9}
&\multirow{4}{*}{$1.5$}&1&1.473&$\bm{1.468}$&(1,95,4)&1.639&$\bm{1.458}$&(26,0,74)\\
&&2&1.123&$\bm{1.117}$&(0,97,3)&1.272&$\bm{1.111}$&(22,0,78)\\
&&3&1.280&$\bm{1.277}$&(0,98,2)&1.382&$\bm{1.258}$&(27,1,72)\\
&&4&1.315&$\bm{1.314}$&(2,95,3)&1.493&$\bm{1.305}$&(30,0,70)\\
\cline{2-9}
&\multirow{4}{*}{$2.0$}&1&1.723&$\bm{1.721}$&(0,97,3)&1.905&$\bm{1.715}$&(34,0,66)\\
&&2&1.371&$\bm{1.370}$&(0,97,3)&1.458&$\bm{1.369}$&(30,1,69)\\
&&3&1.556&$\bm{1.554}$&(1,96,3)&1.598&$\bm{1.539}$&(33,2,65)\\
&&4&1.595&$\bm{1.593}$&(0,97,3)&1.797&$\bm{1.592}$&(31,0,69)\\
\hline
\multirow{12}{*}{2}&\multirow{4}{*}{$1.0$}&1&0.990&$\bm{0.984}$&(1,91,8)&1.077&$\bm{0.978}$&(22,0,78)\\
&&2&0.815&$\bm{0.814}$&(2,94,4)&0.918&$\bm{0.813}$&(24,0,76)\\
&&3&0.892&$\bm{0.891}$&(2,93,5)&0.979&$\bm{0.888}$&(28,0,72)\\
&&4&0.972&$\bm{0.965}$&(0,94,6)&1.122&$\bm{0.962}$&(19,0,81)\\
\cline{2-9}
&\multirow{4}{*}{$1.5$}&1&1.283&$\bm{1.281}$&(1,92,7)&1.327&$\bm{1.276}$&(30,0,70)\\
&&2&1.060&$\bm{1.059}$&(0,98,2)&1.132&$\bm{1.054}$&(31,0,69)\\
&&3&1.179&$\bm{1.179}$&(0,98,2)&1.242&$\bm{1.175}$&(28,0,72)\\
&&4&1.252&$\bm{1.251}$&(0,97,3)&1.386&$\bm{1.246}$&(31,0,69)\\
\cline{2-9}
&\multirow{4}{*}{$2.0$}&1&1.552&$\bm{1.550}$&(1,96,3)&1.581&$\bm{1.549}$&(37,0,63)\\
&&2&1.365&$\bm{1.359}$&(0,96,4)&1.420&$\bm{1.346}$&(33,1,66)\\
&&3&1.484&$\bm{1.483}$&(0,98,2)&1.540&$\bm{1.474}$&(37,0,63)\\
&&4&1.505&$\bm{1.504}$&(1,97,2)&1.603&$\bm{1.493}$&(37,0,63)\\
\bottomrule
\end{tabular}
\end{center}
\end{table}

Table \ref{tab:5} summarizes the results for Cases 5--8.
Compared with Cases 1--4, the number of vertices in the graph is greater, meaning these are settings where more complex models might sometimes perform better.
In fact, when comparing WAIC2 and PIIC2, the differences are not as large as in Cases 1--4, and there are three settings in which, although the differences are small, WAIC2 yields smaller risk.
Here, it should be noted that when $\sigma^2$ is larger, there is a tendency for the more complex model to be more appropriate.
Furthermore, when comparing PIIC1 and PIIC2, the effect of allowing the number of hyperparameters to be selected becomes evident, and it can be confirmed that PIIC2 outperforms PIIC1 for all experimental settings by a larger margin than in Cases 1--4.

Let us summarize what we wish to emphasize in Table \ref{tab:5}.
For these experimental settings, sometimes the simpler Model 1 is more appropriate, and sometimes the more complex Model 2 is more appropriate.
PIIC1 only considers Model 1, and WAIC2 always selects Model 2, but PIIC2 generally yields smaller risk than either of them.
More precisely, settings are considered in which always selecting Model 2 becomes preferable, but even then PIIC2 provides results comparable to those of WAIC2.
In other words, PIIC2 appropriately selects the number of hyperparameters according to the data and provides predictive distributions with good performance.

\begin{table}[p]
\begin{center}
\caption{Comparison of information criteria (Cases 5--8). Each value for WAIC1, PIIC1, WAIC2, and PIIC2 is the average of evaluations of Kullback-Leibler divergence between the true and Bayesian predictive distributions. For pairs WAIC1 and PIIC1, and WAIC2 and PIIC2, the smaller of the two values is shown in bold. Each vector for Rate 1 (Rate 2) indicates the rates at which the evaluation for WAIC1 (WAIC2) is smaller than, equal to, and larger than the evaluation for PIIC1 (PIIC2).}
\label{tab:5}
\begin{tabular}{ccccccccc}
\toprule
Setting&$\sigma^2$&Case&WAIC1&PIIC1&Rate1&WAIC2&PIIC2&Rate2\\
\hline
\multirow{12}{*}{1}&\multirow{4}{*}{$1.0$}&5&7.309&7.309&(0,100,0)&7.545&$\bm{7.098}$&(23,5,72)\\
&&6&9.450&9.450&(0,100,0)&9.644&$\bm{9.265}$&(26,0,74)\\
&&7&8.805&$\bm{8.799}$&(0,96,4)&9.289&$\bm{8.737}$&(8,1,91)\\
&&8&7.040&$\bm{7.034}$&(0,92,8)&7.392&$\bm{6.994}$&(8,1,91)\\
\cline{2-9}
&\multirow{4}{*}{$1.5$}&5&5.854&$\bm{5.854}$&(0,99,1)&5.880&$\bm{5.651}$&(33,4,63)\\
&&6&7.069&7.069&(0,100,0)&6.971&$\bm{6.898}$&(37,4,59)\\
&&7&6.692&$\bm{6.692}$&(0,99,1)&6.884&$\bm{6.601}$&(22,2,76)\\
&&8&5.463&$\bm{5.462}$&(0,99,1)&5.661&$\bm{5.376}$&(12,5,83)\\
\cline{2-9}
&\multirow{4}{*}{$2.0$}&5&5.356&$\bm{5.355}$&(0,99,1)&5.318&$\bm{5.184}$&(38,3,59)\\
&&6&5.997&5.997&(0,100,0)&5.892&$\bm{5.881}$&(40,7,53)\\
&&7&5.699&5.699&(0,100,0)&5.753&$\bm{5.582}$&(25,7,68)\\
&&8&4.777&4.777&(0,100,0)&4.847&$\bm{4.690}$&(27,8,65)\\
\hline
\multirow{12}{*}{2}&\multirow{4}{*}{$1.0$}&5&4.402&$\bm{4.401}$&(0,98,2)&4.670&$\bm{4.344}$&(24,0,76)\\
&&6&4.026&4.026&(0,100,0)&4.037&$\bm{3.963}$&(38,0,62)\\
&&7&5.502&5.502&(0,100,0)&5.605&$\bm{5.453}$&(36,0,64)\\
&&8&4.275&4.275&(0,100,0)&4.339&$\bm{4.238}$&(38,0,62)\\
\cline{2-9}
&\multirow{4}{*}{$1.5$}&5&3.814&3.814&(0,100,0)&3.944&$\bm{3.725}$&(31,0,69)\\
&&6&3.306&$\bm{3.306}$&(0,99,1)&3.279&$\bm{3.233}$&(41,3,56)\\
&&7&4.394&4.394&(0,100,0)&$\bm{4.280}$&4.292&(49,2,49)\\
&&8&3.480&3.480&(0,100,0)&3.478&$\bm{3.396}$&(33,2,65)\\
\cline{2-9}
&\multirow{4}{*}{$2.0$}&5&3.694&3.694&(0,100,0)&3.781&$\bm{3.620}$&(34,1,65)\\
&&6&$\bm{3.193}$&3.193&(2,98,0)&3.090&$\bm{3.087}$&(42,6,52)\\
&&7&3.944&3.944&(0,100,0)&$\bm{3.857}$&3.866&(45,5,50)\\
&&8&3.277&3.277&(0,100,0)&$\bm{3.184}$&3.193&(51,5,44)\\
\bottomrule
\end{tabular}
\end{center}
\end{table}

\section{Real data analysis}\label{sec5}

In order to check how much difference the proposed PIIC makes compared with the existing WAIC in an actual analysis, we conducted an analysis using dust data collected from houses in the United States.
This dataset was gathered in the Wild Life of Our Homes project, and multiple analyses have been conducted, starting with \cite{barberan2015continental}.
Among those studies, the one having the greatest influence on the present paper is \cite{zhao2020solution}, in which the generalized fused lasso was applied, although no choice was made about how the regularization parameters were assigned.

\begin{figure}[t]
    \centering
    \includegraphics[scale=0.2]{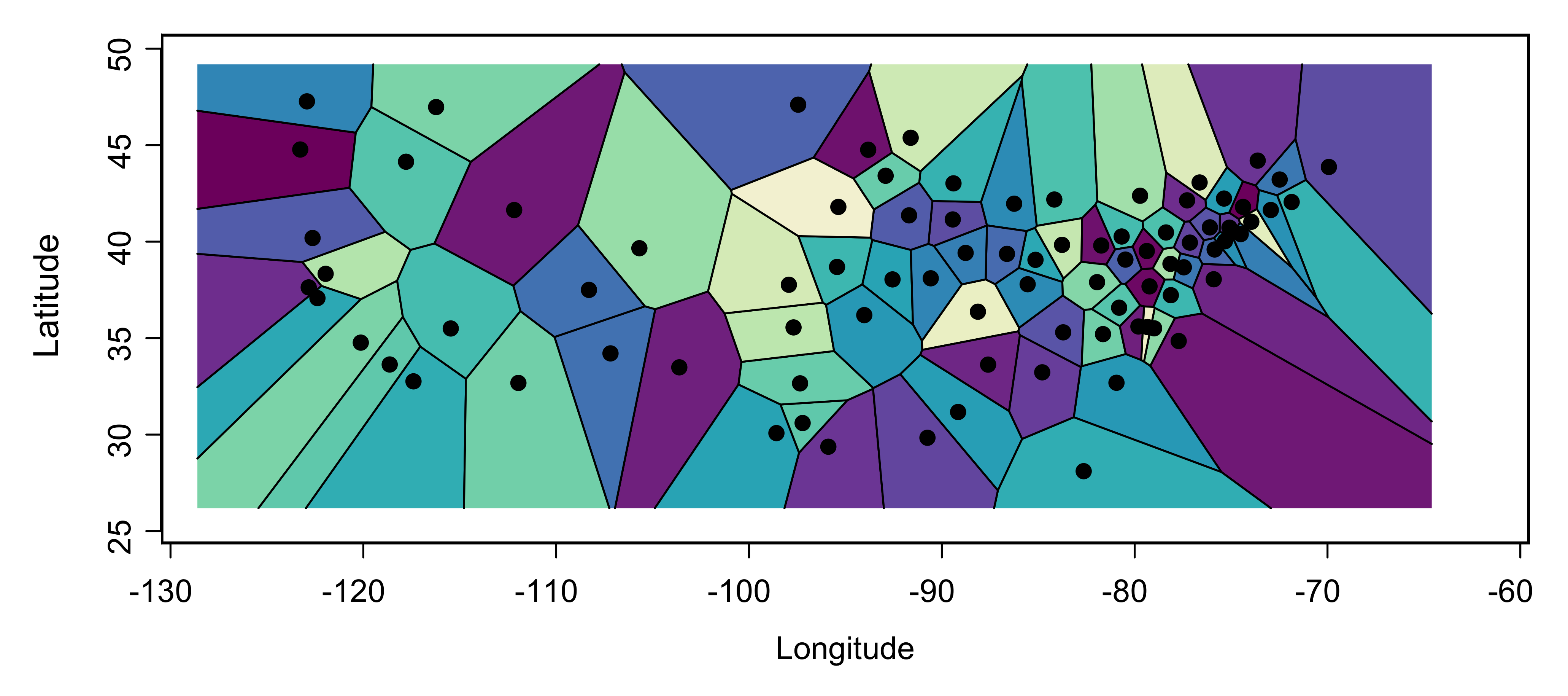}
    \caption{Adjacency structure of regions determined by Voronoi tessellation.}
    \label{volonoi}
\end{figure}

For the analysis present here, we use samples from 1,070 houses, excluding samples containing missing values and data from Alaska and Hawaii.
The response variable is fungal diversity (the proportion of the number of fungal species in each sample to the total number of species, 763), and we consider two explanatory variables.
Specifically, for Setting 1, Variables 1 and 2 are mean annual temperature and elevation, and for Setting 2, they are mean annual precipitation and net primary productivity (NPP), respectively.
Applying the k-means method, we divide the data into 85 clusters according to the latitude and longitude of the houses, and regard these clusters as regions.
Then, based on the centroids of each cluster, we perform a Voronoi tessellation (see Figure \ref{volonoi}), which determines the adjacency relationships of the regions.
As in Section \ref{sec5}, we focus on the regularization for fusing adjacent regression coefficients, and do not perform variable selection.
As the models to be fitted to the data, we consider Model 1 and Model 2 explained in Section \ref{sec:2-3}.
In addition, for WAIC1, we compute it in Model 1 with 20 candidate values of $\lambda_2$, and for WAIC2 and PIIC2, we compute them in Model 2 by repeatedly generating 20 candidate values of $\lambda_{2,j}$ as $j$ varies.
Note that we consider computing PIIC1 to not be necessary, based on the results of the numerical experiments.

\begin{figure}[t]
  \centering

  \begin{minipage}[b]{0.49\linewidth}
    \centering
    \includegraphics[width=\linewidth]{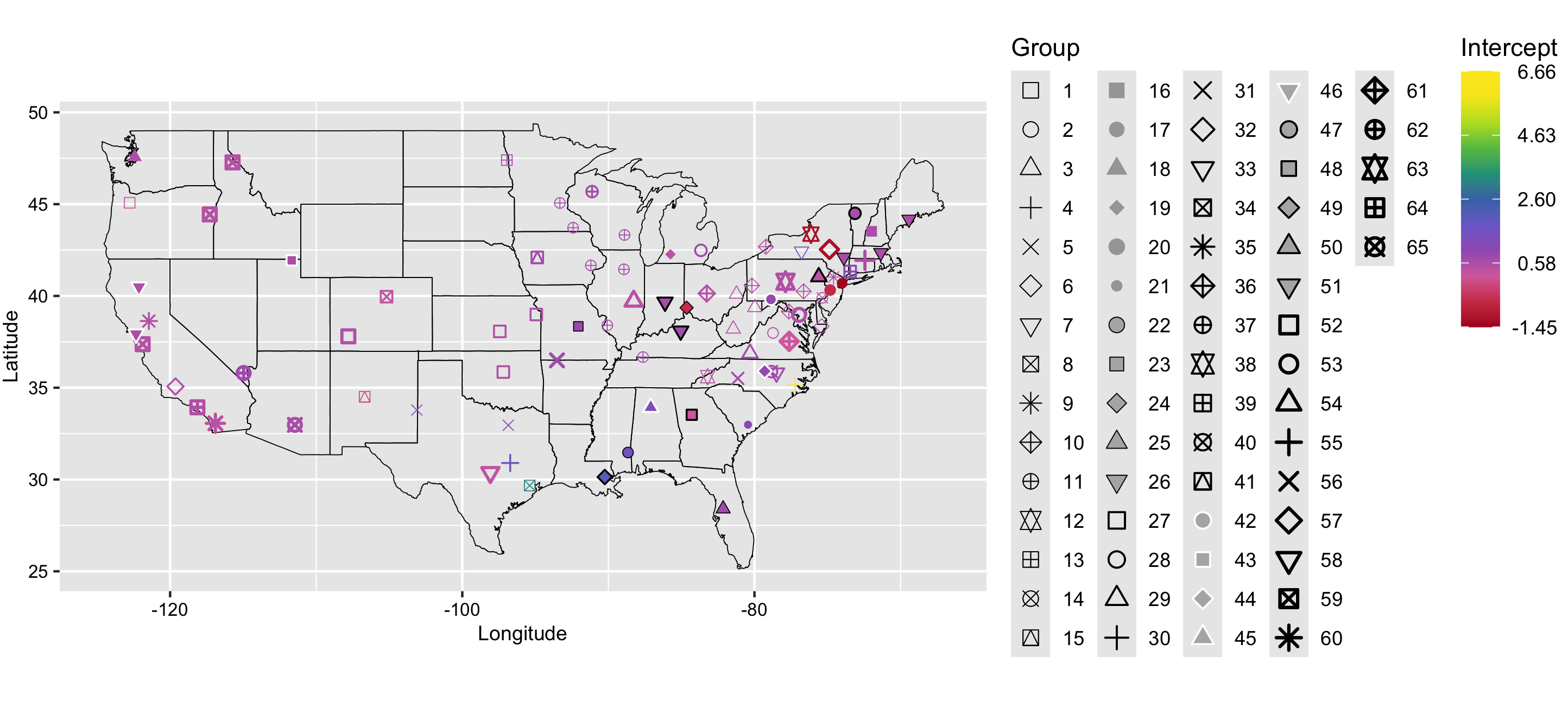}
    \subcaption{WAIC1 for Setting 1}
  \end{minipage}%
  \hspace{0.01\linewidth}%
  \begin{minipage}[b]{0.49\linewidth}
    \centering
    \includegraphics[width=\linewidth]{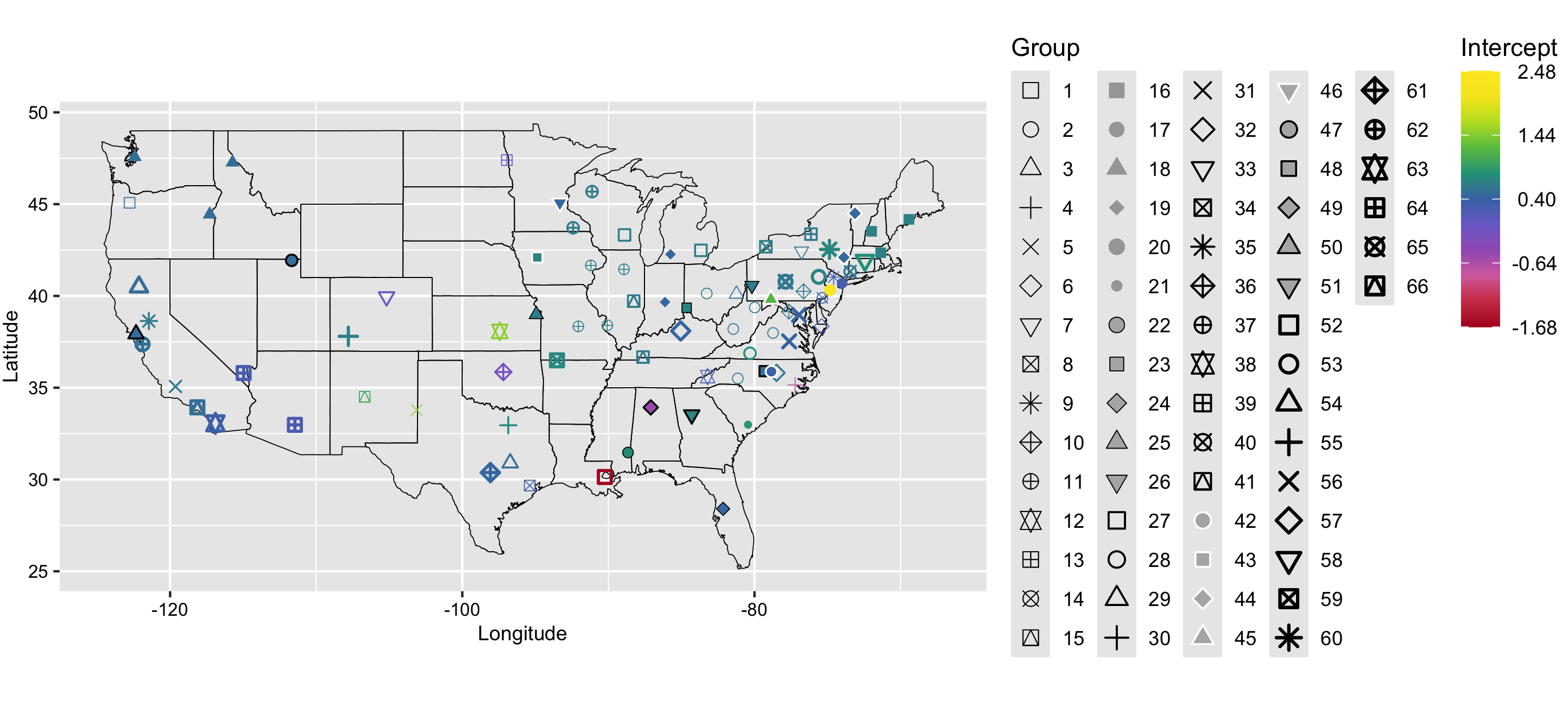}
    \subcaption{WAIC1 for Setting 2}
  \end{minipage}

  \vspace{1.2em}

  \begin{minipage}[b]{0.49\linewidth}
    \centering
    \includegraphics[width=\linewidth]{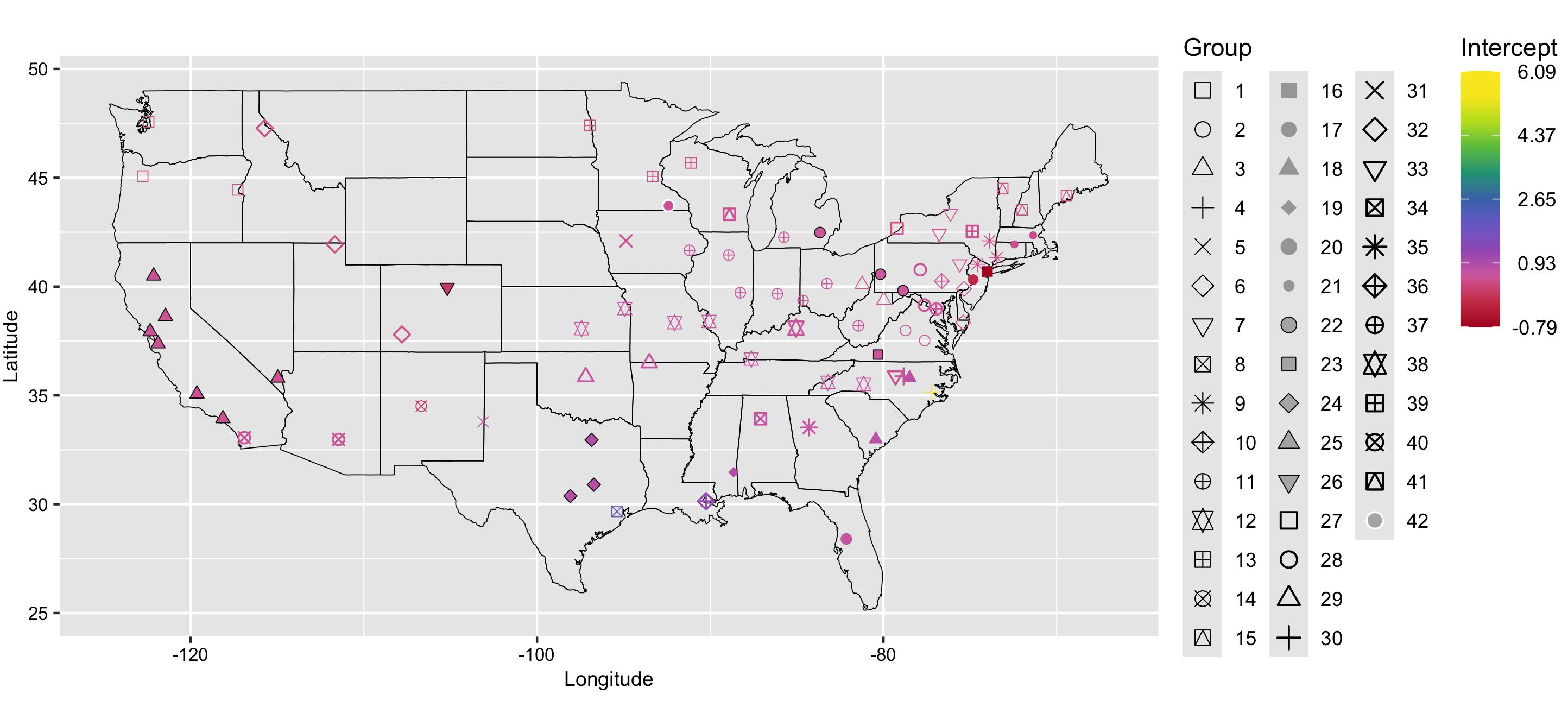}
    \subcaption{WAIC2 for Setting 1}
  \end{minipage}%
  \hspace{0.01\linewidth}%
  \begin{minipage}[b]{0.49\linewidth}
    \centering
    \includegraphics[width=\linewidth]{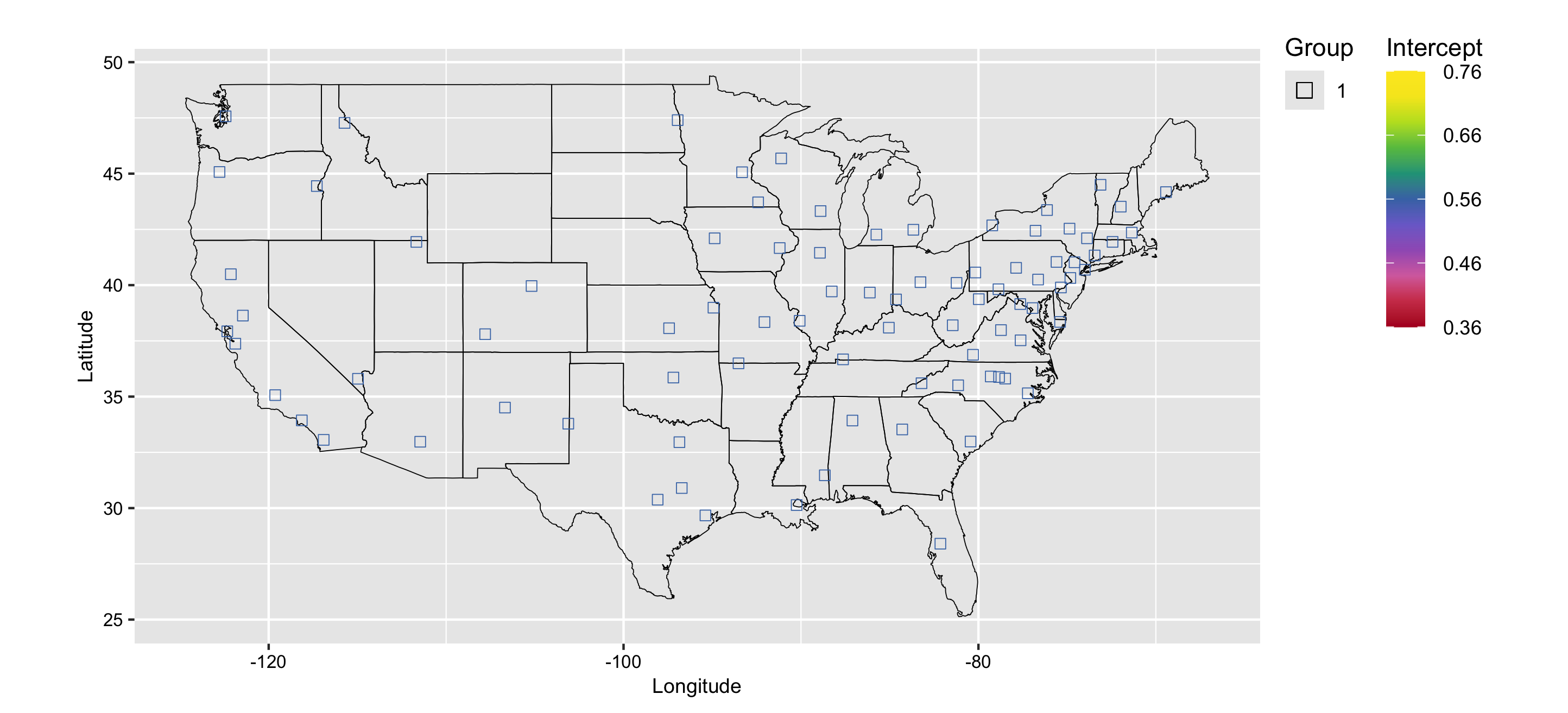}
    \subcaption{WAIC2 for Setting 2}
  \end{minipage}

  \vspace{1.2em}

  \begin{minipage}[b]{0.49\linewidth}
    \centering
    \includegraphics[width=\linewidth]{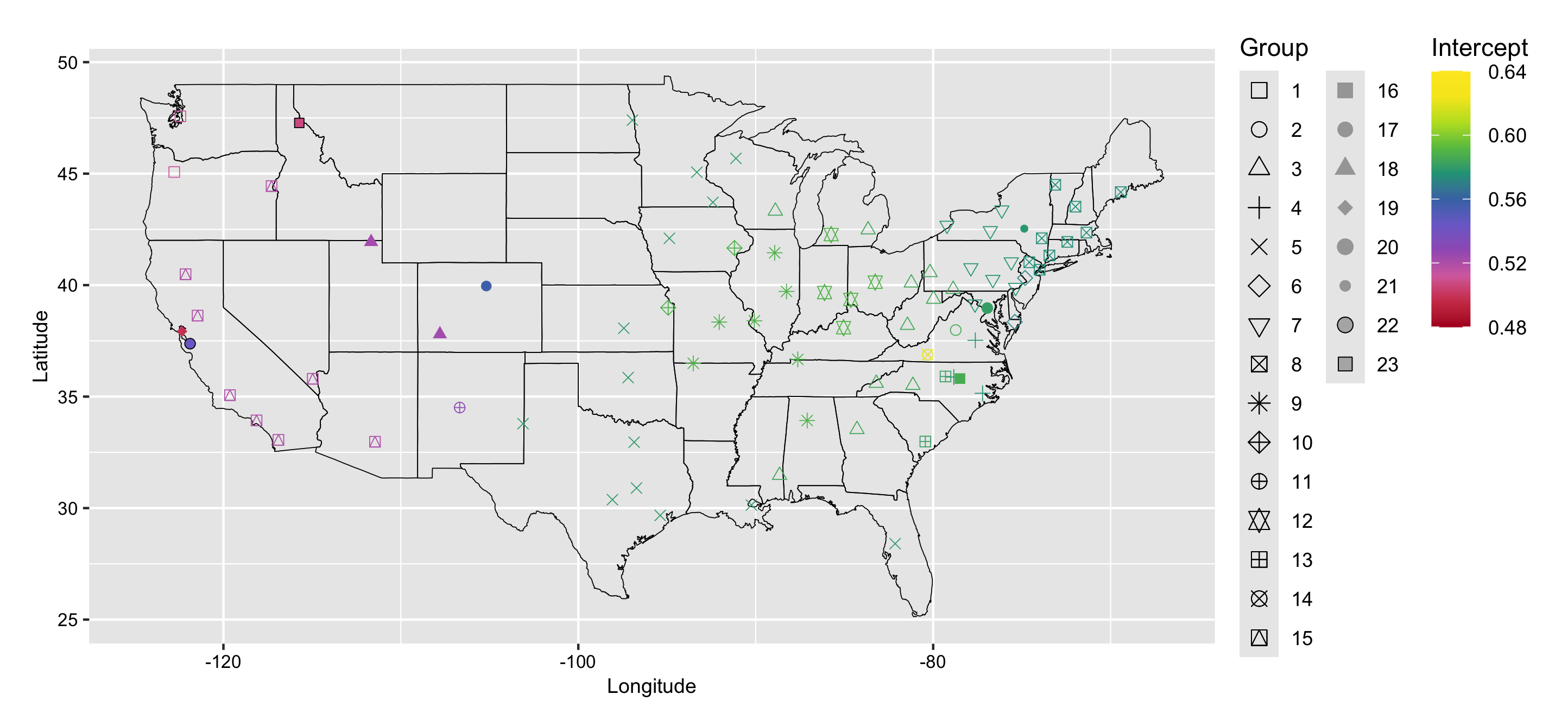}
    \subcaption{PIIC2 for Setting 1}
  \end{minipage}%
  \hspace{0.01\linewidth}%
  \begin{minipage}[b]{0.49\linewidth}
    \centering
    \includegraphics[width=\linewidth]{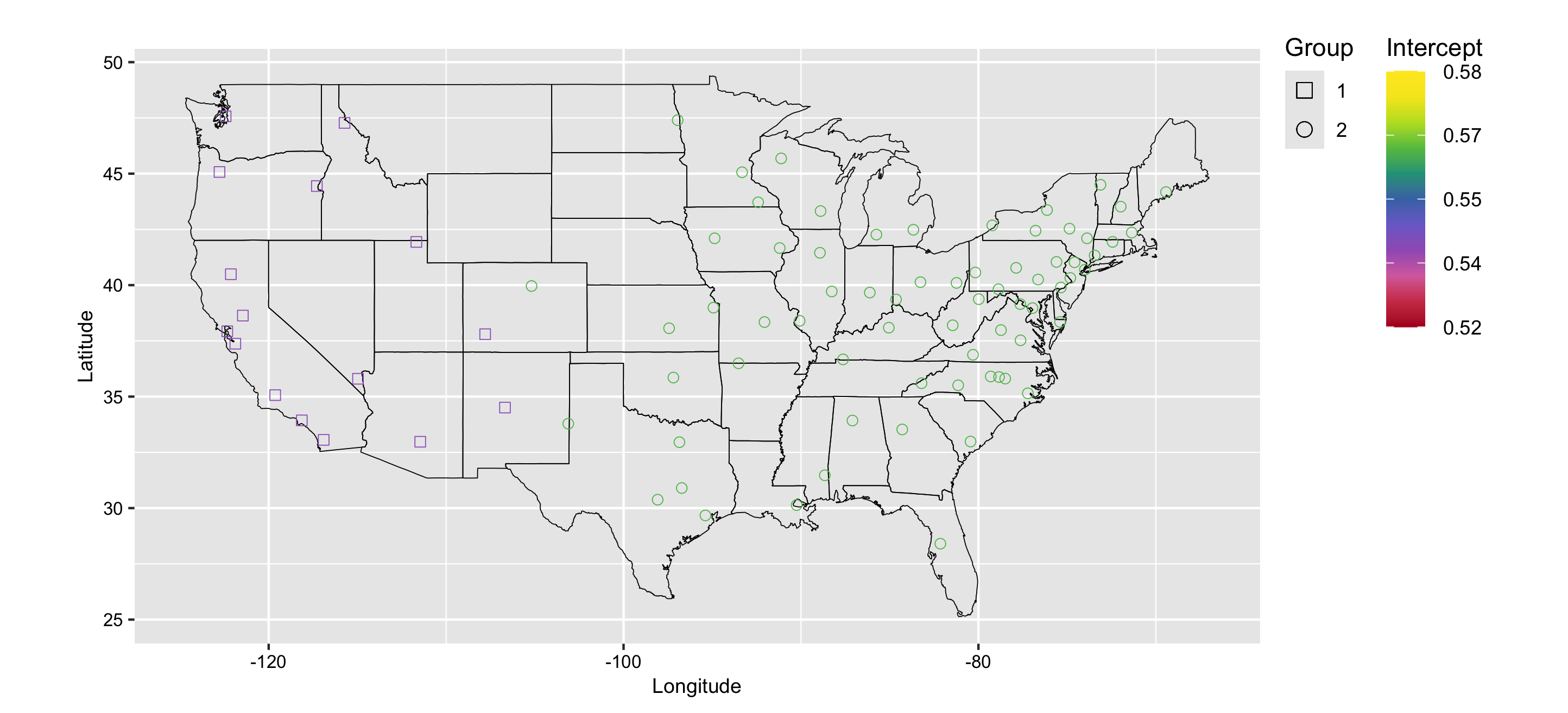}
    \subcaption{PIIC2 for Setting 2}
  \end{minipage}

\caption{Group structure in intercept of model selected by each information criterion.}
\label{intercept}
\end{figure}

\begin{figure}[t]
  \centering

  \begin{minipage}[b]{0.49\linewidth}
    \centering
    \includegraphics[width=\linewidth]{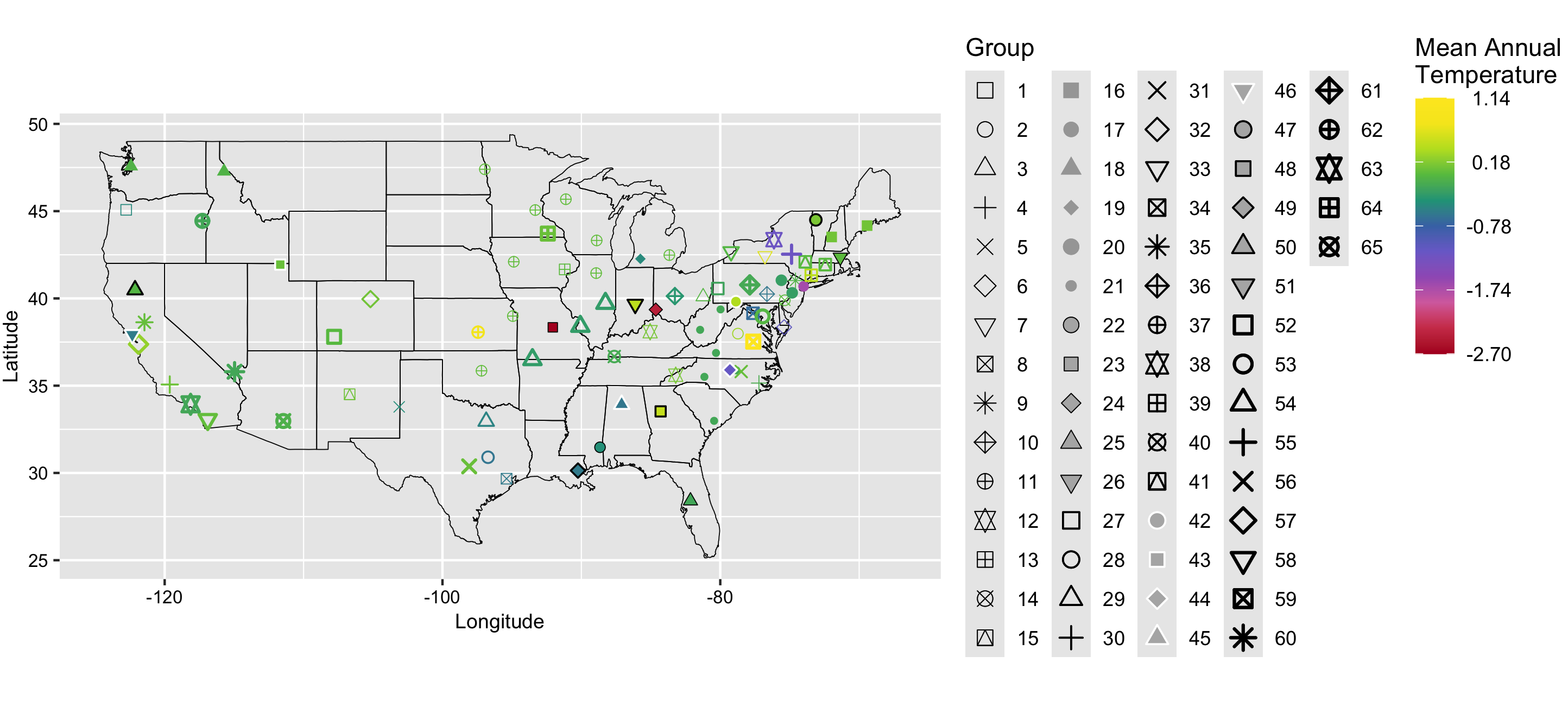}
    \subcaption{WAIC1 for Setting 1}
  \end{minipage}%
  \hspace{0.01\linewidth}%
  \begin{minipage}[b]{0.49\linewidth}
    \centering
    \includegraphics[width=\linewidth]{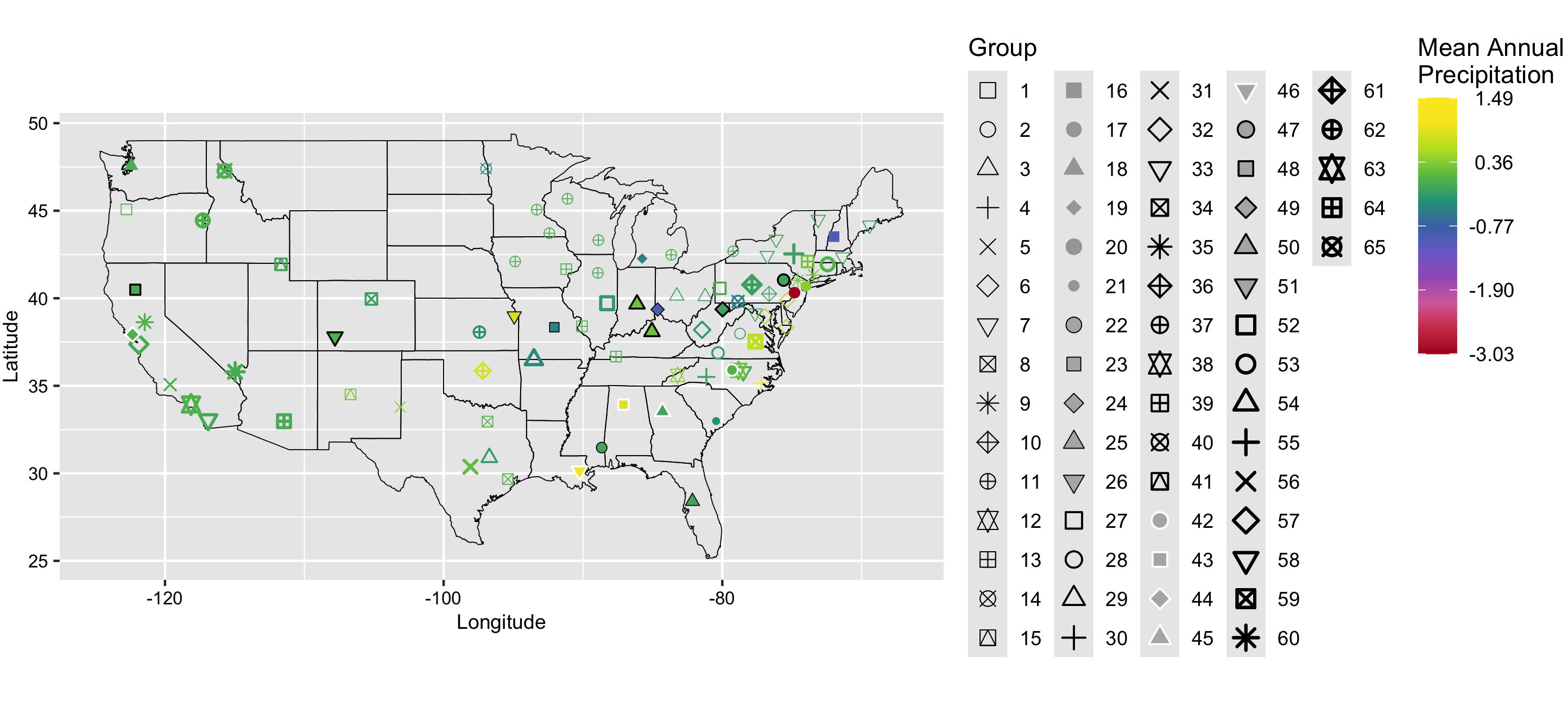}
    \subcaption{WAIC1 for Setting 2}
  \end{minipage}

  \vspace{1.2em}

  \begin{minipage}[b]{0.49\linewidth}
    \centering
    \includegraphics[width=\linewidth]{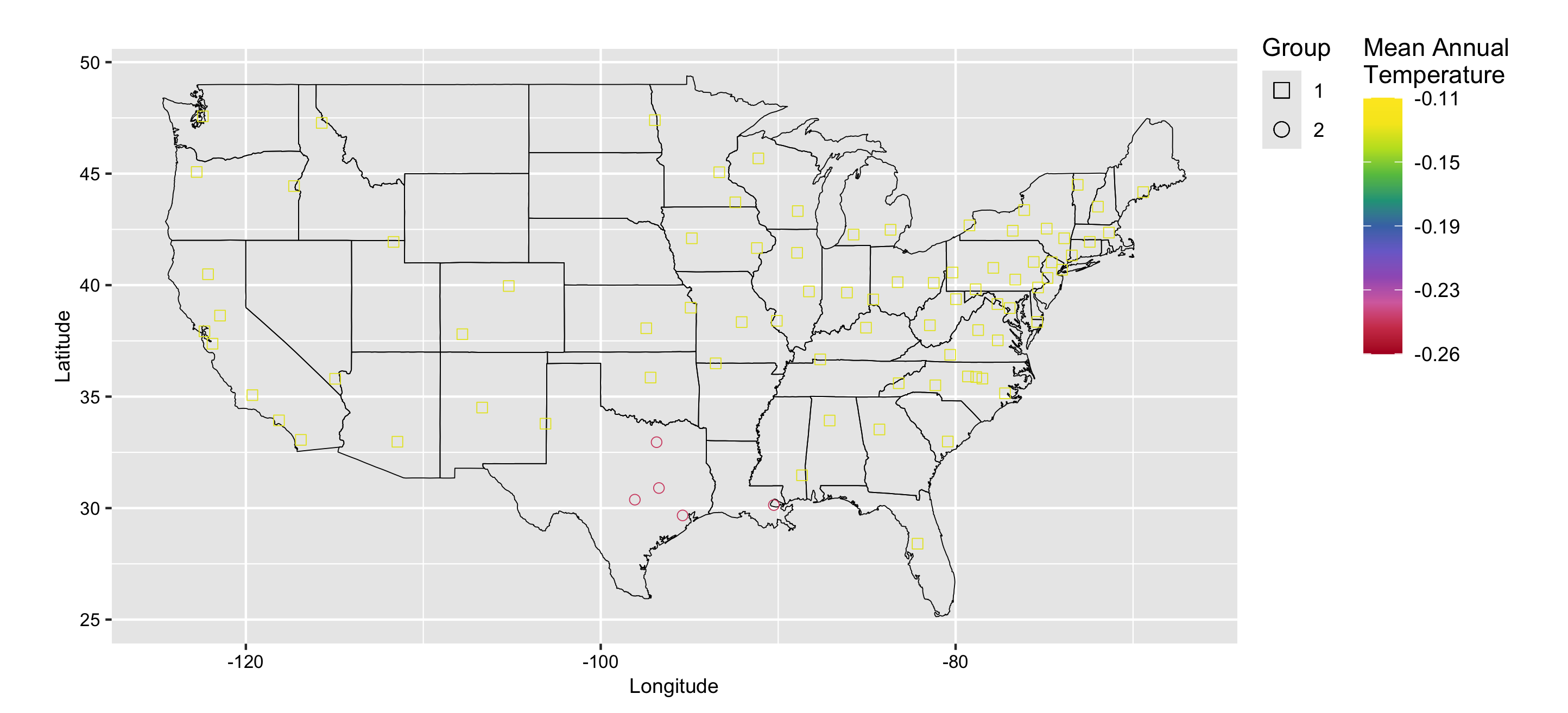}
    \subcaption{WAIC2 for Setting 1}
  \end{minipage}%
  \hspace{0.01\linewidth}%
  \begin{minipage}[b]{0.49\linewidth}
    \centering
    \includegraphics[width=\linewidth]{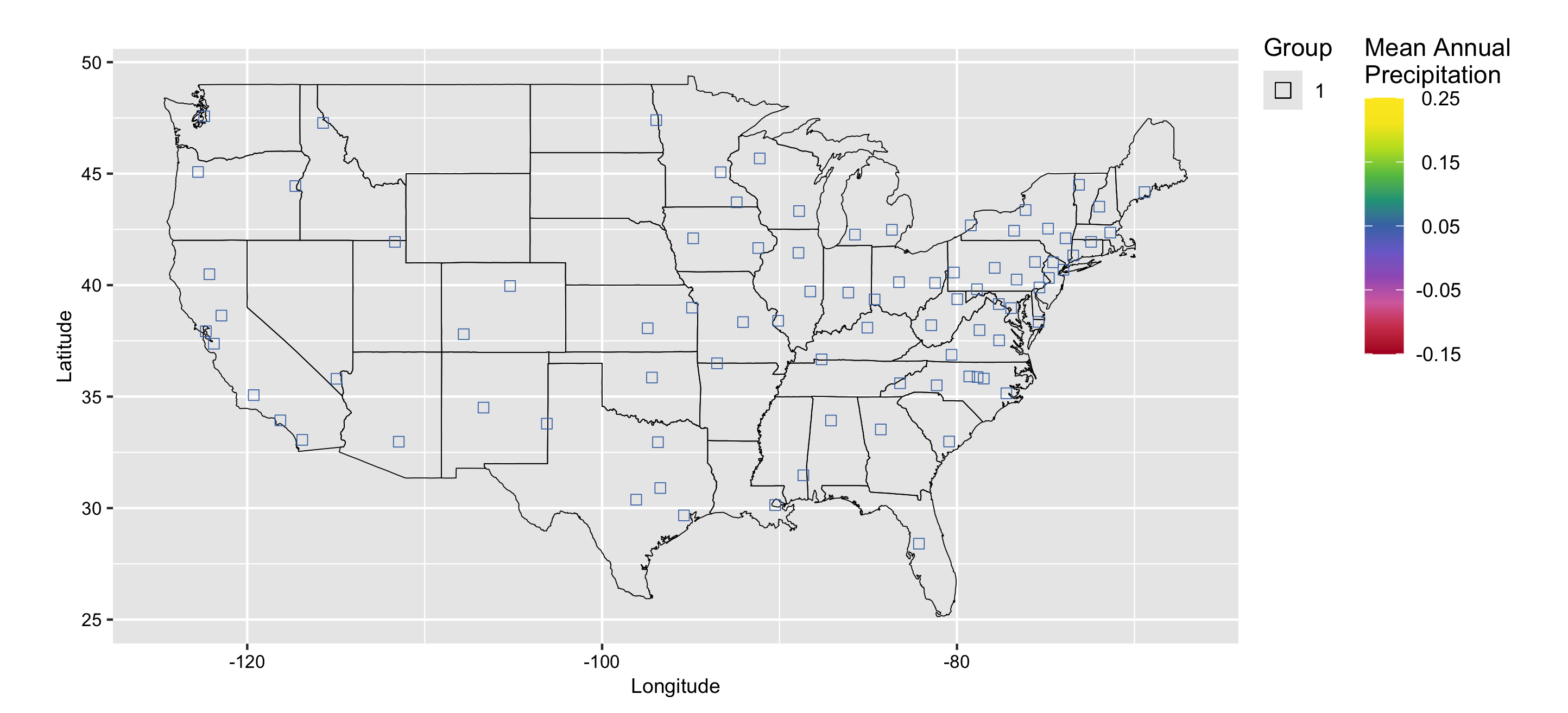}
    \subcaption{WAIC2 for Setting 2}
  \end{minipage}

  \vspace{1.2em}

  \begin{minipage}[b]{0.49\linewidth}
    \centering
    \includegraphics[width=\linewidth]{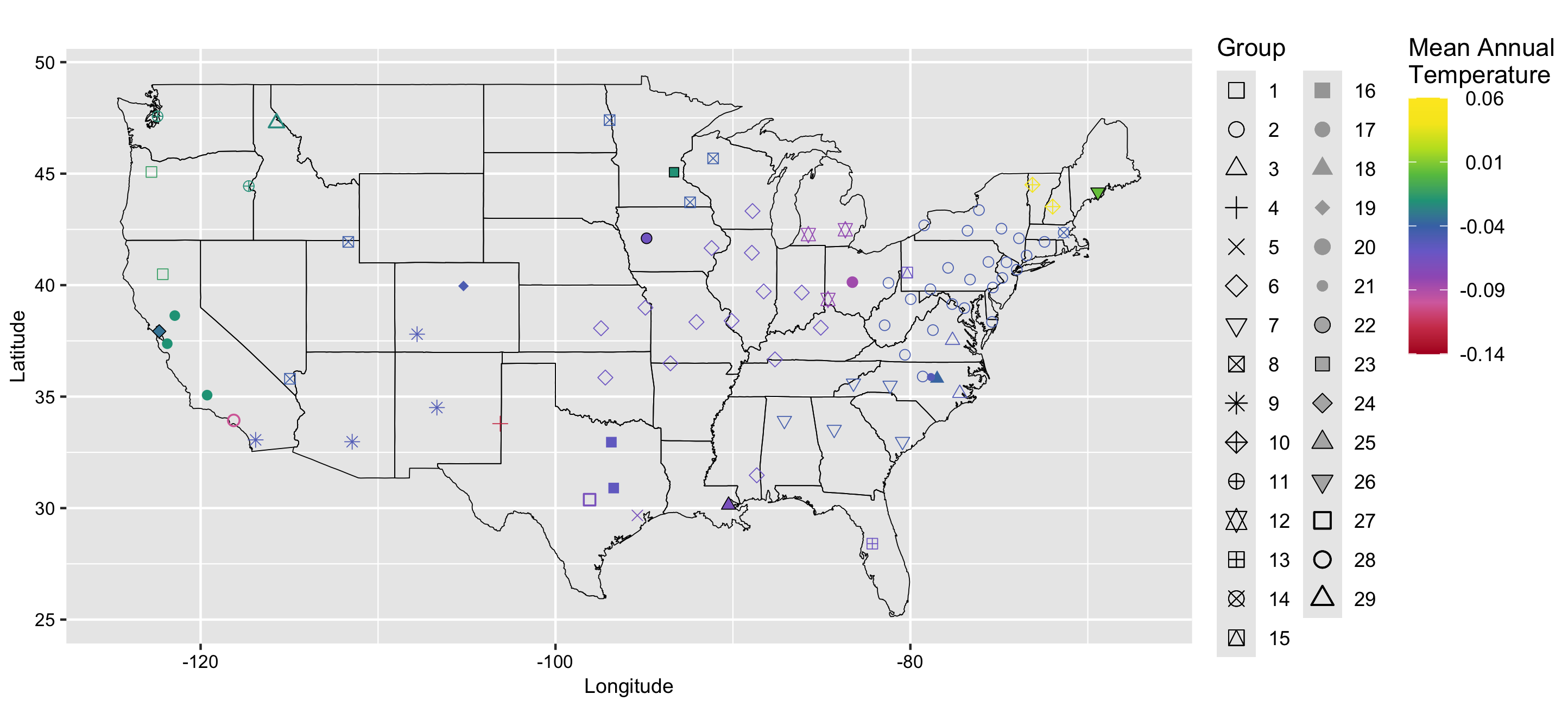}
    \subcaption{PIIC2 for Setting 1}
  \end{minipage}%
  \hspace{0.01\linewidth}%
  \begin{minipage}[b]{0.49\linewidth}
    \centering
    \includegraphics[width=\linewidth]{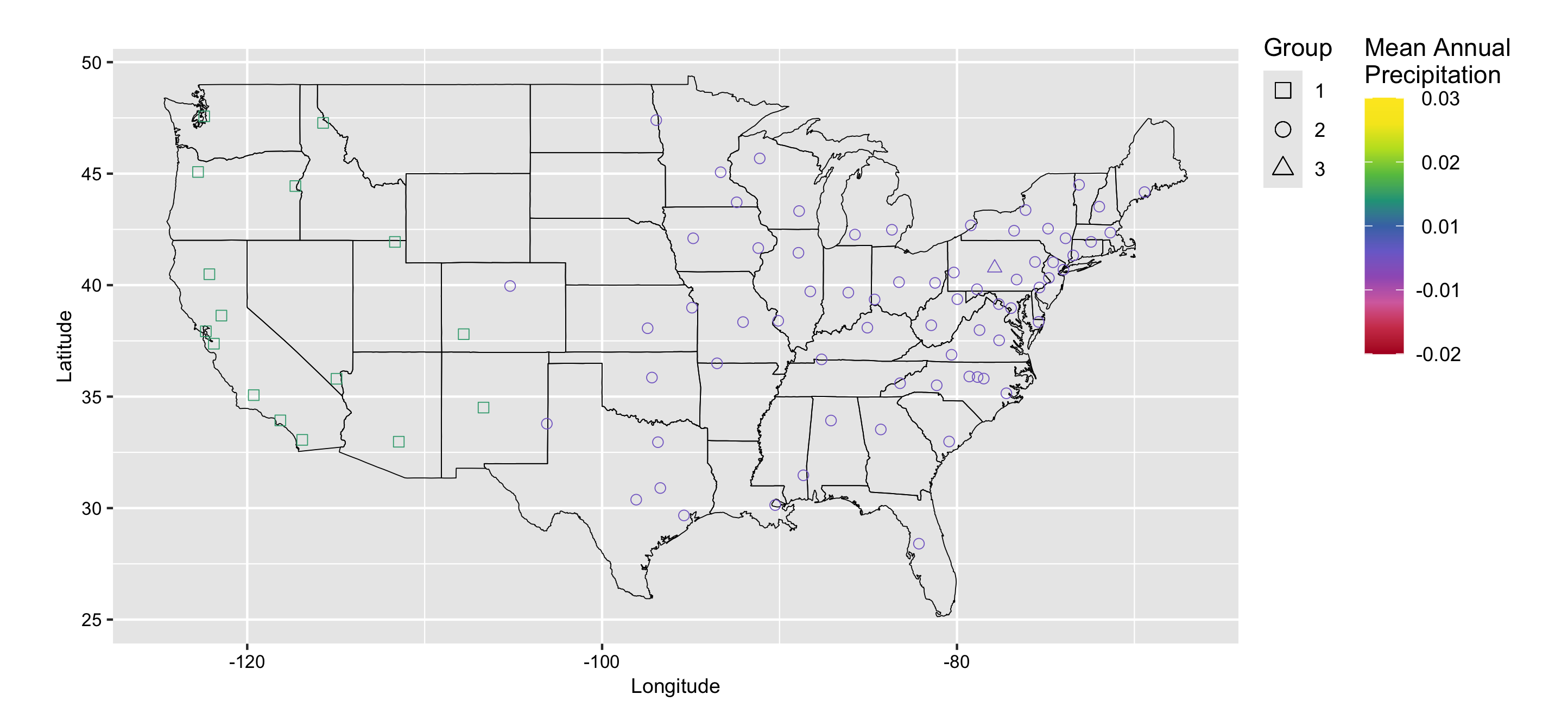}
    \subcaption{PIIC2 for Setting 2}
  \end{minipage}

\caption{Group structure in regression coefficient for Variable 1 of model selected by each information criterion.}
\label{v1}
\end{figure}

\begin{figure}[t]
  \centering

  \begin{minipage}[b]{0.49\linewidth}
    \centering
    \includegraphics[width=\linewidth]{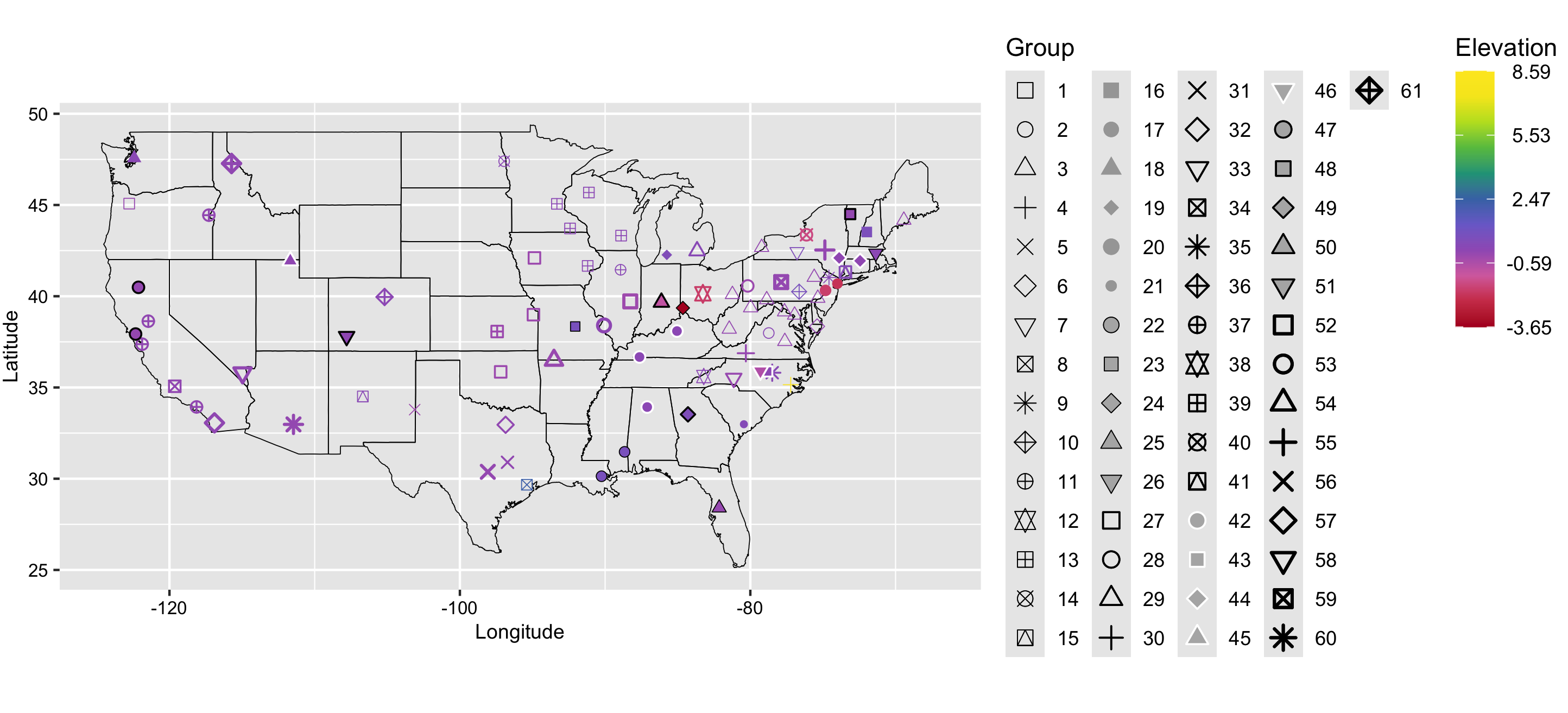}
    \subcaption{WAIC1 for Setting 1}
  \end{minipage}%
  \hspace{0.01\linewidth}%
  \begin{minipage}[b]{0.49\linewidth}
    \centering
    \includegraphics[width=\linewidth]{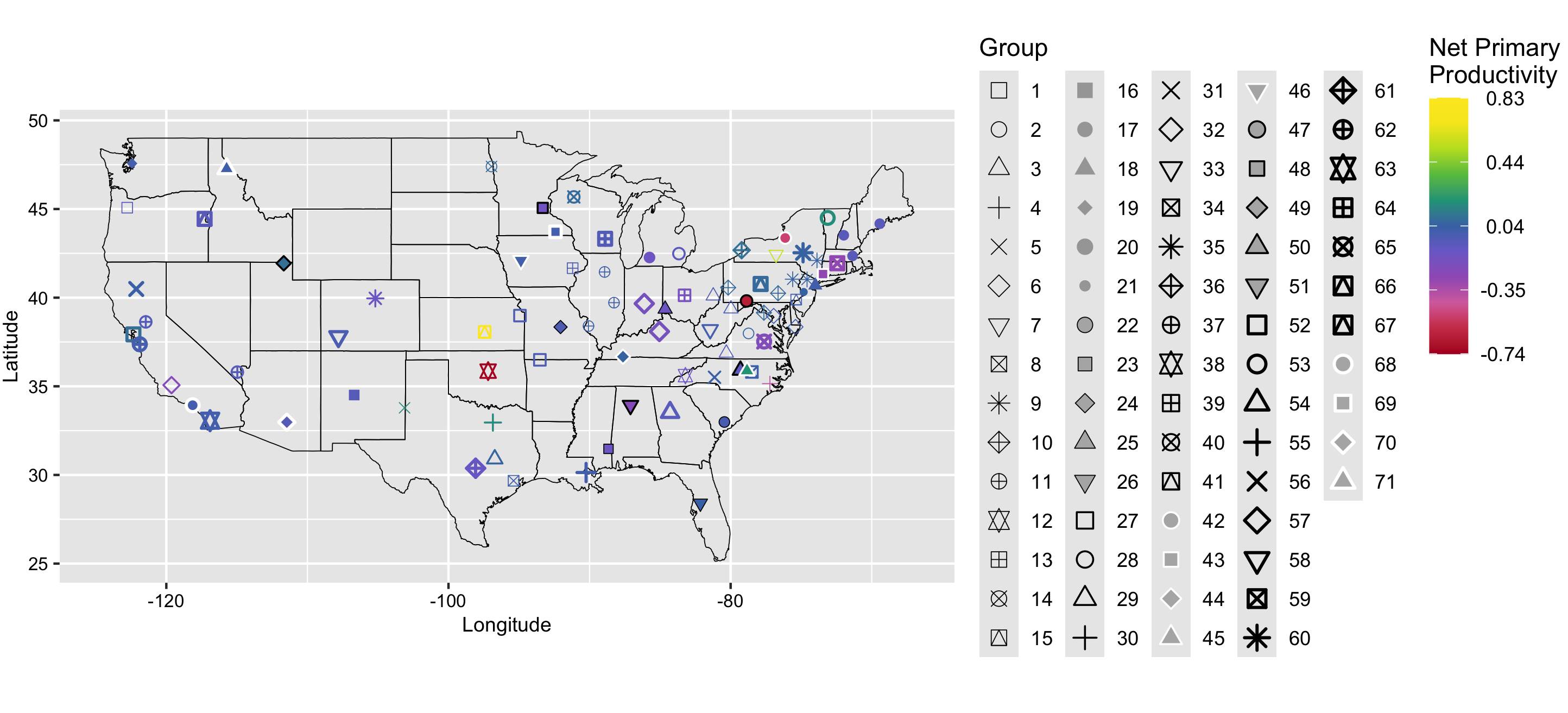}
    \subcaption{WAIC1 for Setting 2}
  \end{minipage}

  \vspace{1.2em}

  \begin{minipage}[b]{0.49\linewidth}
    \centering
    \includegraphics[width=\linewidth]{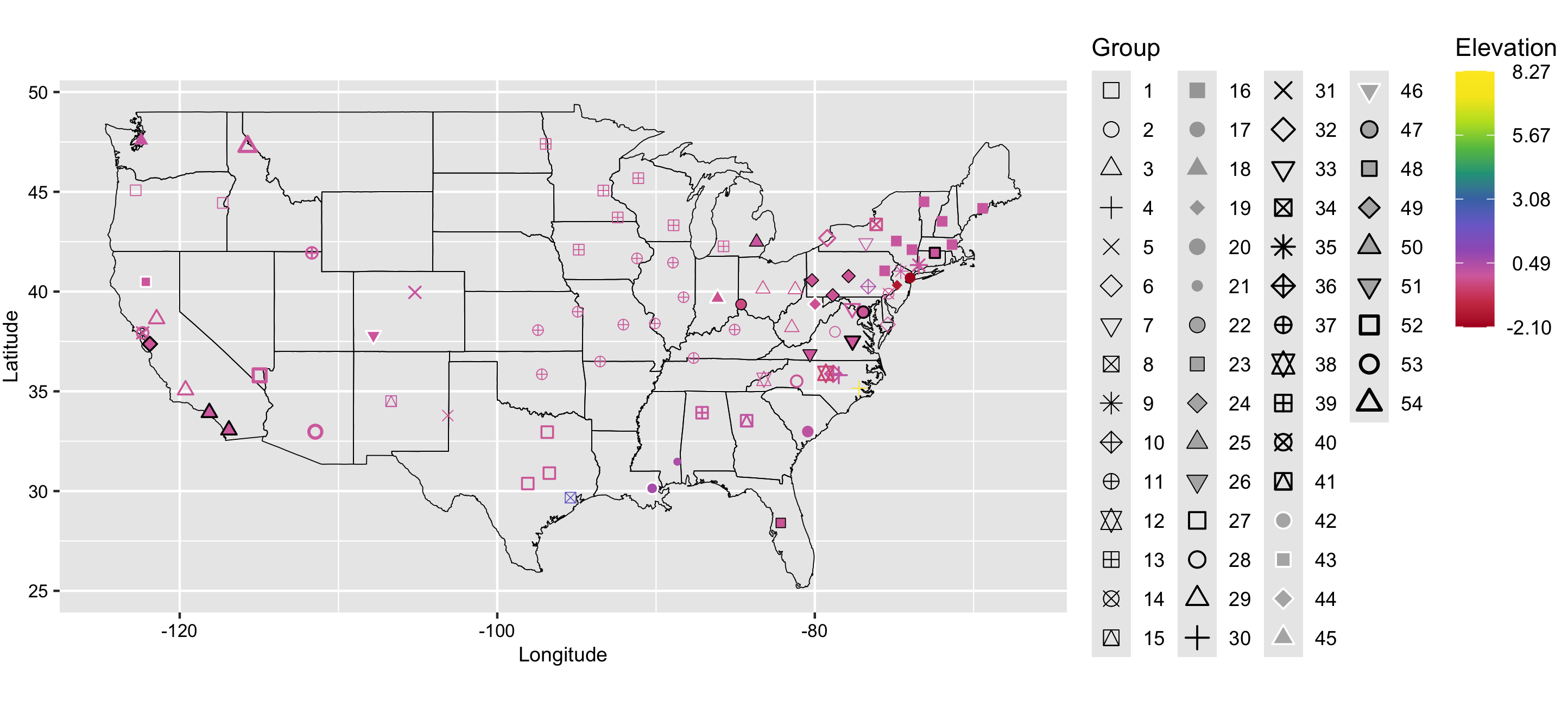}
    \subcaption{WAIC2 for Setting 1}
  \end{minipage}%
  \hspace{0.01\linewidth}%
  \begin{minipage}[b]{0.49\linewidth}
    \centering
    \includegraphics[width=\linewidth]{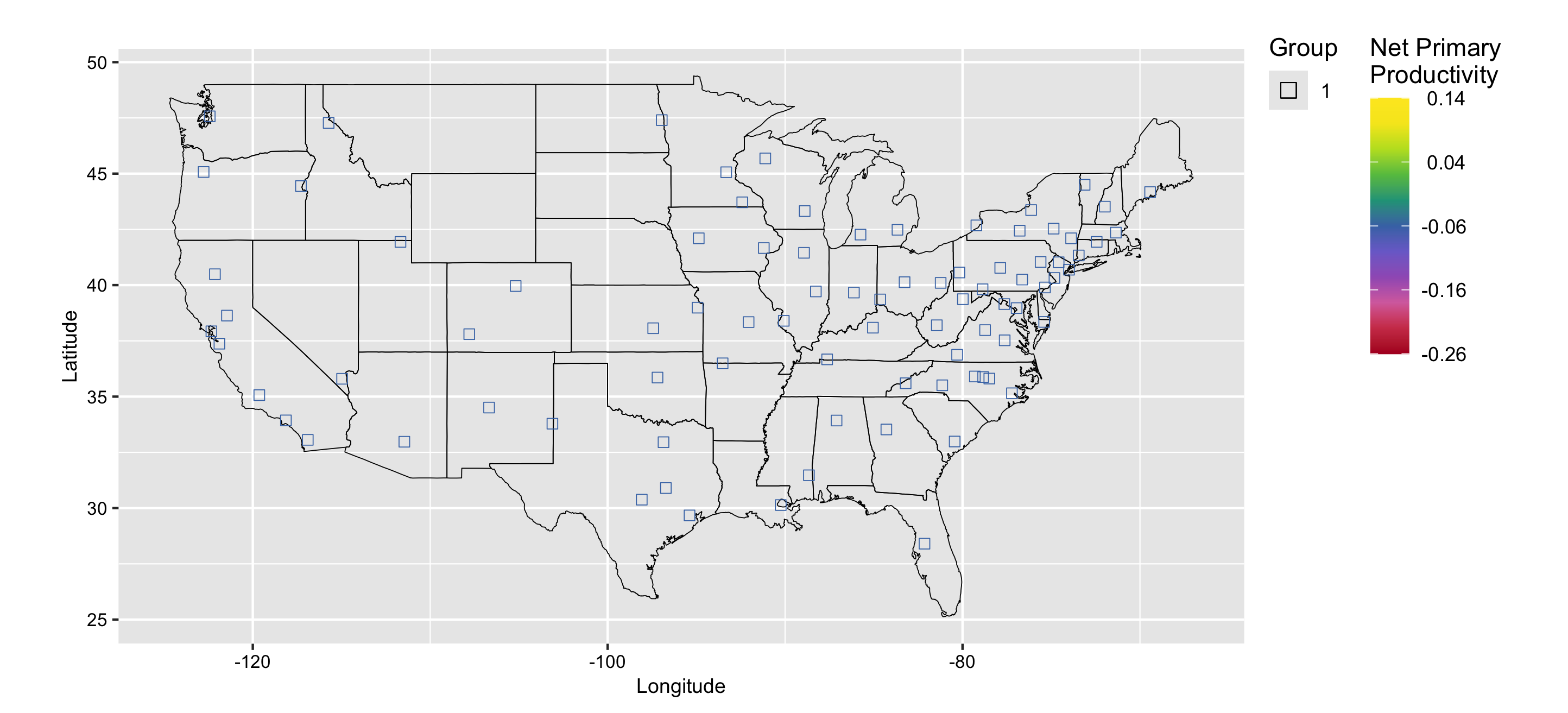}
    \subcaption{WAIC2 for Setting 2}
  \end{minipage}

  \vspace{1.2em}

  \begin{minipage}[b]{0.49\linewidth}
    \centering
    \includegraphics[width=\linewidth]{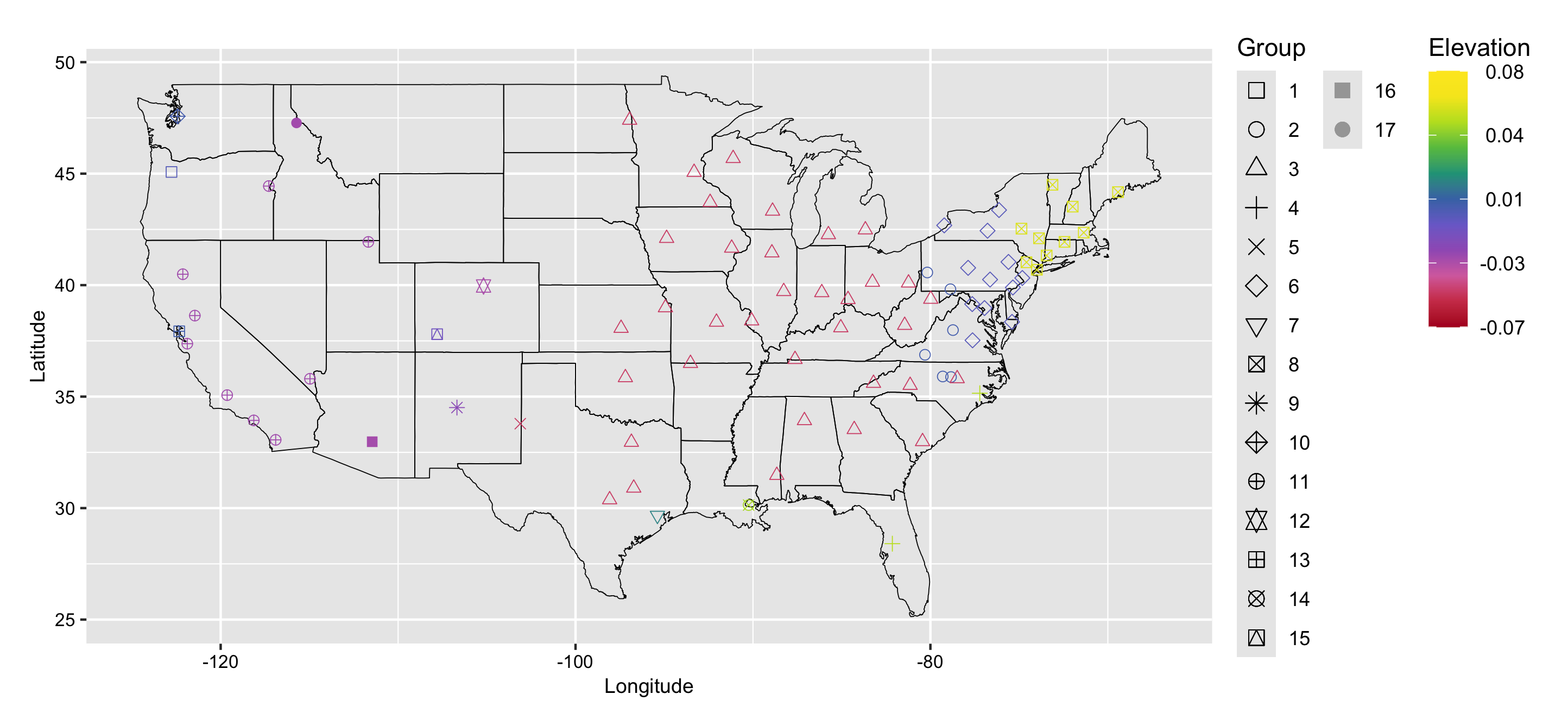}
    \subcaption{PIIC2 for Setting 1}
  \end{minipage}%
  \hspace{0.01\linewidth}%
  \begin{minipage}[b]{0.49\linewidth}
    \centering
    \includegraphics[width=\linewidth]{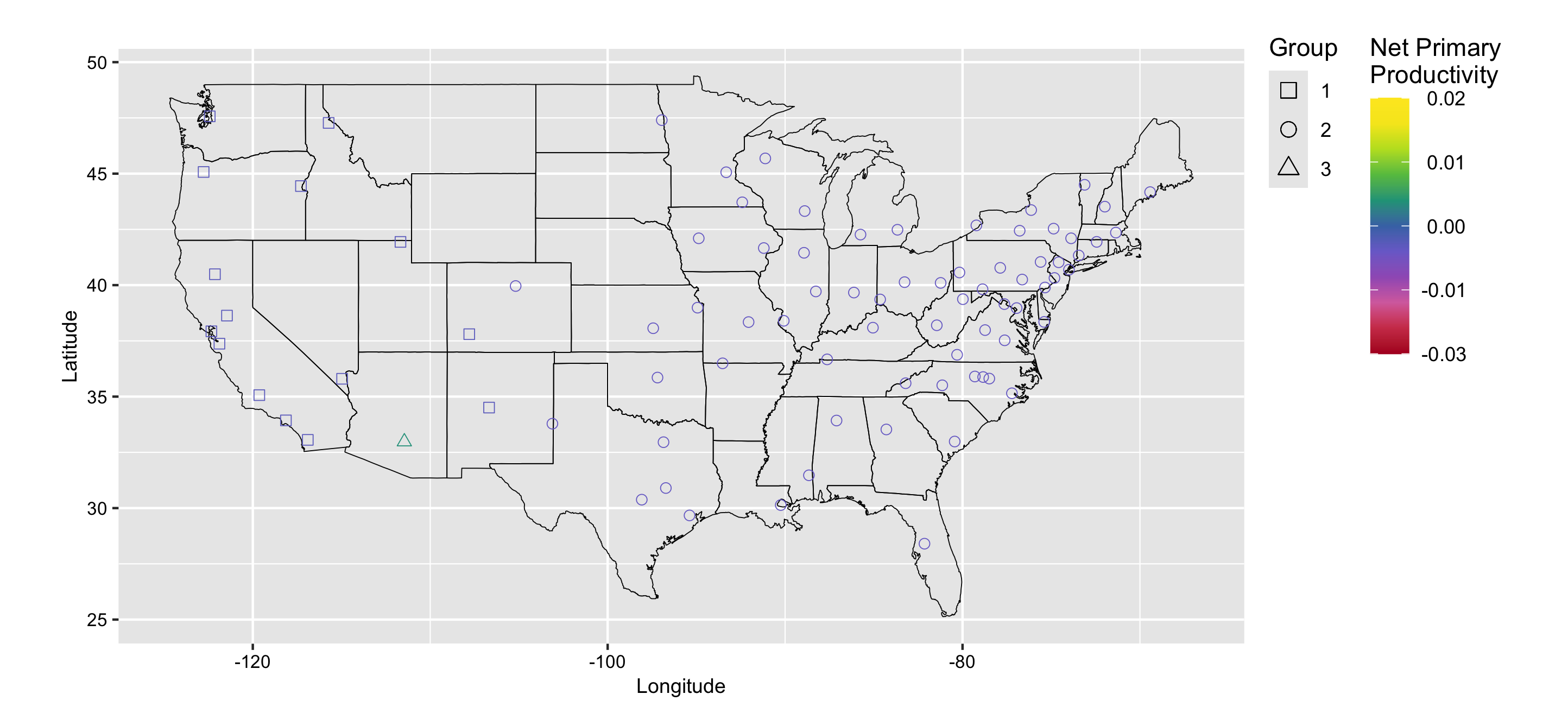}
    \subcaption{PIIC2 for Setting 2}
  \end{minipage}

\caption{Group structure in regression coefficient for Variable 2 of model selected by each information criterion.}
\label{v2}
\end{figure}

Figures \ref{intercept}, \ref{v1}, and \ref{v2} show the group structures of the intercept, the regression coefficients of Variable 1, and the regression coefficients of Variable 2 in the models selected by each information criterion, respectively.
In the figures, regions displayed as belonging to the same group represent those whose regression coefficients have been fused.
First, regarding WAIC1, in both Setting 1 and Setting 2, more than 60 groups are selected for the intercept, Variable 1, and Variable 2.
As a result, despite the application of sparse regularization, the estimation produces results with almost no sparsity.
Next, regarding WAIC2, for Setting 1, more than 40 groups are selected for the intercept and Variable 2, whereas for Setting 2, all regression coefficients are fused, so the number of groups becomes 1.
In the case of Setting 2, perhaps due to excessive regularization, the estimation produces results with no spatial heterogeneity, despite using the SVC model for capturing spatial heterogeneity.
PIIC2 gives moderate results: for Setting 1, it selects around 20 groups, and for Setting 2, it selects 2 or 3 groups.
When the number of groups is 2 or 3, it is easier to make interpretations regarding spatial heterogeneity.
In fact, the groups are roughly divided into the western United States and all other regions, and this is considered to relate to fungal diversity, since the West has low precipitation.
Moreover, because the West has high net primary productivity (NPP) due to soil characteristics and other reasons, the vegetation types are limited, which is also considered to relate to fungal diversity.

For the real data analysis, since the true structure is unknown, it is difficult to discuss which---the proposed method and the existing method---is superior.
What we would like to emphasize in this section is that the results differ considerably between WAIC and PIIC.
Although not in the sense of being inappropriate, WAIC provides models in which the selected number of groups is either too large or too small, making it difficult to interpret spatial heterogeneity, whereas PIIC gives models with moderate numbers of groups.

\section{Conclusion}\label{sec6}

The motivation behind this paper was a desire to examine whether PIIC---which was developed to overcome certain concerns with DIC and WAIC, both of which brought breakthroughs in model selection using the same concept as AIC, despite being Bayesian approaches---is truly useful in practically meaningful settings.
Accordingly, within spatial data analysis, we considered a setting where the Bayesian generalized fused lasso, one of the most appropriate estimation methods for providing model interpretation, is employed for SVC models in the broad sense, which have been developed in various directions to flexibly capture spatial heterogeneity.

The selection consistency of the generalized fused lasso estimator has been shown under an asymptotic setting in which the effect of regularization becomes almost negligible. However, in the derivation of information criteria, an asymptotic setting that instead retains the effect of regularization should be considered for the approximation of bias evaluation.
Under the latter asymptotic setting, we derived the properties of the generalized fused lasso estimator, and used them to adapt PIIC for the SVC model, in consideration of selecting classes of prior distributions with different complexities.
Through numerical experiments, we confirmed that, in settings where the class of prior distributions is also being selected, PIIC was superior to WAIC.
Specifically, in the natural problem setting of SVC models, whether to assume a common Laplace distribution for the regression coefficients of all explanatory variables or to assume different Laplace distributions for each regression coefficient, PIIC generally provided smaller risk than WAIC.
Furthermore, applying the methods to real data on house dust in the United States, we divided the regions of the United States into groups based on the Bayesian generalized fused lasso and confirmed that the number of groups estimated via WAIC and that via PIIC differ substantially.

In Bayesian regularization methods, it is becoming increasingly required to appropriately utilize prior distributions such as the horseshoe prior (\citealt{carvalho2010horseshoe}) and the normal-exponential-gamma distribution (\citealt{griffin2011bayesian}), which are said to improve predictive performance compared to the Laplace distribution, in accordance with the data (\citealt{polson2011}).
In addition, in applications where the explanatory variables have group structures, Bayesian versions of both group lasso, which attempts selection at the group level, and bi-level selection, which simultaneously attempts selection within groups (\citealt{xu2015bayesian}; \citealt{liquet2017bayesian}), have been developed.
Even when incorporating such methods, which are in a sense more complex than the Bayesian lasso, information criteria for hyperparameter selection are necessary, and in particular, information criteria for determining the number of hyperparameters are currently insufficient.
This is an important issue for future research.

\appendix

\section{Proof of Lemma \ref{lemma1}}\label{lemma1proof}

To develop the first-order asymptotics for the generalized fused lasso estimator $\hat{\bm{\xi}}$, we define the following random function:
\begin{align*}
u_n(\bm{\xi}) \equiv -\frac{1}{n} \sum_{i=1}^n\log f(y_i,\bm{x}_i\mid\bm{\xi})+\lambda_1\sum_{j\in V}|\xi_{j}|+\lambda_2\sum_{(j,k)\in E}|\xi_{j}-\xi_{k}|.
\end{align*}
It is obvious that $\hat{\bm{\xi}}=\argmin_{\bm{\xi}\in\Xi}u_n(\bm{\xi})$, and $u_n(\bm{\xi})$ converges in probability for each $\bm{\xi}$ to $h(\bm{\xi})+\lambda_1\sum_{j\in V}|\xi_j|+\lambda_2\sum_{(j^{\dagger},j^{\ddagger})\in E}|\xi_{j^{\dagger}}-\xi_{j^{\ddagger}}|$ from (R1).
Since $u_n(\bm{\xi})$ is convex with respect to $\bm{\xi}$, the convexity lemma of \cite{andersen1982cox} or \cite{pollard1991asymptotics} can be applied, and the conclusion is obtained.

\section{Proof of Theorem \ref{th1}}\label{theorem1proof}

The approach of the proof is close to that of \cite{viallon2016robustness}, but since the order of the penalty term of the generalized fused lasso is $\O(n)$ and the limit of the estimator is not even the true value, we also incorporate the approach of \cite{ninomiya2016aic}.
Let
$\mathcal{K}_1=\{(j^{\dagger},j^{\ddagger}):(j^{\dagger},j^{\ddagger})\in E,\ \xi_{j^{\dagger}}^*=\xi_{j^{\ddagger}}^*=0\}$,
$\mathcal{K}_2=\{(j^{\dagger},j^{\ddagger}):(j^{\dagger},j^{\ddagger})\in E,\ \xi_{j^{\dagger}}^*=0\neq\xi_{j^{\ddagger}}^*\}$,
$\mathcal{K}_3=\{(j^{\dagger},j^{\ddagger}):(j^{\dagger},j^{\ddagger})\in E,\ \xi_{j^{\dagger}}^*=\xi_{j^{\ddagger}}^*\neq0\}$, $\mathcal{K}_4=\{(j^{\dagger},j^{\ddagger}):(j^{\dagger},j^{\ddagger})\in E,\ 0\neq\xi_{j^{\dagger}}^*\neq\xi_{j^{\ddagger}}^*\neq 0\}$, $\mathcal{K}_5=\{(j^{\dagger},j^{\ddagger}):(j^{\dagger},j^{\ddagger})\in E,\ \xi_{j^{\dagger}}^*\neq0=\xi_{j^{\ddagger}}^*\}$.
Also, let $\mathcal{K}_{1,j}=\{k:(j,k)\in \mathcal{K}_1\}\cup\{k:(k,j)\in \mathcal{K}_1\}$, $\mathcal{K}_{2,j}=\{k:(j,k)\in \mathcal{K}_2\}\cup\{k:(k,j)\in \mathcal{K}_5\}$, $\mathcal{K}_{3,j}=\{k:(j,k)\in \mathcal{K}_3\}\cup\{k:(k,j)\in \mathcal{K}_3\}$, $\mathcal{K}_{4,j}=\{k:(j,k)\in \mathcal{K}_4\}\cup\{k:(k,j)\in \mathcal{K}_4\}$, $\mathcal{K}_{5,j}=\{k:(j,k)\in \mathcal{K}_5\}\cup\{k:(k,j)\in \mathcal{K}_2\}$.
Since $h(\bm{\theta})$ is a convex and differentiable function, from the Karush–Kuhn–Tucker (KKT) conditions and (C2), we obtain
\begin{align}
\xi_j^*=0 \quad \Rightarrow \quad \frac{\partial h}{\partial\xi_j}(\bm{\xi}^*) = & -\lambda_1 a_j-\lambda_2\Bigg(\sum_{k\in\mathcal{K}_{1,j}}\sgn(k-j)b_{j,k}-\sum_{k\in\mathcal{K}_{2,j}}\sgn(\xi_k^*)\Bigg),
\label{eq:kktzero} \\
\xi_j^*\neq 0 \quad \Rightarrow \quad \frac{\partial h}{\partial\xi_j}(\bm{\xi}^*) = & -\lambda_1\sgn(\xi_j^*)
\notag \\
& -\lambda_2\Bigg(\sum_{k\in\mathcal{K}_{3,j}}\sgn(k-j)b_{j,k}+\sum_{k\in\mathcal{K}_{4,j}\cup \mathcal{K}_{5,j}}\sgn(\xi_j^*-\xi_k^*)\Bigg).
\label{eq:kktnonzero}
\end{align}
In addition, $-1< a_j< 1$ and $-1< b_{j,k}< 1$.
Furthermore, letting $\mathcal{J}_1=\{j:\exists k;\ (j,k)\in \mathcal{K}_1\ \lor \ (k,j)\in \mathcal{K}_1\}$ and $\mathcal{J}_3=\{j:\exists k;\ (j,k)\in\mathcal{K}_3\ \lor\ (k,j)\in\mathcal{K}_3\}$, it holds that
\begin{align}
& b_{j,k}=b_{k,j} \quad (j\in\mathcal{J}_{1},\ k\in\mathcal{K}_{1,j}),
\label{eq:subdiff1}\\
& b_{j,k}=b_{k,j} \quad (j\in\mathcal{J}_3,\ k\in\mathcal{K}_{3,j}).
\label{eq:subdiff2}
\end{align}
Since $j\in\mathcal{J}_3$ and $k\in\mathcal{K}_{3,j}$ mean that $k\in\mathcal{J}_3$ and $j\in\mathcal{K}_{3,k}$, \eqref{eq:subdiff2} implies that
\begin{align}
\sum_{j\in\mathcal{J}_3}\sum_{k\in\mathcal{K}_{3,j}}\sgn(k-j)b_{j,k}=0.
\label{eq:kktsum1}
\end{align}

Now, recalling that for a vector $\bm{u}=(u_1,\ldots,u_p)^{\T}$, $\bm{u}^{(\ell)}$ represents $(u_j)_{j\in\mathcal{J}^{(\ell)}}$, we define the following random function:
\begin{align}
& v_n(\bm{u}^{(1)},\bm{u}^{(2)})
\notag \\
& \equiv \sum_{i=1}^{n}\bigg\{\log f(y_i,\bm{x}_i\mid\bm{\xi}^*)-\log f\bigg(y_i,\bm{x}_i\ \bigg|\ \frac{\bm{u}^{(1)}}{n}, \frac{\bm{u}^{(2)}}{\sqrt{n}}+{\bm{\xi}^*}^{(2)}\bigg)\bigg\}
\notag \\
& \phantom{\equiv} + n\lambda_1\bigg\|\frac{\bm{u}^{(1)}}{n}\bigg\|_1+n\lambda_1\bigg\|\frac{\bm{u}^{(2)}}{\sqrt{n}}+{\bm{\xi}^*}^{(2)}\bigg\|_1-n\lambda_1\|\bm{\xi}^*\|_1
\notag \\
& \phantom{\equiv} + n\lambda_2\sum_{(j^{\dagger},j^{\ddagger})\in \mathcal{K}_1}\bigg|\frac{u_{j^{\dagger}}-u_{j^{\ddagger}}}{n}\bigg|+n\lambda_2\sum_{(j^{\dagger},j^{\ddagger})\in\mathcal{K}_2}\bigg|\frac{u_{j^{\dagger}}}{n}-\frac{u_{j^{\ddagger}}}{\sqrt{n}}+{\xi_{j^{\dagger}}^*}-{\xi_{j^{\ddagger}}^*}\bigg|
\notag \\
& \phantom{\equiv} + n\lambda_2\sum_{(j^{\dagger},j^{\ddagger})\in\mathcal{K}_3}\bigg|\frac{u_{j^{\dagger}}-u_{j^{\ddagger}}}{\sqrt{n}}+\xi_{j^{\dagger}}^*-\xi_{j^{\ddagger}}^*\bigg|+n\lambda_2\sum_{(j^{\dagger},j^{\ddagger})\in\mathcal{K}_4}\bigg|\frac{u_{j^{\dagger}}-u_{j^{\ddagger}}}{\sqrt{n}}+\xi_{j^{\dagger}}^*-\xi_{j^{\ddagger}}^*\bigg|
\notag \\
& \phantom{\equiv} +n\lambda_2\sum_{\sgn(j^{\dagger},j^{\ddagger})\in \mathcal{K}_5}\bigg|\frac{u_{j^{\dagger}}}{\sqrt{n}}-\frac{u_{j^{\ddagger}}}{n}+\xi_{j^{\dagger}}^*-\xi_{j^{\ddagger}}^*\bigg|-n\lambda_2\sum_{\sgn(j^{\dagger},j^{\ddagger})\in E}|\xi_{j^{\dagger}}^*-\xi_{j^{\ddagger}}^*|.
\label{eq:v_n}
\end{align}
Note that $\argmin_{(\bm{u}^{(1)}, \bm{u}^{(2)})} v_n(\bm{u}^{(1)}, \bm{u}^{(2)})=(n \hat{\bm{\xi}}^{(1)}, \sqrt{n}(\hat{\bm{\xi}}^{(2)}-\bm{\xi}^{*(2)}))$.
In the following, we perform a Taylor expansion of $v_{n}(\bm{u}^{(1)}, \bm{u}^{(2)})$ around $(\bm{u}^{(1)}, \bm{u}^{(2)})=(\bm{0}, \bm{0})$; we consider the expansion in two cases.

\bigskip
\noindent
(i) The case of $u_{j^{\dagger}}=u_{j^{\ddagger}}$ for any $(j^{\dagger},j^{\ddagger})\in E$ such that $\xi_{j^{\dagger}}^*=\xi_{j^{\ddagger}}^*\neq 0$.
\bigskip

Let $\partial\log f(y_i,\bm{x}_i\mid\bm{\xi})/\partial\bm{\xi}$ be denoted by $\bm{g}_{y_i,\bm{x}_i}^{\prime}(\bm{\xi})$, and $\partial^2\log f(y_i,\bm{x}_i\mid\bm{\xi})/\partial\bm{\xi}\partial\bm{\xi}^{\T}$ be denoted by $\bm{G}_{y_i,\bm{x}_i}^{\prime\prime}(\bm{\xi})$, and recalling also that $\bm{G}_{y_i,\bm{x}_i}^{\prime\prime(\ell^{\dagger}\ell^{\ddagger})}(\bm{\xi})$ represents $(\bm{G}_{y_i,\bm{x}_i}^{\prime\prime}(\bm{\xi})_{j^{\dagger}j^{\ddagger}})_{j^{\dagger}\in\mathcal{J}^{(\ell^{\dagger})},j^{\ddagger}\in\mathcal{J}^{(\ell^{\ddagger})}}$, \eqref{eq:v_n} can be rewritten as
\begin{align}
& v_n(\bm{u}^{(1)}, \bm{u}^{(2)})
\notag \\
& = -\sum_{i=1}^n\bigg\{\bm{g}_{y_i,\bm{x}_i}^{\prime(1)}(\bm{\xi}^*)^{\T} \frac{\bm{u}^{(1)}}{n}+\bm{g}_{y_i,\bm{x}_i}^{\prime(2)}(\bm{\xi}^*)^{\T} \frac{\bm{u}^{(2)}}{\sqrt{n}}\bigg\} + \lambda_1\sum_{j\in \mathcal{J}^{(1)}}|u_j| + \sqrt{n} \lambda_1 \bm{u}^{(2) \T} \sgn(\bm{\xi}^{*(2)})
\notag \\
& \ \phantom{=} +\lambda_2\sum_{(j^{\dagger},j^{\ddagger})\in \mathcal{K}_1}|u_{j^{\dagger}}-u_{j^{\ddagger}}|+\lambda_2\sum_{(j^{\dagger},j^{\ddagger})\in \mathcal{K}_2}\sgn(\xi_{j^{\dagger}}^*-\xi_{j^{\ddagger}}^*)u_{j^{\dagger}}-\sqrt{n}\lambda_2 \sum_{(j^{\dagger},j^{\ddagger})\in \mathcal{K}_2}\sgn(\xi_{j^{\dagger}}^*-\xi_{j^{\ddagger}}^*)u_{j^{\ddagger}}
\notag \\
& \ \phantom{=} +\sqrt{n}\lambda_2 \sum_{(j^{\dagger},j^{\ddagger})\in \mathcal{K}_4}\sgn(\xi_{j^{\dagger}}^*-\xi_{j^{\ddagger}}^*)(u_{j^{\dagger}}-u_{j^{\ddagger}})
\notag \\
& \ \phantom{=} +\sqrt{n}\lambda_2\sum_{(j^{\dagger},j^{\ddagger})\in \mathcal{K}_5}\sgn(\xi_{j^{\dagger}}^*-\xi_{j^{\ddagger}}^*)u_{j^{\dagger}}-\lambda_2 \sum_{(j^{\dagger},j^{\ddagger})\in \mathcal{K}_5}\sgn(\xi_{j^{\dagger}}^*-\xi_{j^{\ddagger}}^*)u_{j^{\ddagger}}
\notag \\
& \ \phantom{=} -\sum_{i=1}^n\bigg[\frac{\bm{u}^{(1)\T}}{n}\bigg\{\bm{G}_{y_i,\bm{x}_i}^{\prime\prime(11)}(\bm{\xi}^*) \frac{\bm{u}^{(1)}}{2n}+\bm{G}_{y_i,\bm{x}_i}^{\prime\prime(12)}(\bm{\xi}^*) \frac{\bm{u}^{(2)}}{\sqrt{n}}\bigg\}+\frac{1}{2} \frac{\bm{u}^{(2)\T}}{\sqrt{n}} \bm{G}_{y_i,\bm{x}_i}^{\prime\prime(22)}(\bm{\xi}^*) \frac{\bm{u}^{(2)}}{\sqrt{n}}\bigg] + \oP(1),
\label{taylor1}
\end{align}
where $\sgn(\bm{\xi}^{(2)})$ is a vector whose elements are $\sgn(\xi_j^*)\ (j\in \mathcal{J}^{(2)})$.
Now, defining $\mathcal{J}_{2}=\{j:\exists k;\ (j,k)\in \mathcal{K}_2\ \lor \ (k,j)\in \mathcal{K}_5\}$, $\mathcal{J}_{4}=\{j:\exists k;\ (j,k)\in \mathcal{K}_4\ \lor \ (k,j)\in \mathcal{K}_4\}$, and $\mathcal{J}_{5}=\{j:\exists k;\ (k,j)\in \mathcal{K}_2\ \lor \ (j,k)\in \mathcal{K}_5\}$, the fifth and ninth terms on the right-hand side can be combined as
\begin{align*}
\lambda_2\sum_{(j^{\dagger},j^{\ddagger})\in \mathcal{K}_2}\sgn(-\xi_{j^{\ddagger}}^*)u_{j^{\dagger}} - \lambda_2\sum_{(j^{\dagger},j^{\ddagger})\in \mathcal{K}_5}\sgn(\xi_{j^{\dagger}}^*)u_{j^{\ddagger}} = -\lambda_2\sum_{j\in\mathcal{J}_{2}}\sum_{k\in \mathcal{K}_{2,j}}\sgn(\xi_k^*)u_j,
\end{align*}
the sixth and eighth terms can be combined as
\begin{align*}
-\sqrt{n}\lambda_2\sum_{(j^{\dagger},j^{\ddagger})\in \mathcal{K}_2}\sgn(-\xi_{j^{\ddagger}}^*)u_{j^{\ddagger}} + \sqrt{n}\lambda_2\sum_{(j^{\dagger},j^{\ddagger})\in \mathcal{K}_5}\sgn(\xi_{j^{\dagger}}^*)u_{j^{\dagger}} = \sqrt{n}\lambda_2\sum_{j\in\mathcal{J}_{5}}\sum_{k\in\mathcal{K}_{5,j}}\sgn(\xi_j^*)u_j,
\end{align*}
and the seventh term can be rewritten as
$\sqrt{n}\lambda_2\sum_{j\in\mathcal{J}_{4}}\sum_{k\in \mathcal{K}_{4,j}}\allowbreak\sgn(\xi_j^*-\xi_k^*)u_j$.
By also using the fact that the quadratic term involving $\bm{u}^{(1)}$ is $\oP(1)$, after rearrangement, \eqref{taylor1} becomes
\begin{align}
& v_n(\bm{u}^{(1)}, \bm{u}^{(2)})
\notag \\
& = -\sum_{i=1}^n\bigg\{\bm{g}_{y_i,\bm{x}_i}^{\prime(1)}(\bm{\xi}^*)^{\T} \frac{\bm{u}^{(1)}}{n}+\bm{g}_{y_i,\bm{x}_i}^{\prime(2)}(\bm{\xi}^*)^{\T} \frac{\bm{u}^{(2)}}{\sqrt{n}}\bigg\} + \lambda_1\sum_{j\in \mathcal{J}^{(1)}}|u_j| + \sqrt{n} \lambda_1 \bm{u}^{(2) \T} \sgn(\bm{\xi}^{*(2)}) 
\notag \\
& \ \phantom{=} + \lambda_2\sum_{(j^{\dagger},j^{\ddagger})\in\mathcal{K}_1}|u_{j^{\dagger}}-u_{j^{\ddagger}}|-\lambda_2\sum_{j\in\mathcal{J}_{2}}\sum_{k\in \mathcal{K}_{2,j}}\sgn(\xi_k^*)u_j + \sqrt{n}\lambda_2\sum_{j\in\mathcal{J}_{4}}\sum_{k\in \mathcal{K}_{4,j}}\sgn(\xi_j^*-\xi_k^*)u_j
\notag \\
& \ \phantom{=} + \sqrt{n}\lambda_2\sum_{j\in\mathcal{J}_{5}}\sum_{k\in\mathcal{K}_{5,j}}\sgn(\xi_j^*)u_j - \frac{1}{2}\sum_{i=1}^n\frac{\bm{u}^{(2)\T}}{\sqrt{n}} \bm{
G}_{y_i,\bm{x}_i}^{\prime\prime(22)}(\bm{\xi}^*) \frac{\bm{u}^{(2)}}{\sqrt{n}} + \oP(1).
\label{taylor2}
\end{align}
Next, from (R2), we have that $-\sum_{i=1}^n\bm{g}_{y_i,\bm{x}_i}^{\prime(1)}(\bm{\xi}^*)/n$ converges in probability to $(\partial h/\partial \bm{\xi}^{(1)})(\bm{\xi}^*)$. 
In addition, from (R2), (R3), and \eqref{eq:kktnonzero}, it follows that there exists a $|\mathcal{J}^{(2)}|$-dimensional random vector $\bm{s}^{(2)}$ following $\N(\bm{0}, \bm{J}^{(22)})$, and that
$\sum_{i=1}^n\bm{g}_{y_i,\bm{x}_i}^{\prime(2)}(\bm{\xi}^*)/\sqrt{n}-\sqrt{n}\{\lambda_1\sgn(\bm{\xi}^{*(2)})+\lambda_2(\sum_{k\in\mathcal{K}_{3,j}} \allowbreak \sgn(k-j)b_{j,k})_{j\in\mathcal{J}^{(2)}}+\lambda_2(\sum_{k\in\mathcal{K}_{4,j}\cup\mathcal{K}_{5,j}}\sgn(\xi_j^*-\xi_k^*))_{j\in\mathcal{J}^{(2)}}\}$ converges in distribution to $\bm{s}^{(2)}$.
Furthermore, from (C1), we also have that $-\sum_{i=1}^n(\bm{u}^{(2)}/\sqrt{n})^{\T}\bm{G}_{y_i,\bm{x}_i}^{\prime\prime(22)}(\bm{\xi}^*)(\bm{u}^{(2)}/\sqrt{n})$ converges to $\bm{u}^{(2)\T}\bm{J}^{(22)}\bm{u}^{(2)}$.
Therefore, for each $(\bm{u}^{(1)},\bm{u}^{(2)})$, \eqref{taylor2} converges in distribution to
\begin{align*}
& v(\bm{u}^{(1)}, \bm{u}^{(2)})
\notag \\
& = \sum_{j\in\mathcal{J}^{(1)}}\bigg\{\frac{\partial h}{\partial \xi_j}(\bm{\xi}^*)u_j+\lambda_1|u_j|\bigg\} + \lambda_2\sum_{(j^{\dagger},j^{\ddagger})\in\mathcal{K}_1}|u_{j^{\dagger}}-u_{j^{\ddagger}}| - \lambda_2\sum_{j\in\mathcal{J}_{2}}\sum_{k\in \mathcal{K}_{2,j}}\sgn(\xi_k^*)u_j
\notag \\
& \ \phantom{=} -\bm{u}^{(2)\T}\bm{s}^{(2)} - \sqrt{n}\lambda_2\sum_{j\in\mathcal{J}^{(2)}}\sum_{k\in\mathcal{K}_{3,j}}\sgn(k-j)b_{j,k}u_j
+\frac{1}{2}\bm{u}^{(2)\T}\bm{J}^{(22)}\bm{u}^{(2)}.
\end{align*}
By transforming the sum of the first and third terms on the right-hand side using \eqref{eq:kktzero}, we obtain
\begin{align}
& v(\bm{u}^{(1)}, \bm{u}^{(2)})
\notag \\
& = -\lambda_1\sum_{j\in\mathcal{J}^{(1)}}a_ju_j - \lambda_2\sum_{j\in\mathcal{J}^{(1)}}\sum_{k\in\mathcal{K}_{1,j}}\sgn(k-j)b_{j,k}u_j + \lambda_1\sum_{j\in\mathcal{J}^{(1)}}|u_j| + \lambda_2\sum_{(j^{\dagger},j^{\ddagger})\in\mathcal{K}_1}|u_{j^{\dagger}}-u_{j^{\ddagger}}|
\notag \\
& \ \phantom{=} -\bm{u}^{(2)\T}\bm{s}^{(2)} - \sqrt{n}\lambda_2\sum_{j\in\mathcal{J}^{(2)}}\sum_{k\in\mathcal{K}_{3,j}}\sgn(k-j)b_{j,k}u_j
+\frac{1}{2}\bm{u}^{(2)\T}\bm{J}^{(22)}\bm{u}^{(2)}.
\label{eq:conv1}
\end{align}
From \eqref{eq:subdiff1}, the second term on the right-hand side can be rewritten as $-\lambda_2\sum_{(j^{\dagger},j^{\ddagger})\in \mathcal{K}_1}b_{j^{\dagger},j^{\ddagger}}(u_{j^{\dagger}}-u_{j^{\ddagger}})$.
Regarding the sixth term, $\sum_{j\in\mathcal{J}^{(2)}}$ can be replaced with $\sum_{j\in\mathcal{J}_3}$, and in the present case, all $u_j\ (j\in\mathcal{J}_3)$ take the same value; hence, by \eqref{eq:kktsum1}, the sum equals $0$.
By also applying this property of $u_j$ to the fifth and seventh terms, \eqref{eq:conv1} can be rewritten using the matrix $\bm{A}$ defined in \eqref{defA} as
\begin{align}
& v(\bm{u}^{(1)}, \bm{u}^{(2)})
\notag \\
& = \lambda_1\sum_{j\in\mathcal{J}^{(1)}}(|u_j|-a_ju_j) + \lambda_2\sum_{(j^{\dagger},j^{\ddagger})\in\mathcal{K}_1}\{|u_{j^{\dagger}}-u_{j^{\ddagger}}|-b_{j^{\dagger},j^{\ddagger}}(u_{j^{\dagger}}-u_{j^{\ddagger}})\}
\notag \\
& \ \phantom{=} - \bm{u}^{(3)\T}\bm{A}^{(32)}\bm{s}^{(2)} + \frac{1}{2}\bm{u}^{(3)\T}\bm{A}^{(32)}\bm{J}^{(22)}\bm{A}^{(23)}\bm{u}^{(3)}. 
\label{eq:conv2}
\end{align}
Noting that $-1 < a_j < 1$ and $-1 < b_{j^{\dagger}, j^{\ddagger}} < 1$, it follows that $v(\bm{u}^{(1)}, \bm{u}^{(2)})$ has a unique minimum at $\bm{u}^{(1)}=\bm{0}$ and $\bm{u}^{(3)}=(\bm{A}^{(32)}\bm{J}^{(22)}\bm{A}^{(23)})^{-1}\bm{A}^{(32)}\bm{s}^{(2)}$.

\bigskip
\noindent
(ii) The case other than (i)
\bigskip

Equation \eqref{eq:v_n} can be rewritten as \eqref{taylor1} with the addition of $\sqrt{n}\lambda_2\sum_{(j^{\dagger},j^{\ddagger})\in \mathcal{K}3}|u{j^{\dagger}}-u_{j^{\ddagger}}|$, which, according to \eqref{eq:conv1}, converges in distribution to \begin{align}
v(\bm{u}^{(1)}, \bm{u}^{(2)}) = & -\lambda_1\sum_{j\in\mathcal{J}^{(1)}}a_ju_j - \lambda_2\sum_{j\in\mathcal{J}^{(1)}}\sum_{k\in\mathcal{K}_{1,j}}\sgn(k-j)b_{j,k}u_j + \lambda_1\sum_{j\in\mathcal{J}^{(1)}}|u_j|
\notag \\
& + \lambda_2\sum_{\sgn(j^{\dagger},j^{\ddagger})\in \mathcal{K}_1}|u_{j^{\dagger}}-u_{j^{\ddagger}}|+\sqrt{n}\lambda_2\sum_{\sgn(j^{\dagger},j^{\ddagger})\in \mathcal{K}_3}|u_{j^{\dagger}}-u_{j^{\ddagger}}|
\notag \\
& - \bm{u}^{(2)\T}\bm{s}^{(2)} - \sqrt{n}\lambda_2\sum_{j\in\mathcal{J}^{(2)}}\sum_{k\in\mathcal{K}_{3,j}}\sgn(k-j)b_{j,k}u_j
+\frac{1}{2}\bm{u}^{(2)\T}\bm{J}^{(22)}\bm{u}^{(2)}.
\label{eq:conv3}
\end{align}
The seventh term on the right-hand side can also be expressed as $-\sqrt{n}\lambda_2\sum_{(j^{\dagger},j^{\ddagger})\in \mathcal{K}_3} b_{j^{\dagger},j^{\ddagger}} (u_{j^{\dagger}} - u_{j^{\ddagger}})$ according to \eqref{eq:subdiff2}. Therefore, just as in deriving \eqref{eq:conv2}, \eqref{eq:conv3} can be rewritten as
\begin{align}
& v(\bm{u}^{(1)}, \bm{u}^{(2)})
\notag \\
& = \lambda_1\sum_{j\in\mathcal{J}^{(1)}}(|u_j|-a_ju_j) + \lambda_2\sum_{(j^{\dagger},j^{\ddagger})\in\mathcal{K}_1}\{|u_{j^{\dagger}}-u_{j^{\ddagger}}|-b_{j^{\dagger},j^{\ddagger}}(u_{j^{\dagger}}-u_{j^{\ddagger}})\}
\notag \\
& \ \phantom{=} + \sqrt{n}\lambda_2\sum_{(j^{\dagger},j^{\ddagger})\in\mathcal{K}_3}\{|u_{j^{\dagger}}-u_{j^{\ddagger}}|-b_{j^{\dagger},j^{\ddagger}}(u_{j^{\dagger}}-u_{j^{\ddagger}})\}
\notag \\
& \ \phantom{=} - \bm{u}^{(3)\T}\bm{A}^{(32)}\bm{s}^{(2)} + \frac{1}{2}\bm{u}^{(3)\T}\bm{A}^{(32)}\bm{J}^{(22)}\bm{A}^{(23)}\bm{u}^{(3)}. 
\label{eq:conv5}
\end{align}
In the present case, there exists $(j^{\dagger},j^{\ddagger}) \in \mathcal{K}_3$ such that $u_{j^{\dagger}} \neq u_{j^{\ddagger}}$. Therefore, the third term on the right-hand side is positive and $\OP(\sqrt{n})$. On the other hand, all other terms are $\OP(1)$; hence, it follows that \eqref{eq:conv5} diverges to positive infinity.

\bigskip

From (i) and (ii), it follows that $v(\bm{u}^{(1)},\bm{u}^{(2)})$ has the unique minimizer given by 
\begin{align}
& \bm{u}^{(1)} = \bm{0},
\notag \\
& j^{\dagger}\in\mathcal{J}^{(2)} \quad \Rightarrow \quad \exists j^{\ddagger}\in\mathcal{J}^{(3)};\ u_{j^{\dagger}} = u_{j^{\ddagger}},
\notag \\
& \bm{u}^{(3)} = (\bm{A}^{(32)}\bm{J}^{(22)}\bm{A}^{(23)})^{-1}\bm{A}^{(32)}\bm{s}^{(2)}.
\label{base}
\end{align}
Since $v_n(\bm{u}^{(1)},\bm{u}^{(2)})$ is convex, it follows from the convexity lemma of \cite{hjort2011asymptotics} or \cite{geyer1996asymptotics} that 
\begin{align*}
\argmin_{\bm{u}^{(1)},\bm{u}^{(2)}} v_{n}(\bm{u}^{(1)},\bm{u}^{(2)})&\xrightarrow{\rm d}\argmin_{\bm{u}^{(1)},\bm{u}^{(2)}} v(\bm{u}^{(1)},\bm{u}^{(2)}),
\end{align*}
and hence the conclusion follows.

\section*{Acknowledgment}

Yuko Kakikawa acknowledges support from JST SPRING (grant number JPMJSP2104), and Yoshiyuki Ninomiya acknowledges support from JSPS KAKENHI (grant numbers 23H00809 and 23K18471).

\section*{Conflict of interest}
The authors declare no potential conflict of interests.

\section*{Data Availability Statement}
The data analyzed in this study were collected through the Wild Life of Our Homes project and are openly available from https://figshare.com/articles/dataset/1000homes/1270900.

\bibliographystyle{sn-basic}
\bibliography{ref}

\end{document}